\documentclass[prd,aps,twocolumn,preprintnumbers, showpacs, nofootinbib,superscriptaddress,notitlepage]{revtex4-2}
\usepackage{mathrsfs}
\usepackage{amsfonts}
\usepackage{amsmath}
\usepackage{slashed}
\usepackage{array}
\usepackage{verbatim}
\usepackage{epsfig}
\usepackage{graphicx}
\usepackage{color}
\usepackage[dvipsnames]{xcolor}
\usepackage[colorlinks,linkcolor=red,anchorcolor=green,citecolor=blue,CJKbookmarks=True]{hyperref}
\definecolor{red}{rgb}{1,0,0}

\usepackage{multirow}

\newcommand{\beq}{\begin{eqnarray}}
\newcommand{\eeq}{\end{eqnarray}}

\def\be{\begin{equation}}
\def\ee{\end{equation}}
\def\bea{\begin{eqnarray}}
\def\eea{\end{eqnarray}}
\def\bal#1\eal{\begin{align}#1\end{align}}

\begin{document}

\title{Accurate B meson and Bottomonium masses and decay constants from the tadpole improved clover ensembles}

\collaboration{\bf{CLQCD Collaboration}}

\author{
\includegraphics[scale=0.30]{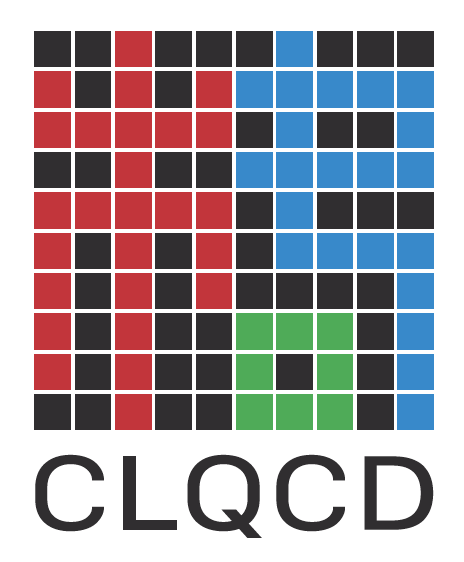}\\
Mengchu Cai}
\affiliation{CAS Key Laboratory of Theoretical Physics, Institute of Theoretical Physics, Chinese Academy of Sciences, Beijing 100190, China}

\author{
Hai-Yang Du}
\affiliation{CAS Key Laboratory of Theoretical Physics, Institute of Theoretical Physics, Chinese Academy of Sciences, Beijing 100190, China}
\affiliation{University of Chinese Academy of Sciences, School of Physical Sciences, Beijing 100049, China}

\author{
Xiangyu Jiang}
\affiliation{CAS Key Laboratory of Theoretical Physics, Institute of Theoretical Physics, Chinese Academy of Sciences, Beijing 100190, China}

\author{Peng Sun}
\email[Corresponding author: ]{pengsun@impcas.ac.cn}
\affiliation{Institute of Modern Physics, Chinese Academy of Sciences, Lanzhou, 730000, China}

\author{
Wei Sun}
\affiliation{Institute of High Energy Physics, Chinese Academy of Sciences, Beijing 100049, China}

\author{
Ji-Hao Wang}
\affiliation{CAS Key Laboratory of Theoretical Physics, Institute of Theoretical Physics, Chinese Academy of Sciences, Beijing 100190, China}
\affiliation{University of Chinese Academy of Sciences, School of Physical Sciences, Beijing 100049, China}

\author{Yi-Bo Yang}
\email[Corresponding author: ]{ybyang@itp.ac.cn}
\affiliation{University of Chinese Academy of Sciences, School of Physical Sciences, Beijing 100049, China}
\affiliation{CAS Key Laboratory of Theoretical Physics, Institute of Theoretical Physics, Chinese Academy of Sciences, Beijing 100190, China}
\affiliation{School of Fundamental Physics and Mathematical Sciences, Hangzhou Institute for Advanced Study, UCAS, Hangzhou 310024, China}
\affiliation{International Centre for Theoretical Physics Asia-Pacific, Beijing/Hangzhou, China}

\date{\today}

\begin{abstract}
We present a determination of the bottom quark mass, the masses of S-wave bottom mesons, and their decay constants using an anisotropic clover fermion discretization for the heavy quark, on $2+1$ flavor isotropic QCD ensembles. Our analysis is based on 16 ensembles spanning 6 lattice spacings, with pion masses in the range of 135--350 MeV and several values of the strange quark mass. We demonstrate that the effective anisotropy parameter for the heavy quark approaches unity with controllable $\mathcal{O}(a^2)$ corrections. A non-perturbative renormalization procedure is developed and validated through predictions of the bottom quark mass and decay constants. This framework enables calculations at the physical $b$-quark mass even on lattices with spacing $a \sim 0.1$ fm, where $m_b a \sim 2.5$, while keeping discretization errors in hadronic matrix elements at the $\sim 10$\% level which can be eliminated properly through the continuum extrapolation. Using the physical $\Upsilon$ mass as input, we obtain $m_b^{\overline{\mathrm{MS}}}(m_b) = 4.185(37)$ GeV and the full spectrum of S-wave bottom mesons with 0.1\% uncertainty or less. Pseudoscalar and vector decay constants and their ratios for all kinds of S-wave bottom mesons are also provided.
\end{abstract}

\maketitle


{\it Introduction: }The study of bottom quark physics plays a pivotal role in probing the origins of the matter-antimatter asymmetry in the universe, one of the most profound open questions in fundamental physics. This asymmetry, a necessary condition for our existence, is believed to have arisen via the mechanism of CP violation within and potentially beyond the Standard Model. Thus precision stand model predictions are essential for constraining the unitarity triangle from the CP violated bottom meson and baryon decays~\cite{LHCb:2025ray} and isolating subtle signals of new physics. Thus non-perturbative Lattice QCD determinations of bottom physics with controlled uncertainties are therefore indispensable inputs for interpreting experimental results from facilities like LHCb and Belle II, and for ultimately understanding whether the observed CP violation is sufficient to explain the cosmic dominance of matter.

Direct lattice QCD calculations involving the physical bottom quark are, however, numerically challenging. The primary obstacle is the heavy quark mass itself: to maintain a small discretization error of order \((m_b a)^2\), one would naively require a lattice spacing of \(a \lesssim 0.02\) fm—finer than those typically available. Consequently, most of the high-precision lattice determinations~\cite{Bazavov:2017lyh,Hughes:2017spc,Lubicz:2017asp,Mathur:2018epb,FermilabLattice:2018est,Hatton:2021dvg} rely not on direct simulation at the physical \(b\)-quark mass, but on extrapolation from lighter masses using effective theories such as Heavy Quark Effective Theory (HQET)~\cite{Kronfeld:2000ck,Harada:2001fi,Harada:2001fj,Heitger:2003nj} or Non-Relativistic QCD (NRQCD)~\cite{Lepage:1992tx,Thacker:1990bm}. These approaches introduce additional non-perturbative low-energy constants that must be determined, adding layers of complexity to the analysis.

Anisotropic lattice discretizations~\cite{Burgers:1987mb,Alford:1996nx,Klassen:1998ua,Chen:2000ej} provide alternative pathway to overcome this challenge, utilizing a finer lattice spacing in the temporal direction to better resolve the energy scales—dominantly set by the heavy quark mass—involved in hadronic correlators. This approach, however, introduces additional complexity in the renormalization of quark bilinear operators and requires careful determination of the bare quark and gluon anisotropy parameters. Simplified implementations that apply an anisotropic heavy-quark action to isotropic gauge ensembles have been developed to suppress heavy-quark discretization errors~\cite{El-Khadra:1996wdx,Christ:2006us,Liu:2009jc,Brown:2014ena}. While these have been successfully employed in calculations of B-meson (semi-)leptonic decays~\cite{Christ:2014uea,Detmold:2015aaa,Flynn:2015mha,Leskovec:2025gsw}, the renormalization of the axial current in such setups has so far relied on lattice perturbation theory~\cite{Hashimoto:1999yp,El-Khadra:2001wco}.

In this work, we present the first-principles, high-precision determination of the bottom quark mass, S-wave bottom meson masses, and their decay constants using an anisotropic clover fermion action. Our formulation introduces a single additional parameter—the effective heavy-quark anisotropy—which we find approaches unity with corrections scaling as $a^2 m_{\Upsilon}^2$. Crucially, we develop and validate a fully non-perturbative renormalization procedure for the local vector, axial-vector, and pseudoscalar currents, including those relevant for heavy-light flavor-changing transitions. This systematic framework enables us to obtain the decay constants and $b$-quark mass with controlled systematic uncertainties.


\begin{table}[ht!]                   
\caption{Information of lattice spacing, pion mass, $\eta_s$ mass and lattice size. The uncertainties of the lattice spacing is estimated through the Akaike Information Criterion (AIC) analysis procedure~\cite{Borsanyi:2020mff, BMW:2014pzb}, and the second uncertainty is from the uncertainty of $w_0$ scale.}  
\renewcommand\arraystretch{1.2}
\begin{tabular}{ccccc } 
\hline\hline
ensemble &$a\,(\mathrm{fm})$  &$m_{\pi}\,(\mathrm{MeV})$ &$m_{\eta_s}\,(\mathrm{MeV})$ &$\tilde{L}^3\times\tilde{T}$ \\
\hline 
C24P34 &\multirow{6}{*}{0.10542(17)(62)} &340.6(1.7) &749.2(0.7) &$24^3\times64$ \\
C24P29 & &292.4(1.0) &658.5(0.6) &$24^3\times72$ \\
C32P29 & &293.4(0.8) &659.5(0.4) &$32^3\times64$ \\
C32P23 & &228.1(1.2) &644.6(0.5) &$32^3\times64$ \\
C48P23 & &224.3(1.2) &644.8(0.6) &$48^3\times96$ \\
C48P14 & &136.5(1.7) &707.3(0.4) &$48^3\times96$ \\
\hline
E28P35 &\multirow{3}{*}{0.09013(25)(53)} &345.4(1.1) &710.7(1.0) &$28^3\times64$ \\
E32P29 & &285.4(1.8) &698.2(0.9) &$32^3\times64$ \\
E32P22 & &215.1(2.2) &685.8(0.7) &$32^3\times96$ \\
\hline
F32P30 &\multirow{4}{*}{0.07760(07)(46)} &301.0(1.2) &677.3(1.0) &$32^3\times96$ \\
F32P21 & &210.5(2.2) &660.1(1.0) &$32^3\times64$ \\
F48P21 & &207.9(1.1) &663.7(0.6) &$48^3\times96$ \\
F64P14 & &135.6(1.2) &681.0(0.5) &$64^3\times128$ \\
\hline
G36P29 &0.06895(17)(41) &295.7(1.1) &692.6(0.5) &$36^3\times108$ \\
\hline
H48P32 &0.05235(11)(31) &316.1(1.0) &690.6(0.7) &$48^3\times144$ \\
\hline
I64P31 &0.03761(08)(22) &312.2(1.6) &671.4(1.3) &$64^3\times128$ \\
\hline\hline
\end{tabular}  
\renewcommand\arraystretch{1.}
\label{tab:ensem}
\end{table}

{\it Numerical setup:} We perform our calculations on the CLQCD gauge ensembles, which employ a tadpole-improved Symanzik gauge action and 2+1 flavors of stout-smeared, tadpole-improved clover fermions. The ensembles span six lattice spacings; key parameters, including the lattice spacing, pion mass \(m_{\pi}\), mass of the fictitious \(\eta_s\) meson (which sets the strange quark mass), and grid size $\tilde{L}^3\times\tilde{T}$, are listed in Table~\ref{tab:ensem}. Details on the discretized action, lattice-spacing determinations and Monte Carlo simulations~\cite{CLQCD:2023sdb,CLQCD:2024yyn,Hasenbusch:2001ne} are provided in the supplemental material~\cite{supplemental}. The light, strange, and charm hadron spectrum and decay constants computed on subsets of these ensembles~\cite{CLQCD:2023sdb,CLQCD:2024yyn} show excellent agreement with literature values and, achieve superior precision for low-energy constants and charmed vector meson decay constants.

Now we extend our calculation to the bottom quark using the anisotropic clover fermion action introduced in Ref.~\cite{Liu:2009jc},
\begin{align}\label{eq:aniso}
    S_Q = a^4 &\sum_x \bar{Q} \Big[ m_Q + \gamma_4 \nabla_4 - \frac{a}{2} \nabla_4^2
        + \nu \sum_{i=1}^3 \Big( \gamma_i \nabla_i - \frac{a}{2} \nabla_i^2 \Big) \nonumber \\
        &- c_E\, \frac{a}{2} \sum_{i=1}^{3} \sigma_{i4} F_{i4}
        - c_B\, \frac{a}{2} \sum_{i>j=1}^{3} \sigma_{ij} F_{ij} \Big]Q,
\end{align}
which is simulated on the isotropic gauge ensembles. Here, \(m_Q\) is the bare heavy-quark mass, \(\nu\) is the bare anisotropy parameter, and \(u_0\) is the tadpole improvement factor. The couplings for the clover terms are tree-level
tadpole-improved values \(c_E=(1+\nu)/(2u_0^3)\) and \(c_B = \nu/(u_0^3)\) from Ref.~\cite{Chen:2000ej}. The parameters \(m_Q\) and \(\nu\) are tuned non-perturbatively by requiring the vector meson mass to match the physical \(\Upsilon\) mass \(m_\Upsilon\) and the effective speed of light to equal unity. While the discretization errors could be further suppressed by tuning \(c_{E/B}\) with the physical value of hyperfine splitting~\cite{RBC:2012pds,Christ:2014uea}, doing so reduces the predictivity of the setup and complicates the parameter tuning. For alternative formulations of anisotropic heavy-quark actions, see Refs.~\cite{El-Khadra:1996wdx,Aoki:2001ra,Christ:2006us} and the appendix of Ref.~\cite{FlavourLatticeAveragingGroup:2019iem}.

\begin{figure}[thb]
    \centering
    \includegraphics[width=1.0\linewidth]{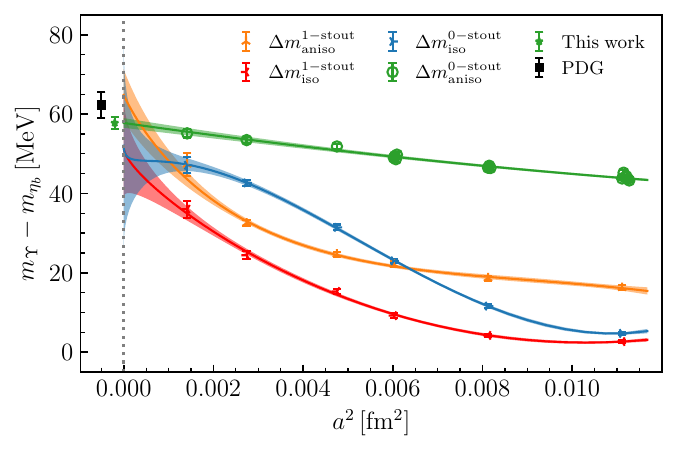}
    \caption{The hyperfine splitting \(\Delta^{\bar{b}b}_{\mathrm{HFS}} \equiv m_{\Upsilon} - m_{\eta_b}\), computed using different heavy-quark fermion actions at different lattice spacing. The discretization error is significantly suppressed by employing the anisotropic clover action without stout smearing, and both the usage of the anisotropic action and original link are relevant.}
    \label{fig:comp_mhfs}
\end{figure}

Fig.~\ref{fig:comp_mhfs} compares the bottomonium hyperfine splitting, \(\Delta^{\bar{b}b}_{\mathrm{HFS}} \equiv m_{\Upsilon} - m_{\eta_b}\), computed using three heavy-quark actions: (1) the isotropic clover action with smeared gauge links, identical to that used for the lighter flavors (red crosses); (2) the same isotropic clover fermion action with unsmeared (original) gauge links (orange boxes),  (3) the anisotropic clover action defined in Eq.~(\ref{eq:aniso}), with smeared gauge links (blue triangles);  and (4) the same anisotropic action but with unsmeared gauge links (green circles). Choice (4) yields 57.8(1.2)(1.0) MeV with 1 MeV systematic uncertainty estimated for the missing disconnected diagram~\cite{Hatton:2021dvg}, a value with substantially smaller discretization errors than the other choices. It is also consistent with the PDG value of 62.3(3.2) MeV~\cite{ParticleDataGroup:2024cfk} and the previous highest-precision lattice QCD result of 57.5(2.3)(1.0) MeV~\cite{Hatton:2021dvg}, while achieving a significantly smaller uncertainty.

\begin{figure}[thb]
    \centering
    \includegraphics[width=1.0\linewidth]{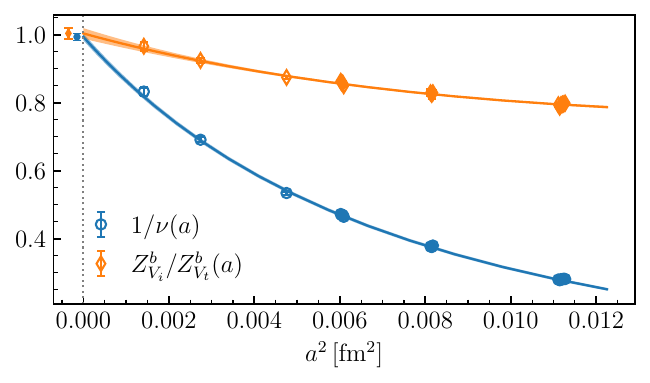}
    \caption{Lattice-spacing dependence of the anisotropy parameter \(1/\nu\) and the renormalization-factor ratio \(Z^b_{V_i} / Z^b_{V_t}\). Both quantities are well described by polynomials in \(a^2\) and extrapolate to unity in the continuum limit, consistent with the restoration of Euclidean isotropy.}
    \label{fig:aniso_RZVit}
\end{figure}

The anisotropy parameter $\nu$, determined across ensembles with different lattice spacings and sea quark masses, is well described by the empirical form as shown as the blue circles in Fig.~\ref{fig:aniso_RZVit},
\begin{align}\label{eq:nu}
&\nu(am_h) = \frac{\sinh(c_{\nu}  am_h)}{c_{\nu}  am_h}\Big[ \nu_0 + \sum_{{\mathrm{PS}}=\pi,\eta_s}c_{\mathrm{PS}}\,\delta m^{2}_{\mathrm{PS},\mathrm{sea}} \Big],
\end{align}
where $m_h$ is the quarkonium mass used to tune the dispersion relation and $\delta m^2_{\mathrm{PS},\mathrm{sea}}=m^2_{\mathrm{PS},\mathrm{sea}}-m^2_{\mathrm{PS},\mathrm{phys}}$. The coefficient $c_{\nu} = 0.6119(25)$ is close to the free‑field estimate of $0.5$, $\nu_0 = 1.0063(95)$ ensures that $\nu \to 1$ in the continuum limit with ${\cal O}(m_h^2 a^2)$ corrections, and  $c_{\pi}=0.250(95)~\mathrm{GeV}^{-2}$ and $c_{\eta_s}=0.012(66)~\mathrm{GeV}^{-2}$ encode the light‑quark mass dependence. The consistency between our data and the empirical form further indicates that $\nu\sim1$ would be obtained if a similar anisotropic action were used for the light quarks, up to ${\cal O}(\Lambda_{\rm QCD}^2a^2)$ discretization error. 

Beyond the hadron spectrum, precise determinations of hadronic matrix elements hae additional complication from the non-perturbative renormalization. For instance, a renormalized quark mass can be defined via the partially conserved axial current (PCAC) relation~\cite{JLQCD:2007xff,CLQCD:2023sdb,CLQCD:2024yyn},
\begin{align}
m_q^{\rm PC} = \left. \frac{m_{\rm PS}\, \sum_{\vec{x}} \langle A_4(\vec{x},t) P^\dagger(\vec{0},0) \rangle}{2 \sum_{\vec{x}} \langle P(\vec{x},t) P^\dagger(\vec{0},0) \rangle} \right|_{t \to \infty} \label{eq:mass},
\end{align}
where \(A_\mu = \bar{\psi}\gamma_5\gamma_\mu\psi\) and \(P = \bar{\psi}\gamma_5\psi\) are the axial-vector and pseudoscalar currents, and \(m_{\rm PS}\) is the pseudoscalar meson mass. The fully renormalized quark mass is then \(m_q^R = (Z_A / Z_P) \, m_q^{\rm PC}\), with \(Z_{A,P}\) denoting the renormalization constants of the respective currents. Similarly, the decay constants of pseudoscalar (\(P\)) and vector (\(V\)) mesons are defined through the vacuum-to-meson matrix elements
\begin{align}
Z_V\langle 0 | V^\mu | V_i \rangle &= f_V \, m_V \, \epsilon^\mu_i\,, \\[2mm]
Z_A\langle 0 | A^0 | P \rangle &= f_{PS} \, m_{PS},
\end{align}
where \(V^\mu = \bar{\psi} \gamma^\mu \psi\), and \(\epsilon^\mu_i\) is the polarization vector of the vector meson.

The heavy-quark vector current \(V^\mu \equiv \bar b \gamma^\mu b\) can be normalized non-perturbatively via its hadronic matrix element,
\begin{align}
     Z^b_{V_\mu} = 2p_\mu/{\langle \eta_b|V_\mu|\eta_b\rangle},
\end{align}
where \(p\) is the four-momentum of $\eta_b$. For an anisotropic heavy-quark action, the temporal and spatial normalization factors \(Z^b_{V_t}\) and \(Z^b_{V_i} \equiv Z^b_{V_{x/y/z}}\) may differ at finite lattice spacing. As shown by the orange diamonds in Fig.~\ref{fig:aniso_RZVit}, the ratio \(Z^b_{V_i} / Z^b_{V_t}\), extrapolated to the continuum limit, is consistent with unity up to ${\cal O}(a^2)$ polynomial correction, confirming the restoration of Euclidean isotropy.

As demonstrated for the charm quark in Ref.~\cite{CLQCD:2024yyn}, heavy-quark bilinear currents with the clover fermion action receive sizable quark-mass-dependent corrections. These corrections are, however, strongly suppressed in ratios of renormalization constants between different currents, allowing such ratios to be computed reliably in the chiral limit. We apply the same strategy to the anisotropic heavy-quark action. Specifically, we determine $Z^b_{V_t}$ and $Z^b_{V_i}$ separately—each of which can be ${\cal O}(10)$ for the physical $b$ quark mass at $a \sim 0.1$ fm—while all ratios $Z_X / Z_V$ (with $X = A, P,...$) are evaluated in the chiral limit nonperturbatively~\cite{Sturm:2009kb,Martinelli:1994ty,He:2022lse,Kniehl:2006bg,Baikov:2016tgj,Baikov:2014qja,Bednyakov:2020ugu,Kniehl:2020sgo}. 

Because the bottom quark action uses unsmeared gauge links while the light flavors (\(q=u/d/s/c\)) employ smeared ones, the heavy-light currents \(\bar b \Gamma q\) relevant for \(B\)-meson decay constants require mixed-action renormalization constants $Z^{ns,s}_{\bar{q} \Gamma q}$ in the chiral limit~\cite{supplemental}. 
Similar to the charmed meson cases~\cite{CLQCD:2024yyn}, we further define the renormalization constant of the heavy-light current aas,
\begin{align}
    Z_{\bar b \Gamma q}
    \equiv \sqrt{Z^b_{V_{t(i)}}Z^q_V}\,\left.\frac{Z^{ns,s}_{\bar{q} \Gamma q}}
        {Z^{ns,s}_{\bar{q}\gamma_\mu q}}\right|_{m_q\rightarrow 0}\,,
        \label{eq:ZO_ZV}
\end{align}
where \(Z^q_V\) is the vector-current renormalization factor for the lighter flavors quark field with smeared gauge links~\cite{CLQCD:2024yyn}, and $Z_{V_{t(i)}}^b$ denotes mass dependent vector-current renormalization factor for bottom quark field with unsmeared gauge links.

Further details on the determination of anisotropic parameters, the extraction of bare lattice quantities, and the full renormalization strategy are provided in the supplemental material~\cite{supplemental}.

\begin{figure}[thb]
    \centering
    \includegraphics[width=1.0\linewidth]{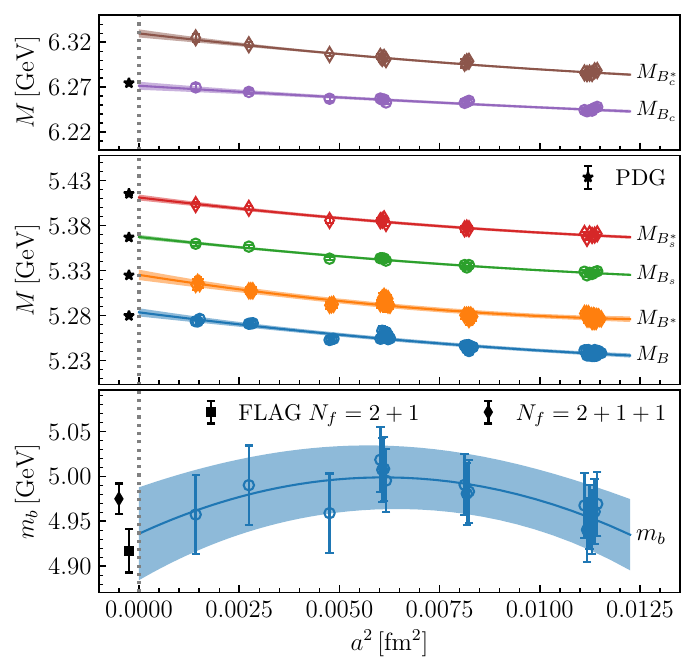}
    \caption{Lattice spacing dependence of ${B^{(*)}_{l/s/c}}$ masses and bottom quark mass $m_b^{\overline{\mathrm{MS}}}(2~\mathrm{GeV})$. In the upper two panels, the points at the continuum limit display the results from PDG~\cite{ParticleDataGroup:2024cfk}. The black points in the lowest panel show the results of $N_f$=2+1 and 2+1+1 from FLAG~\cite{FlavourLatticeAveragingGroupFLAG:2024oxs} average of $m_b^{\overline{\mathrm{MS}}}(2~\mathrm{GeV})$. Data points in the panels correspond to values after the extrapolation to the physical point using the fit parameters, and the optical bands show the discretization effects (see \cite{supplemental} for further details).}
    \label{fig:masses}
\end{figure}

{\it Results:} For hadron masses we use the combined continuum and chiral fit,
\begin{align}
    X(m_{\pi},&m_{\eta_s},a) = X_{\mathrm{phys}}\left[1 \;+\; c^X_{\mathrm{val}}\delta m^2_{\mathrm{PS}, \mathrm{val}}\right. \nonumber\\
    &+\left.c^X_{\mathrm{sea}}\delta m^2_{\mathrm{PS}, \mathrm{sea}}+d_a(a^2,a^2\delta m^2_{\mathrm{PS}, \mathrm{sea}})\right]\, , \label{eq:x_dep}
\end{align}
where \(X_{\mathrm{phys}} \equiv X(m_{\pi,\mathrm{phys}}, \, m_{\eta_s,\mathrm{phys}}, 0)\) is the physical value at the continuum limit. The detailed discussion on $\delta m^2_{\mathrm{PS}, \mathrm{val}}$, $\delta m^2_{\mathrm{PS}, \mathrm{sea}}$, and the discretization effect term $d_a$ can be found in the supplemental materials~\cite{supplemental}. We take \(m_{\pi,\mathrm{phys}} = 134.98\;\mathrm{MeV}\) and, to fix the strange sea-quark mass, \(m_{\eta_s,\mathrm{phys}} = 689.89(49)\;\mathrm{MeV}\)~\cite{Borsanyi:2020mff}. The strange and charm valence-quark mass are tuned separately to their physical value~\cite{CLQCD:2024yyn}. This ansatz describes the data well, with \(\chi^{2}/\mathrm{dof} \sim 1\).

We combine the statistical uncertainty from the two point function and the impacts from those of the lattice spacing, unitary pion and etas masses into the total statistical uncertainty, and take the deviation from different continuum extrapolation ansatz and impact from the scale setting parameter $w_0$ as the systematic uncertainties. 

As shown in Fig.~\ref{fig:masses}, the discretization error on the coarsest lattice (\(a\simeq 0.1\) fm) is at the \(1\%\) level for hadron masses and renormalized bottom-quark mass when the anisotropic action is employed and the bare \(b\)-quark mass is tuned to the physical \(\Upsilon\) mass on each ensemble. Through the SMOM scheme and corresponding perturbative matching, we obtain $m_b^{\overline{\mathrm{MS}}}(2\;\mathrm{GeV}) = 4.936(52)\;\mathrm{GeV}$ which corresponds to 
\begin{align}
m_b^{\overline{\mathrm{MS}}}(m_b) = 4.185(37)\;\mathrm{GeV},
\end{align}
and represents the most precise lattice determination to date that uses a relativistic heavy-quark action free of both residual fermion doubling of the staggered fermion and any heavy-quark expansion.

\begin{figure}[thb]
    \centering
    \includegraphics[width=1.0\linewidth]{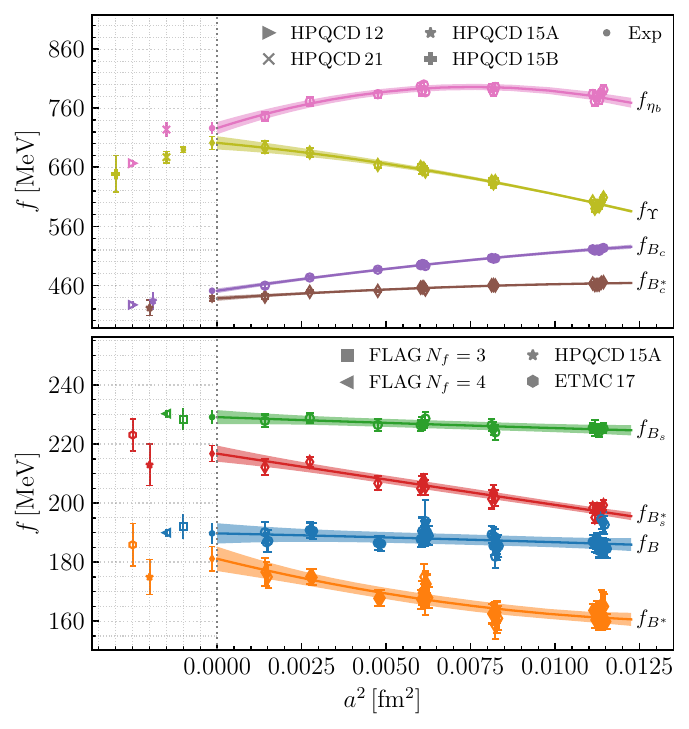}
    \caption{Lattice spacing dependence of the decay constants of $\eta_b$, $\Upsilon$ and $B^{(*)}_{l,s,c}$. FLAG~\cite{FlavourLatticeAveragingGroupFLAG:2024oxs} averages (for $f_{B/B_s}$) and values from literature are also shown for comparison. Data points in the panels correspond to values after the extrapolation to the physical point using the fit parameters, and the optical bands show the discretization effects (see \cite{supplemental} for further details).}
     \label{fig:decay_constants}
\end{figure}

The renormalized decay constants show discretization errors of only $\sim$2\% on our finest ensemble ($a \sim 0.04\,\mathrm{fm}$)—markedly smaller than the $\sim$10\% deviation found in earlier HISQ calculations at comparable lattice spacings~\cite{Hatton:2021dvg}, and also well below the expectations of naive power counting. The observed difference between the pseudoscalar decay constant \(f_P\) and the vector decay constant \(f_V\) arises primarily from the use of distinct renormalization factors: \(Z^b_{V_t}\) for the temporal axial current \(A_4\) (in \(f_P\)) and \(Z^b_{V_s}\) for the spatial vector current \(V_i\) (in \(f_V\)). This difference at non-zero $a$ can be further reduced by employing \(Z^b_{V_t}\) for both currents while keep the continuum limit unchanged.

After continuum extrapolation, our results for \(f_{B_{(s)}}\) achieve a precision surpassing all previous \(N_f = 2+1\) lattice determinations, as well as most \(N_f = 2+1+1\) results. The sole exception is the HPQCD 2017 result~\cite{Bazavov:2017lyh}, which uses the HISQ action with higher statistics but reaches the physical \(b\)-quark mass only on its two finest lattice spacings. Using the experimental value \(f_B |V_{ub}| = 0.77 \pm 0.12\ \text{MeV}\)~\cite{FlavourLatticeAveragingGroupFLAG:2024oxs}, our result for \(f_B\) corresponds to $|V_{ub}| = 4.06(64) \times 10^{-3}$.

\begin{table}[h] 
\renewcommand\arraystretch{1.3}
\caption{Summary table of our major predictions. A detailed decomposition of the uncertainties into their statistical and systematic components is available in the supplemental materials~\cite{supplemental}. }
\label{tab:final}
\begin{tabular}{c|cccc}
\hline\hline
        &$\bar{b}l$ &$\bar{b}s$ &$\bar{b}c$ &$\bar{b}b$ \\
\hline 
$f_{PS}\,[\mathrm{MeV}]$       &189.8(3.5) &229.1(2.4) &451.0(5.2) &726.4(10.5)  \\
$f_{V}\,[\mathrm{MeV}]$        &181.2(4.2) &216.8(2.7) &437.9(4.3) &701.4(11.3)  \\
$f_V/f_{PS}$                   &0.957(20) &0.946(12) &0.969(09) &0.955(16)   \\
$m_{PS}\,[\mathrm{GeV}]$       &5.2835(46) &5.3673(23) &6.2715(45) &9.4026(12)  \\ 
$\Delta m\,[\mathrm{MeV}]$     &41.9(4.4) &43.6(2.0) &58.1(1.0) &57.8(1.2)(1.0)   \\
\hline\hline
\end{tabular}
\end{table}

We summarize our results for $f_{PS}$, $f_V$, $f_{PS}/f_V$, $m_{PS}$, and $\Delta m\equiv m_V-m_{PS}$ in Table \ref{tab:final}. Additionally, we find \(f_{B_s}/f_B = 1.207(17)\), which is slightly larger than our earlier determination of \(f_K/f_\pi\). Full comparison of our results with previous results in the literature~\cite{FlavourLatticeAveragingGroupFLAG:2024oxs,Gamiz:2009ku,FermilabLattice:2011njy,McNeile:2011ng,Na:2012kp,Witzel:2013sla,Aoki:2014nga,Christ:2014uea,Dowdall:2013tga,Carrasco:2013naa,ETM:2016nbo,Hughes:2017spc,Bazavov:2017lyh,Frezzotti:2024kqk,Albertus:2010nm,Boyle:2018knm,Hollitt:2022exk,McNeile:2010ji,Lee:2013mla,Maezawa:2016vgv,Petreczky:2019ozv,Colquhoun:2014ica,Gambino:2017vkx,FermilabLattice:2018est,Hatton:2021syc,Hatton:2021dvg,Colquhoun:2015oha,McNeile:2012qf,Becirevic:2018qlo,Lubicz:2017asp,Conigli:2023rod,Lubicz:2017asp} can be found in the supplemental materials~\cite{supplemental}.

{\it Summary:} In summary, we have performed a precise calculation of bottom-quark observables—including the quark mass, S-wave meson masses, and decay constants—by simulating at the physical bottom-quark mass across six lattice spacings. Using an anisotropic heavy-quark action, we suppress discretization errors to the ~1\% level even on our coarsest ensembles, where \(a \sim 0.1\;\text{fm}\) and \(m_b^{\mathrm{phys}} a \sim 2.5\). We further develop and apply the necessary non-perturbative renormalization, enabling a rigorous determination of the bottom-quark mass and hadronic matrix elements. As a result, we obtain the most precise lattice-QCD determination to date of the hyperfine splittings of bottomonium, as well as a prediction for that of $B^{(*)}_c$ meson, together with the decay constants $f_{B^*}$, $f_{B_s^*}$ $f_{B^{(*)}_c}$ and $f_{\eta_b}$. All systematic uncertainties from chiral and continuum extrapolations are kept under control, a feat made possible by working directly at the physical mass and thereby avoiding any reliance on heavy-quark effective theory extrapolations.

In the present work, the bottom quark mass is tuned using the physical \(\Upsilon\) mass, but the calculation omits quark-disconnected diagrams and QED corrections. A complementary approach that avoids this systematic uncertainty is to define \(m_b\) via the physical \(B_s\) mass after subtracting electromagnetic corrections. In the so call Dashen's scheme, where the QED contribution to the \(\eta_q\) mass is set to zero~\cite{Horsley:2015vla}, the electromagnetic shifts of the \(B_{l/s/c}^{(*)}\) masses are estimated to be \(\mathcal{O}(1\text{--}2)\;\mathrm{MeV}\) for charged states and even smaller for neutral ones such as $B_s$ and \(\Upsilon\)~\cite{CLQED_26}; these corrections are thus negligible compared to our current total uncertainties. Using this scheme, the difference between our connected-diagram predictions for \(m_{\eta_b}\) and \(m_{\Upsilon}\) and their physical values would be reduced by approximately \(-1(5)\;\mathrm{MeV}\)—a shift that can be attributed to the missing quark-disconnected contributions while consistent with zero within the uncertainty. 

The relativistic heavy-quark action employed here further enables the calculation of semileptonic \(B\)-decay form factors directly in the recoil frame. This approach significantly extends the accessible kinematic range while maintaining a favorable signal‑to‑noise ratio~\cite{Ding:2024lfj}. Our work therefore establishes a foundation for predictions of \(B\)-physics observables--including CKM unitarity tests and semileptonic decay rates--with percent‑level precision, where systematic uncertainties from heavy‑quark discretization and non‑perturbative renormalization are brought under control.

\section*{Acknowledgement}
We thank the CLQCD collaborations for providing us their gauge configurations with dynamical fermions~\cite{CLQCD:2023sdb,CLQCD:2024yyn}, which are generated on HPC Cluster of ITP-CAS, IHEP-CAS and CSNS-CAS, ORISE Supercomputer, the Southern Nuclear Science Computing Center(SNSC) and the Siyuan-1 cluster supported by the Center for High Performance Computing at Shanghai Jiao Tong University. 
The calculations were performed using the PyQUDA~\cite{Jiang:2024lto} and Chroma~\cite{Edwards:2004sx} packages with QUDA~\cite{Clark:2009wm,Babich:2011np,Clark:2016rdz} through HIP programming model~\cite{Bi:2020wpt}. We also thank Ying Chen and the other CLQCD members for helpful discussion and comments. The numerical calculations were carried out on the ORISE Supercomputer, HPC Cluster of ITP-CAS, and Advanced Computing East China Sub-center. This work is supported in part by National Key R\&D Program of China No.2024YFE0109800, NSFC grants No. 12525504, 12435002, 12293060, 12293062, 12293061, 12293065 and 12447101, the Strategic Priority Research Program of Chinese Academy of Sciences, Grant No. YSBR-101.

\bibliography{ref}

\begin{widetext}

\section*{Appendix}

\subsection{Information of CLQCD ensembles}

\begin{table}[ht!]                   
\caption{Gauge coupling $\hat{\beta}=10/(g^2u_0^4)$, lattice spacing $a$ (the second uncertainty is estimated from the uncertainty of the scale parameter $w_0$), tadpole improvement factors $u^I_0$ and $v^I_0$, dimensionless bare quark mass parameters $\tilde{m}^{\rm b}_{l,s}$, lattice size $\tilde{L}^3\times \tilde{T}$, pion and $\eta_s$ masses for different ensembles. The ensembles labeled by `` * " are used in the global fit for the determination of lattice spacings, but they are not included in the bottom physics calculations.} 
\label{tab:ensemble_info}
\renewcommand\arraystretch{1.2}
\begin{tabular}{l| c c c c c c c c l l | cc } 
\hline
\hline
ensemble &$\hat{\beta}$ &$a$ (fm) &$u^I_0$ &$v^I_0$  &$\tilde{m}^b_{l}$ &$\tilde{m}^b_s$ &$\tilde{L}^3\times \tilde{T}$ &$m_\pi L$ &$m_{\pi}\,(\mathrm{MeV})$ &$m_{\eta_s}\,(\mathrm{MeV})$ & $N_{\rm cfg}$ & $N_{\rm src}$\\
\hline 
C24P34 &\multirow{6}{*}{6.200} &\multirow{6}{*}{0.10542(17)(62)} &0.855453 &0.951479 &-0.2770 &-0.2310 &$24^3\times64$ &4.37 &340.6(1.7) &749.2(0.7)&200 &32 \\
C24P29 & & &0.855453 &0.951479 &-0.2770 &-0.2400 &$24^3\times72$ &3.75 &292.4(1.0) &658.5(0.6)&760 &3 \\
C32P29 & & &0.855453 &0.951479 &-0.2770 &-0.2400 &$32^3\times64$ &5.02 &293.42(82) &659.5(0.4)&489 &3 \\
C32P23 & & &0.855520 &0.951545 &-0.2790 &-0.2400 &$32^3\times64$ &3.90 &228.1(1.2) &644.6(0.5)&400 &3 \\
C48P23 & & &0.855520 &0.951545 &-0.2790 &-0.2400 &$48^3\times96$ &5.75 &224.3(1.2) &644.8(0.6)&62 &3 \\
C48P14 & & &0.855548 &0.951570 &-0.2825 &-0.2310 &$48^3\times96$ &3.50 &136.5(1.7) &707.3(0.4)&188 &3 \\
\hline
E28P35 &\multirow{3}{*}{6.308} &\multirow{3}{*}{0.09013(25)(53)} &0.859645 &0.954385 &-0.2490 &-0.2170 &$28^3\times64$ &4.42 &345.4(1.1) &710.7(1.0)&152 &4 \\
E32P29 & & &0.859727 &0.954467 &-0.2510 &-0.2170 &$32^3\times64$ &4.17 &285.4(1.8) &698.2(1.0)&100 &4 \\
E32P22 & & &0.859737 &0.954469 &-0.2530 &-0.2170 &$32^3\times96$ &3.14 &215.1(2.2) &685.8(0.7)&143 &3 \\
\hline
F32P30 &\multirow{5}{*}{6.410} &\multirow{5}{*}{0.07760(07)(46)} &0.863437 &0.956942 &-0.2295 &-0.2050 &$32^3\times96$ &3.79 &301.0(1.2) &677.3(1.0)&250 &3 \\
F48P30$^*$ & & &0.863473 &0.956984 &-0.2295 &-0.2050 &$48^3\times96$ &5.73 &303.49(67) &676.1(0.5)&- &- \\
F32P21 & & &0.863488 &0.957017 &-0.2320 &-0.2050 &$32^3\times64$ &2.65 &210.5(2.2) &660.1(0.9)&194 &3 \\
F48P21 & & &0.863499 &0.957006 &-0.2320 &-0.2050 &$48^3\times96$ &3.93 &207.9(1.1) &663.7(0.6)&82 &12 \\
F64P14 & & &0.863533 &0.957033 &-0.2334 &-0.2030 &$64^3\times128$ &3.41 &135.6(1.2) &681.0(0.5)&70 &2 \\
\hline
G32P35$^*$ &\multirow{2}{*}{6.498} &\multirow{2}{*}{0.06895(17)(41)}  &0.866456&0.958910 &-0.2135 &-0.1926 &$32^3\times96$ &3.94 &352.2(2.6) &706.9(1.8)&- &- \\
G36P29 & & &0.866470 &0.958910 
  &-0.2150 &-0.1926 &$36^3\times108$ &3.72 &295.7(1.1) &692.6(0.5)&270 &3\\
\hline
H48P32 &6.720 &0.05235(11)(31) &0.873378 &0.963137 &-0.1850 &-0.1700 &$48^3\times144$ &4.03 &316.07(97) &690.6(0.7)&157 &12 \\
\hline
I64P31 &\multirow{2}{*}{7.020} &\multirow{2}{*}{0.03761(08)(22)} &0.881466 &0.967695 &-0.1569 &-0.1475 &$64^3\times128$ &3.81 &312.2(1.6) &671.4(1.3)&90 &2 \\
I64P19$^*$ & & &0.881474 &0.967703 &-0.1585 &-0.1475 &$64^3\times128$ &2.29 &188.1(3.6) &664.8(1.4) &- &-
\\
\hline
\hline
\end{tabular}  
\renewcommand\arraystretch{1}
\end{table}

The results in this work are based on the 2+1 flavor ensembles from the CLQCD collaboration using the tadpole improved tree level Symanzik (TITLS) gauge action $S_g(U)$ and the tadpole improved tree level Clover (TITLC) fermion action $S_q(V, m)$. $S_g(U)$ is expressed by
\begin{align}
    S_g = \frac{1}{N_c} \, \mathrm{Re} \sum_{x,\mu < \nu} \mathrm{Tr} \left[ 1 - \hat{\beta} \left( \mathcal{P}^{U}_{\mu,\nu}(x) + \frac{c_1 \mathcal{R}^{U}_{\mu,\nu}(x)}{1 - 8 c_1^{0}} \right) \right]~,
\end{align}
where $\mathcal{P}^{U}_{\mu,\nu}(x) = U_{\mu}(x) U_{\nu}(x + a \hat{\mu}) U^{\dagger}_{\mu}(x + a \hat{\nu}) U^{\dagger}_{\nu}(x)$ is the plaquette, $\mathcal{R}^{U}_{\mu,\nu}(x) = U_{\mu}(x) U_{\mu}(x + a \hat{\mu}) U_{\nu}(x + 2 a \hat{\mu})U^{\dagger}_{\mu}(x + a \hat{\mu} + a \hat{\nu}) U^{\dagger}_{\mu}(x + a \hat{\nu}) U^{\dagger}_{\nu}(x)$ is the $1\times2$ wilson loop, $ U_{\mu}(x) = P \left[ \exp \left( i g_0 \int_{x}^{x + \hat{\mu} a} \mathrm{d}y A_{\mu}(y) \right) \right]$ is the gauge link, $\hat{\beta} = 6(1 - 8 c_{1}^{0})/(g_{0}^{2} u_{0}^{4}) \equiv 10/(g_{0}^{2} u_{0}^{4})$ is the gauge coupling with $c_1^0=-1/12$ and $c_1=c_1^0/u_0^2$, $u_0=\left\langle \operatorname{ReTr} \displaystyle\sum_{x,\,\mu < \nu} \mathcal{P}^{U}_{\mu\nu}(x)/\left(6 N_{c} \tilde{V}\right) \right\rangle^{1/4}$  is the tadpole improvement factor, $\tilde{V}=\tilde{L}^3\times\tilde{T}$  is the dimensionless 4D volume of the lattice, and symbol $\tilde{O}$ with tilde is used to represent the dimensionless value of quantity $O$. The TITLC fermion action $S_q(V,m)$ uses 1-step stout smeared link $V_\mu(x)$ with smearing parameter $\rho=0.125$. $S_q(V,m)$ is expressed by
\begin{equation}
    S_q(m) = -\sum_{x} \bar{\psi}(x) \sum_{\eta=\pm 1}\sum_{\mu=1}^{4} \frac{1 - \eta\gamma_\mu}{2} V_{\eta\mu}(x) \psi(x + \eta\hat{\mu} a)
    + \sum_x \psi(x) \left[ (4 + m a) - c_{sw}\frac{a}{2} \sum_{\mu<\nu}\sigma^{\mu \nu} F_{\mu \nu}^V \right] \psi(x)\,,
\end{equation}
where $c_{sw}=1/v_0^3$ with $v_0=\left\langle \operatorname{ReTr} \displaystyle\sum_{x,\,\mu < \nu} \mathcal{P}^{V}_{\mu\nu}(x)/\left(6 N_{c} \tilde{V}\right) \right\rangle^{1/4}$, and 
\begin{align}
    F_{\mu \nu}^{V} = \frac{i}{8 a^2}
    \left(
        \mathcal{P}_{\mu,\nu}^{V}
        - \mathcal{P}_{\nu,\mu}^{V}
        + \mathcal{P}_{\nu,-\mu}^{V}
        - \mathcal{P}_{-\mu,\nu}^{V}
        + \mathcal{P}_{-\mu,-\nu}^{V}
        - \mathcal{P}_{-\nu,-\mu}^{V}
        + \mathcal{P}_{-\nu,\mu}^{V}
        - \mathcal{P}_{\mu,-\nu}^{V}
    \right)\,.
    \label{eq:SqVm}
\end{align}
One can refer to Refs.~\cite{CLQCD:2023sdb} for the details of the setup of configurations. The information of the ensembles including the tadpole improvement factors are displayed in Table~\ref{tab:ensemble_info}. The ensembles labeled by * are used in the determination of lattice spacings but not in the calculation of other quantities in this work. For other ensembles, Table~\ref{tab:ensemble_info} additionally shows the number of configurations used $(N_{\rm cfg})$ and the number of sources per configuration $(N_{\rm src})$.

\begin{table}[htbp]
  \centering
  \caption{The integration step size $\tau$, numbers of the fermion and gauge sub-steps, and also Hasenbusch precondition masses for the degenerated light quarks.}
    \begin{tabular}{c|ccc|c|c|c|c|c|c}
    \hline
    \hline
     & \multirow{2}{*}{$\tau$} & \multirow{2}{*}{$N_{\rm step,f}$} & \multirow{2}{*}{$N_{\rm step,g}$} & 
     \multicolumn{6}{c}{Hasenbusch masses} \\
     \cline{5-10}
     & & & & $m_0$ & $m_1$ & $m_2$ & $m_3$ & $m_4$ & $m_5$ \\
    \hline
    C24P34    & 1.000 & 8 & 2 & $-0.2770$ & $-0.2765$ & $-0.2755$ & $-0.2735$ & $-0.2675$ & $-0.2495$ \\
    C24P29    & 0.707 & 10 & 2 & $-0.2770$ & $-0.2765$ & $-0.2755$ & $-0.2735$ & $-0.2655$ & \\
    C32P29    & 0.700 & 10 & 2 & $-0.2770$ & $-0.2765$ & $-0.2755$ & $-0.2735$ & $-0.2655$ & \\
    C24P23    & 1.000 & 10 & 2 & $-0.2790$ & $-0.2785$ & $-0.2775$ & $-0.2755$ & $-0.2715$ & $-0.2635$ \\
    C32P23    & 0.700 & 10 & 2 & $-0.2790$ & $-0.2785$ & $-0.2775$ & $-0.2755$ & $-0.2715$ & $-0.2635$ \\
    C48P23    & 0.700 & 10 & 2 & $-0.2790$ & $-0.2785$ & $-0.2775$ & $-0.2755$ & $-0.2715$ & $-0.2635$ \\
    C48P14    & 1.000 & 55 & 2 & $-0.2825$ & $-0.2824$ & $-0.2822$ & $-0.2818$ & $-0.2806$ & $-0.2770$ \\
    C64P14    & 1.000 & 25 & 2 & $-0.2825$ & $-0.2820$ & $-0.2810$ & $-0.2790$ & $-0.2750$ & $-0.2670$ \\
    \hline                                                  E28P35    & 1.000 & 8 & 2 & $-0.2490$ & $-0.2485$ & $-0.2475$ & $-0.2455$ & $-0.2395$ & $-0.2215$ \\      
    E32P29    & 1.000 & 10 & 2 & $-0.2510$ & $-0.2505$ & $-0.2495$ & $-0.2475$ & $-0.2435$ & $-0.2355$ \\
    E32P22    & 1.000 & 8 & 3 & $-0.2530$ & $-0.2495$ & $-0.2475$ & $-0.2355$ &  &  \\
    \hline                                                                                          
    F32P30    & 0.500 & 8 & 2 & $-0.2295$ & $-0.2293$ & $-0.2289$ & $-0.2281$ & $-0.2265$ & $-0.2233$ \\
    F48P30    & 0.500 & 10 & 2 & $-0.2295$ & $-0.2293$ & $-0.2289$ & $-0.2281$ & $-0.2265$ & $-0.2233$ \\
    F32P21    & 0.500 & 10 & 2 & $-0.2320$ & $-0.2315$ & $-0.2305$ & $-0.2285$ & $-0.2245$ & $-0.2165$ \\
    F48P21    & 0.500 & 8 & 2 & $-0.2320$ & $-0.2315$ & $-0.2305$ & $-0.2285$ & $-0.2245$ & $-0.2165$ \\
    F64P13    & 1.000 & 20 & 2 & $-0.2334$ & $-0.2329$ & $-0.2319$ & $-0.2299$ & $-0.2259$ & $-0.2179$ \\
    \hline                                                           
    G32P35    & 1.000 & 8 & 3 & $-0.2135$ & $-0.2115$ & $-0.2055$ & $-0.1875$ & & \\
    G36P29    & 1.000 & 10 & 2 & $-0.2150$ & $-0.2145$ & $-0.2135$ & $-0.2115$ & $-0.2055$ & $-0.1875$ \\
    \hline                                                           
    H48P32    & 1.000 & 12 & 2 & $-0.1850$ & $-0.1845$ & $-0.1835$ & $-0.1815$ & $-0.1775$ & $-0.1695$ \\
    \hline 
    I64P31    & 1.000 & 25 & 2 & $-0.1569$ & $-0.1564$ & $-0.1554$ & $-0.1534$ & $-0.1494$ & $-0.1414$ \\
    I64P19    & 1.000 & 25 & 2 & $-0.1585$ & $-0.1580$ & $-0.1570$ & $-0.1550$ & $-0.1510$ & $-0.1430$ \\
    \hline    
    \hline
    \end{tabular}%
  \label{tab:Hasenbusch}%
\end{table}

During the generation of the gauge configurations, we separate the integration time $\tau$ between neighborhood trajectories into \(N_{\mathrm{step},f}\) and \(N_{\mathrm{step},g}\) sub-steps for the fermion and gauge, respectively to suppress the violation of the Hamiltonian. The corresponding substep sizes are \(\delta\tau_f = \tau / N_{\mathrm{step},f}\) for the fermions and \(\delta\tau_g = \tau / N_{\mathrm{step},g}\) for the gauge fields.

To further improve the efficiency of our simulations, we employ Hasenbusch mass preconditioning~\cite{Hasenbusch:2001ne}, which accelerates the Hybrid Monte Carlo (HMC) algorithm by reducing the condition number of the fermion matrix and thereby increasing the acceptance rate. This technique factorizes the original fermion determinant \(\det M(m_0)\) using a sequence of masses \(m_0 < m_1 < \cdots < m_n\):
\begin{align}
    \det M(m_0) = \det M(m_n) \prod_{i=0}^{n-1} \det \big( M(m_i) M^{-1}(m_{i+1}) \big).
\end{align}

In this factorization, the evaluation of \(M(m_0)\) is effectively replaced by that of \(M(m_n)\), which, having a larger mass, is better conditioned. Moreover, each ratio \(M(m_i)M^{-1}(m_{i+1})\) benefits from the cancellation of infrared modes between numerator and denominator, leading to further improvements in conditioning. Each determinant factor is represented by an independent pseudofermion field.

The integration step size \(\tau\), the number of fermion and gauge field substeps, \(N_{\mathrm{step},f}\) and \(N_{\mathrm{step},g}\) and specific choices of Hasenbusch masses \(m_i\)  are listed in Table~\ref{tab:Hasenbusch}.

\begin{table} 
\centering
\renewcommand\arraystretch{1.2}
\caption{ $u^{\rm I}_0$ and $v^{\rm I}_{0}$ are the tadpole improvement factors used in the gauge and fermion actions which are also listed in Table~\ref{tab:ensemble_info}. $u_0$ and $v_{0}$ are the measured values from the realistic configurations. $\tilde{w}_0=w_0/a$ is the dimensionless scale parameter determined from gradient flow.}
\resizebox{1.0\columnwidth}{!}{
\begin{tabular}{c|llllllllll}
\hline\hline
                 &C24P34  &C24P29  &C32P29  &C32P23  &C48P23  &C48P14  &E28P35  &E32P29  &E32P22   \\
\hline
$u_0^\mathrm{I}$ &0.855453  &0.855453  &0.855453  &0.855520  &0.855520  &0.855548  &0.859645  &0.859727  &0.859737 \\
$u_0$            &0.855262(4)  &0.855438(4)  &0.855430(2)  &0.855528(3)  &0.855524(1)  &0.855529(1)  &0.859650(3)  &0.859705(5)  &0.859776(3) \\
$v_0^\mathrm{I}$ &0.951479  &0.951479  &0.951479  &0.951545  &0.951545  &0.951570  &0.954385  &0.954467  &0.954469 \\
$v_0$            &0.951280(4)  &0.951460(3)  &0.951453(2)  &0.951550(3)  &0.951548(1)  &0.951553(1)  &0.954391(3)  &0.954512(4)  &0.954444(4) \\
$\tilde{w}_0$          &1.5485(18)  &1.6040(17)  &1.5996(11)  &1.6328(19)  &1.6333(10)  &1.6424(9)  &1.8359(29)  &1.8706(43)  &1.9135(21) \\
\hline
\hline
                 &F32P30  &F48P30  &F32P21  &F48P21  &F64P14  &G36P29  &G32P35  &H48P32  &I64P19  &I64P30  \\
\hline
$u_0^\mathrm{I}$ &0.863437  &0.863473  &0.863488  &0.863499  &0.863533  &0.866470  &0.866456  &0.873378  &0.881474  &0.881466 \\
$u_0$            &0.863461(1)  &0.863460(1)  &0.863519(2)  &0.863515(1)  &0.863531(1)  &0.866473(1)  &0.866445(2)  &0.873373(1)  &0.881474(1)  &0.881463(1) \\
$v_0^\mathrm{I}$ &0.956942  &0.956984  &0.957017  &0.957006  &0.957033  &0.958910  &0.958910  &0.963137  &0.967703  &0.967695 \\
$v_0$            &0.956970(1)  &0.956968(1)  &0.957024(2)  &0.957020(1)  &0.957033(1)  &0.958914(1)  &0.958888(2)  &0.963135(1)  &0.967703(1)  &0.967694(1) \\
$\tilde{w}_0$          &2.1692(39)  &2.1645(16)  &2.2382(40)  &2.2244(22)  &2.2351(20)  &2.4317(38)  &2.4129(35)  &3.2018(40)  &4.5719(63)  &4.4914(48) \\
\hline\hline
\end{tabular}
}
\renewcommand\arraystretch{1}
\end{table}

In Refs.~\cite{CLQCD:2023sdb, CLQCD:2024yyn}, lattice spacings are determined by a global fitting method using Eq.~(A12) in Ref.~\cite{CLQCD:2023sdb}. Since there are more gauge ensembles at different lattice spacings at present, Eq.~(A12) can be adapted into a more flexible form,
\begin{align}
    \begin{aligned}
        a_{w_0}&(\hat{\beta}, \tilde{m}_{\pi}, \tilde{m}_{\eta_s}, \delta u_0, \delta v_0)
    = \ a(\hat{\beta}) \left[ 1 + \sum_{i=1}^{N_{c_l}} c_{l, i} \left( \frac{\tilde{m}_{\pi}^2}{a(\hat{\beta})^2} - m_{\pi,\text{phys}}^2 \right)^i \right.  + \sum_{i=1}^{N_{c_s}} c_{s, i} \left( \frac{\tilde{m}_{\eta_s}^2}{a(\hat{\beta})^2} - m_{\eta_s,\text{phys}}^2 \right)^i \\
    & + \sum_{i=1}^{N_{u_0}} c_{u_0, i} (u_0 - u_0^I)^i + \sum_{i=1}^{N_{v_0}} c_{v_0, i} (v_0 - v_0^I)^i  + c_L e^{-\tilde{m}_{\pi} \tilde{L}}\\
    & + c^{u_0}_{l}(u_0 - u_0^I)\left( \frac{\tilde{m}_{\pi}^2}{a(\hat{\beta})^2} - m_{\pi,\text{phys}}^2 \right) + \left.c^{v_0}_{l}(v_0 - v_0^I)\left( \frac{\tilde{m}_{\pi}^2}{a(\hat{\beta})^2} - m_{\pi,\text{phys}}^2 \right)\right]~.
    \end{aligned}
\end{align}
It is possible to obtain a variety of global fit ansatzes by setting different coefficients in the above equation to zero. In the global fit, we generate 1000 bootstrap samples of the gradient flow parameter $w_0/a$ for each ensemble. These samples are generated independently since there is no correlation among different ensembles. The bootstrap samples of dimensional $w_0$ are the same for different ensembles and the value $0.17355(92)\,\mathrm{fm}$ from FLAG \cite{FlavourLatticeAveragingGroupFLAG:2024oxs} are used. In the fit for a specific bootstrap sample, we also utilize bootstrap samples for $\tilde{m}_{\pi}$, $\tilde{m}_{\eta_s}$, $u_0$ and $v_0$ to account for the uncertainties from these quantities. The fitting results for 16 different ansatzes are listed in the Table~\ref{tab:a_fit_results} and the coefficients that are set to zero in the fit are indicated by hyphens in the table. 
\begin{table}[h] 
\centering
\renewcommand\arraystretch{1.3}
\caption{Lattice spacings fit parameters for different ansatzes. The uncertainties are the total uncertainties.}
\label{tab:a_fit_results}
\resizebox{1.0\columnwidth}{!}{
\begin{tabular}{c|cccccccc}
\hline\hline
    &fit 1 &fit 2 &fit 3 &fit 4 &fit 5 &fit 6 &fit 7 & fit 8 \\
    \hline
$a(\hat{\beta}=6.200)$ &0.10539(62) &0.10542(63) &0.10538(63) &0.10509(59) &0.10557(64) &0.10550(70) &0.10557(64) &0.10562(65)  \\
$a(\hat{\beta}=6.308)$ &0.09019(55) &0.09014(57) &0.09028(56) &0.09062(56) &0.09000(57) &0.08999(67) &0.09009(59) &0.08994(59)  \\
$a(\hat{\beta}=6.410)$ &0.07761(46) &0.07759(47) &0.07764(46) &0.07766(44) &0.07761(48) &0.07760(51) &0.07764(48) &0.07758(48)  \\
$a(\hat{\beta}=6.498)$ &0.06901(43) &0.06897(44) &0.06903(43) &0.06909(41) &0.06891(45) &0.06879(48) &0.06894(45) &0.06888(45)  \\
$a(\hat{\beta}=6.720)$ &0.05235(32) &0.05230(33) &0.05238(33) &0.05241(31) &0.05237(33) &0.05234(37) &0.05241(34) &0.05232(34)  \\
$a(\hat{\beta}=7.020)$ &0.03761(23) &0.03760(23) &0.03762(23) &0.03757(22) &0.03765(24) &0.03765(25) &0.03766(24) &0.03764(24)  \\
$c_{l,1}\,[\mathrm{GeV^{-2}}]$ &0.425(14) &0.483(61) &0.420(17) &0.412(16) &0.428(15) &0.58(12) &0.422(20) &0.478(76)  \\
$c_{s,1}\,[\mathrm{GeV^{-2}}]$ &0.077(18) &0.081(33) &0.075(23) &0.020(24) &0.107(15) &0.152(67) &0.104(17) &0.107(46)  \\
$c_{u_0,1}$ &-118(29) &-125(33) &-1.2(1.2)e+02 &-3.7(1.0)e+02 &- &- &- &-  \\
$c_{v_0,1}$ &36(21) &25(21) &16(79) &0.5(1.2)e+02 &- &- &- &-  \\
$c_{u_0,2}$ &- &- &-1.2(4.7)e+06 &- &-3.5(1.5)e+06 &-7.1(5.5)e+06 &-4.7(3.8)e+06 &-3.4(1.5)e+06  \\
$c_{v_0,2}$ &- &- &1.0(4.4)e+06 &- &3.5(1.4)e+06 &6.6(5.1)e+06 &4.5(3.7)e+06 &3.5(1.5)e+06  \\
$c_{l,2}\,[\mathrm{GeV^{-4}}]$ &- &-0.59(61) &- &- &- &-1.7(1.4) &- &-0.50(78)  \\
$c_{s,2}\,[\mathrm{GeV^{-4}}]$ &- &-0.58(39) &- &- &- &-0.06(1.07) &- &-0.57(71)  \\
$c_l^{u_0}\,[\mathrm{GeV^{-2}}]$ &- &- &- &3.7(2.3)e+03 &- &1.5(2.8)e+03 &0.4(2.0)e+03 &-  \\
$c_l^{v_0}\,[\mathrm{GeV^{-2}}]$ &- &- &- &-1.7(2.5)e+03 &- &-1.7(2.3)e+03 &-0.5(1.4)e+03 &-  \\
$c_L$ &- &- &- &- &- &- &- &-  \\
$\chi^2/\mathrm{dof}$ &0.57 &0.7 &0.74 &0.61 &0.61 &1.06 &0.79 &0.76 \\
\hline\hline
&fit 9 &fit 10 &fit 11 &fit 12 &fit 13 &fit 14 &fit 15 & fit 16 \\
\hline
$a(\hat{\beta}=6.200)$ &0.10534(62) &0.10536(63) &0.10530(63) &0.10491(59) &0.10552(65) &0.10535(73) &0.10551(65) &0.10555(66)  \\
$a(\hat{\beta}=6.308)$ &0.09013(56) &0.09006(57) &0.09018(58) &0.09048(56) &0.08993(58) &0.08981(71) &0.09001(60) &0.08984(59)  \\
$a(\hat{\beta}=6.410)$ &0.07756(46) &0.07752(47) &0.07757(47) &0.07753(44) &0.07756(48) &0.07748(53) &0.07758(48) &0.07751(49)  \\
$a(\hat{\beta}=6.498)$ &0.06896(44) &0.06890(44) &0.06893(45) &0.06894(41) &0.06884(45) &0.06861(53) &0.06886(46) &0.06880(46)  \\
$a(\hat{\beta}=6.720)$ &0.05232(33) &0.05225(33) &0.05232(33) &0.05231(31) &0.05233(34) &0.05224(38) &0.05236(35) &0.05226(35)  \\
$a(\hat{\beta}=7.020)$ &0.03757(24) &0.03754(24) &0.03755(24) &0.03744(22) &0.03760(25) &0.03754(28) &0.03760(25) &0.03757(25)  \\
$c_{l,1}\,[\mathrm{GeV^{-2}}]$ &0.429(17) &0.488(61) &0.427(19) &0.426(18) &0.433(17) &0.61(13) &0.428(21) &0.489(77)  \\
$c_{s,1}\,[\mathrm{GeV^{-2}}]$ &0.078(18) &0.079(33) &0.074(23) &0.012(24) &0.108(15) &0.164(66) &0.104(17) &0.110(46)  \\
$c_{u_0,1}$ &-120(29) &-129(35) &-1.2(1.2)e+02 &-4.0(1.1)e+02 &- &- &- &-  \\
$c_{v_0,1}$ &38(22) &28(22) &13(83) &0.5(1.2)e+02 &- &- &- &-  \\
$c_{u_0,2}$ &- &- &-1.6(5.1)e+06 &- &-3.8(1.7)e+06 &-8.1(6.1)e+06 &-5.1(4.1)e+06 &-3.7(1.8)e+06  \\
$c_{v_0,2}$ &- &- &1.3(4.7)e+06 &- &3.7(1.6)e+06 &7.4(5.6)e+06 &4.9(4.0)e+06 &3.7(1.7)e+06  \\
$c_{l,2}\,[\mathrm{GeV^{-4}}]$ &- &-0.57(62) &- &- &- &-2.0(1.5) &- &-0.54(78)  \\
$c_{s,2}\,[\mathrm{GeV^{-4}}]$ &- &-0.65(39) &- &- &- &0.04(1.08) &- &-0.57(71)  \\
$c_l^{u_0}\,[\mathrm{GeV^{-2}}]$ &- &- &- &4.1(2.3)e+03 &- &1.7(2.9)e+03 &0.5(2.1)e+03 &-  \\
$c_l^{v_0}\,[\mathrm{GeV^{-2}}]$ &- &- &- &-1.7(2.5)e+03 &- &-2.0(2.5)e+03 &-0.5(1.5)e+03 &-  \\
$c_L$ &0.011(24) &0.018(24) &0.022(27) &0.039(25) &0.015(27) &0.030(34) &0.018(29) &0.018(28)  \\
$\chi^2/\mathrm{dof}$ &0.64 &0.81 &0.86 &0.68 &0.69 &1.32 &0.91 &0.89 \\
\hline\hline
\end{tabular}
}
\renewcommand\arraystretch{1.1}
\end{table}

We follow the strategy in Refs.~\cite{Borsanyi:2020mff, BMW:2014pzb} to evaluate model average values for lattice spacings. A brief overview of the method is provided here. The global fit within a certain ansatz is called an analysis, and a weight $w_i$ given by the Akaike Information Criterion (AIC) can be assigned to each analysis and 
\begin{align}
    w_i = \frac{\exp\left[-\frac{1}{2}\left(\chi_i^2+2n_{i,p}-n_{i,d}\right)\right]}{\sum_j \exp\left[-\frac{1}{2}\left(\chi_j^2+2n_{j,p}-n_{j,d}\right)\right]}~,
\end{align}
where $\chi_i^2$, the number of fit parameters $n_{i,p}$ and the number of the data points $n_{i,d}$ are from the $i$-th analysis. The uncertainties for $a_{w_0}(\hat{\beta})$ of each analysis are assumed to obey a Gaussian distribution $N(a;m_i,\sigma_i)$ where $m_i$ is the central value and $\sigma_i$ is the standard deviation. Then the joint probability distribution is defined by
\begin{align}
    N(a; \lambda) = \sum_i w_i N(a;m_i,\sqrt{\lambda}\sigma_i)
    \label{AICjointdis}
\end{align}
and the cumulative distribution function (CDF) is expressed by $P(a;\lambda)=\int_{-\infty}^{a} \mathrm{d} a' \,N(a'; \lambda)$. The central value of lattice spacing $a$ is given by the median of the CDF and the error is estimated by the difference of values $a$ at $16\%$ and $84\%$ percentiles of the CDF:
\begin{align}
    \sigma^2=\left[(a_{84}-a_{16})/2\right]^2\, \quad \text{where}\; P(a_{16}; 1)=0.16\; \text{and}\; P(a_{84}; 1)=0.84\,.
\end{align}
$\lambda$ in Eq.~(\ref{AICjointdis}) is used to rescale a certain kind of uncertainty and help to separate it from other uncertainties. In the current case, it is assumed that the total error can be decomposed in the form of $\sigma^2_{\mathrm{total}}=\sigma^2_{w_0}+\sigma^2_{\mathrm{stat+ansatz}}$.
If we rescale the error from $w_0$ by $\lambda=2$, we will obtain an total error $\dot{\sigma}^2_{\mathrm{total}}=2\sigma^2_{w_0}+\sigma^2_{\mathrm{stat+ansatz}}$ according to the above model average method. Then $\sigma^2_{w_0}$ is estimated by $\dot{\sigma}^2_{\mathrm{total}} - \sigma^2_{\mathrm{total}}$. $\sigma_{\mathrm{stat+ansatz}}$ includes the statistical uncertainties of $w_0/a$ determined by the gradient flow and the uncertainties from different fit ansatzes. Fig.~\ref{fig:latt_spac_CDF} shows CDFs of lattice spacings at different $\hat{\beta}$. Lattice spacings at different values of $\hat{\beta}$ are summarized in Table~\ref{tab:a_beta}. 

\begin{table}[h] 
\centering
\renewcommand\arraystretch{1.3}
\caption{Lattice spacings at different values of $\hat{\beta}$. The first error is $\sigma_{w_0}$ and the second error is $\sigma_{\mathrm{stat+ansatz}}$.}
\label{tab:a_beta}
\begin{tabular}{cccccc}
\hline\hline
$a(\hat{\beta}=6.200)$  & $a(\hat{\beta}=6.308)$ & $a(\hat{\beta}=6.410)$ & $a(\hat{\beta}=6.498)$ & $a(\hat{\beta}=6.720)$ & $a(\hat{\beta}=7.020)$ \\
$0.10542(17)(62)$  &$0.09013(25)(53)$  &$0.07760(07)(46)$  &$0.06895(17)(41)$  &$0.05235(11)(31)$  &$0.03761(08)(22)$   \\ 
\hline\hline
\end{tabular}
\end{table}
\begin{figure}[thb]
    \centering
    \includegraphics[width=1.0\linewidth]{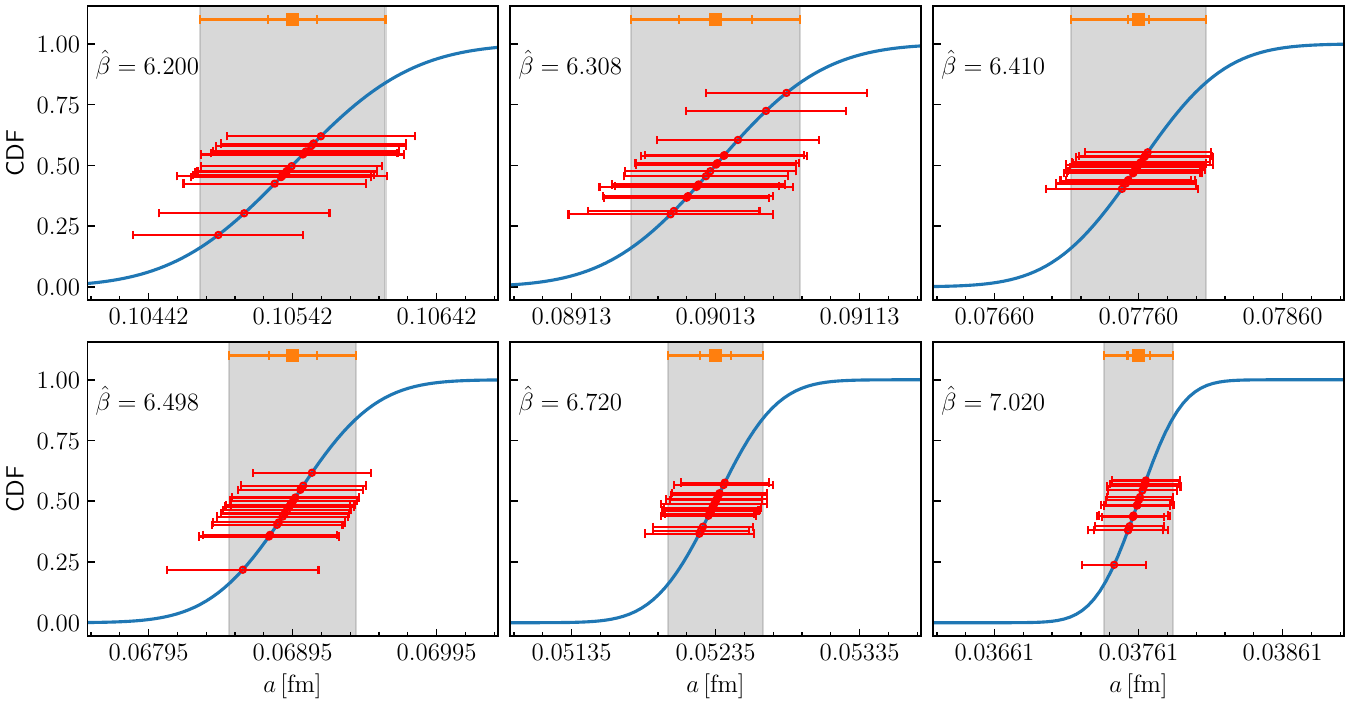}
    \caption{CDFs of lattice spacings at different values of $\hat{\beta}$. The red data points along the CDF curves represent the results obtained from different fit ansatzes. The orange squares above indicate the model average values, where the inner error bar shows $\sigma_{\mathrm{stat+ansatz}}$ and the outer error bar shows the total error.}
    \label{fig:latt_spac_CDF}
\end{figure}

\subsection{Anisotropic lattice heavy quark action}
\label{sec:aniso_action}
The bottom quark is simulated with the anisotropic lattice fermion action~\cite{Chen:2000ej} which is expressed by
\begin{equation}
S = a^4\sum_x \bar{\psi}(x) \left[ m_Q + \gamma_4 \nabla_4 - \frac{a}{2} \nabla_4^2
        + \nu \sum_{i=1}^3 \Big( \gamma_i \nabla_i - \frac{a}{2} \nabla_i^2 \Big) - c_E(\nu,u_0) \, \frac{a}{2} \sum_{i=1}^{3} \sigma_{i4} F_{i4}
        - c_B(\nu,u_0) \, \frac{a}{2} \sum_{i>j=1}^{3} \sigma_{ij} F_{ij}\right] \psi(x),
\end{equation}
where $a$ is the lattice spacing, $\nabla_{0(i)}$ and $\Delta_{0(i)}$ are the first and second order derivatives in the temporal (spatial) direction, $F_{\mu\nu}$ is the field strength, $\nu$ is an anisotropy parameter, and $c_E=(1 + \nu)/(2u_0^3)$ and $c_B=\nu/u_0^3$ are the tadpole-improved clover coefficients in the temporal and spatial direction~\cite{Chen:2000ej}.

\begin{figure}
    \centering
    \includegraphics[width=1.0\linewidth]{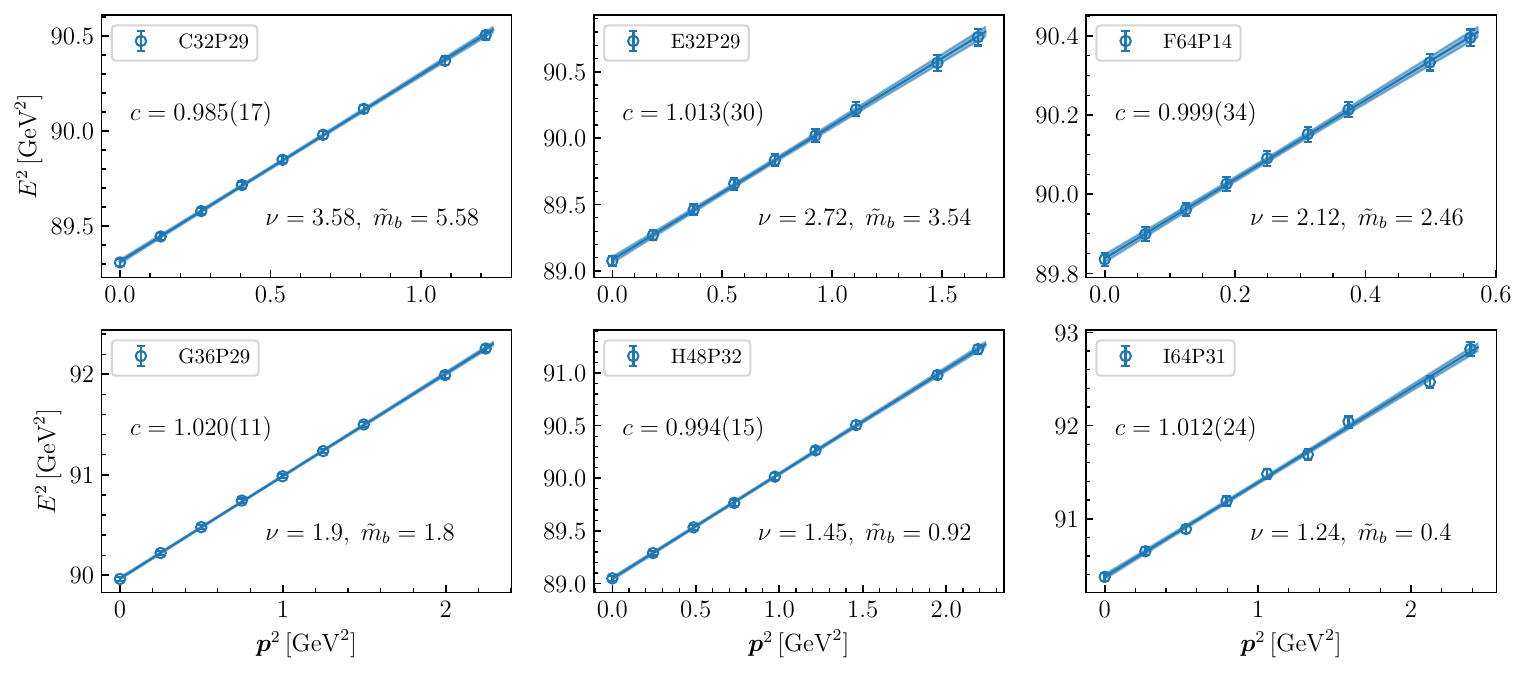}
    \caption{Dispersion relation of $\Upsilon$ meson on different ensembles at specified values of $(\nu, \tilde{m}_b)$. The analogous plots for other values of $(\nu, \tilde{m}_b)$ and other ensembles look similar. The values of $c$ extracted from fits, together with the effective mass $m_V$, are used to tune the target parameters $\nu$ and $\tilde{m}_b$, as explained in the text.}
    \label{fig:dispersion_relation}
\end{figure}

Since the calculation is performed on the $2+1$ flavor ensembles, there is no bottom quark in the sea and the bare bottom quark mass $\tilde{m}_b$ is an undetermined parameter. We jointly tuning the values of $\nu$ and $\tilde{m}_b$ in the following procedure. We initially select three values of $\nu$ and three values of $\tilde{m}_b$ which are indicated by superscripts $k=1, 2, \text{and}\,3$ and $l=1, 2, \text{and}\, 3$, respectively. Under each combination of $(\nu^k,\, \tilde{m}^l_b)$, we extract the energy $E(\boldsymbol{p})$ of vector meson from the two-point function constructed by vector interpolating operator
\begin{align}
    C(t, \boldsymbol{p}) = \frac{1}{3}\sum_i\sum_{\boldsymbol{x}} e^{-i\boldsymbol{p}\cdot\boldsymbol{x}}\langle \bar{\psi}(\boldsymbol{x},t)\gamma^i\psi(\boldsymbol{x},t)\bar{\psi}(0)\gamma^i\psi(0)\rangle~,
\end{align}
using point source propagators and project to the sink momentum $a\boldsymbol{p} = 2\pi\boldsymbol{n}/L$ with
\begin{align}
    \boldsymbol{n}=(0,0,0),\, (0,0,1),\, (0,1,1),\, (1,1,1),\, (0,0,2),\, (0,1,2),\, (1,1,2),\, (0,2,2),\;\text{and}\;(1,2,2)\, .
\end{align}
We then fit the dispersion relation $E^{kl}(\boldsymbol{p})^2 = (m^{kl}_V)^2 + (c^{kl})^2\boldsymbol{p}^2$ using  $E(\boldsymbol{p})$, and $\boldsymbol{p}$ to get nine corresponding $c^{kl}$ values. Some fit results are shown in Fig.~\ref{fig:dispersion_relation}. Next, we construct a joint fitting formula as
\begin{align}
    \begin{aligned}
    m^{kl}_V &= m_V^i + d_1 (\nu^k-\nu^i) + d_2 (\tilde{m}^l_b - \tilde{m}^i_b)\, ,\\
    (c^{kl})^2 &= 1 + d_3 (\nu^k-\nu^i) + d_4 (\tilde{m}^l_b - \tilde{m}^i_b)\, .
\end{aligned}
\label{eq:joint_nu_mb}
\end{align}
where $d_1$, $d_2$, $d_3$, $d_4$, $\nu^i$ and $\tilde{m}^i_b$ are viewed as fitting parameters, once the value of $m_V^i$ is given. It is expected that  $c$ will be $1$ for the vector meson with mass $m_V^i$, when parameter $(\nu^i, \tilde{m}^i_b)$ is utilized in the propagator calculations.

On each ensemble, we can determine an unitary $(\nu, m_b)$ with which the vector meson mass is $\Upsilon$ meson mass $9460.40(10)\,\mathrm{MeV}$, and the dispersion relation $E(\boldsymbol{p})^2 = M_\Upsilon^2 + \boldsymbol{p}^2$ is satisfied. For the numerical evaluation, we select three values of $m_V^i$ in the vicinity of $m_\Upsilon$ and determine the corresponding sets $(\nu^i, \tilde{m}_b^i)$ from Eq.~(\ref{eq:joint_nu_mb}). At each set of $(\nu^i, \tilde{m}_b^i)$, we calculate vector meson mass $m^i_V(\nu^i, \tilde{m}_b^i)$ and other quantities $X^i(\nu^i, \tilde{m}_b^i)$. The difference between $M^i$ and $M_\Upsilon$ is constrained within a suitable range, therefore a linear function is sufficient to fit the relation between these quantities and $\tilde{m}_b$,
\begin{align}
    \tilde{m}_V^i = d^0_{\tilde{m}_V} \tilde{m}^i_b + d^1_{\tilde{m}_V}\, ,\qquad
    \tilde{X}^i = d^0_{\tilde{X}} \tilde{m}^i_b + d^1_{\tilde{X}}\, . \label{eq:MV_X_mb}
\end{align}
The coefficients $d^0_{\tilde{m}_V(\tilde{X})}$ and $d^1_{\tilde{m}_V(\tilde{X})}$, obtained from the linear fit, include the statistical uncertainties associated with the effective masses (quantity $\tilde{X}$). The physical bare bottom quark mass $\tilde{m}_b(m_\Upsilon)$ corresponding to $\Upsilon$ meson mass, determined using $(am_\Upsilon - d_{\tilde{m}_V}^1)/d_{\tilde{m}_V}^0$, also incorporates the experimental uncertainty in $m_\Upsilon$ and uncertainties arising from the lattice spacing. However, for $X(m_\Upsilon)$ derived via $ d^0_{\tilde{X}} \tilde{m}_b(m_\Upsilon) + d^1_{\tilde{X}}$, its uncertainties due to the lattice spacing largely cancel out if the quantity $X$ has the same dimension as $m_V$.

\begin{table}[h] 
\centering
\renewcommand\arraystretch{1.3}
\caption{Interpolated anisotropy parameter $\nu$ and bare bottom quark mass $\tilde{m}_b$ on different ensembles in the form of $(\nu, \tilde{m}_b)$ when $M^i$ in Eq.~(\ref{eq:joint_nu_mb}) is equal to Upsilon mass. }
\label{tab:unitary_nu_mb}
\begin{tabular}{c|cccc}
\hline\hline
&C24P34 &C24P29 &C32P29 &C32P23 \\
$[\nu, \tilde{m}_b]$&[3.666(16), 5.533(11)] &[3.658(9), 5.554(7)] &[3.620(14), 5.580(10)] &[3.598(12), 5.601(9)] \\
\hline\hline
&C48P23 &C48P14 &E28P35 &E32P29 \\
$[\nu, \tilde{m}_b]$&[3.567(36), 5.625(25)] &[3.554(26), 5.634(18)] &[2.720(9), 3.572(6)] &[2.707(16), 3.576(10)] \\
\hline\hline
&E32P22 &F32P30 &F32P21 &F48P21 \\
$[\nu, \tilde{m}_b]$&[2.651(18), 3.620(11)] &[2.154(9), 2.422(5)] &[2.143(8), 2.434(4)] &[2.141(20), 2.433(11)] \\
\hline\hline
&F64P14 &G36P29 &H48P32 &I64P30 \\
$[\nu, \tilde{m}_b]$&[2.155(20), 2.426(10)] &[1.904(6), 1.783(3)] &[1.476(7), 0.918(3)] &[1.225(9), 0.398(3)] \\
\hline\hline
\end{tabular}
\renewcommand\arraystretch{1.0}
\end{table}

The interpolated $\nu$ and $\tilde{m}_b$ which yield physical $\Upsilon$ meson mass on each ensemble are collected in Table~\ref{tab:unitary_nu_mb}. The bare quark mass $(\tilde{m}_b-\tilde{m}_{\rm crti})/a$ changes by a factor of 3 from $a=0.105$ fm to $a=0.04$ fm due to the discreitzaiton error, while the corresponding PCAC mass is much more stable. For the anisotropy parameter $\nu$, a global fitting with the following ansatz,
\begin{align}
\nu(m_h a, m_{\pi}, m_{\eta_s}) = \frac{\sinh(c_{\nu}  m_h a)}{c_{\nu}  m_h a}\Big[ \nu_0 + \sum_{PS=\pi,\eta_s} c_{PS}\big(m_{PS}^2 - m_{PS,\mathrm{phys}}^2\big) \Big],
\label{eq:nu_empirical_sm}
\end{align}
gives $c_{\nu}=0.6119(25)$, $\nu_0 = 1.0063(95)$, and the coefficient $c_{\pi}=0.250(95)~\mathrm{GeV}^{-2}$ and $c_{\eta_s}=0.012(66)~\mathrm{GeV}^{-2}$ are introduced to take account of the quark mass dependence of the lattice spacing determination. We have rescaled the uncertainty of $\nu$ by a factor of $1.8$ to bring the $\chi^2/\mathrm{dof}$ of the fit to approximately unity. Fig.~\ref{fig:anisotropic_vs_a2} shows the variance of $\nu$ with lattice spacing. The data points show the values after the extrapolation to physical $\pi$ and $\eta_s$ meson masses, which are explicitly expressed as
\begin{align}
    \nu(m_h a, m^{\mathrm{phys}}_{\pi}, m^{\mathrm{phys}}_{\eta_s}) = \nu(m_h a, m_{\pi}, m_{\eta_s}) - \frac{\sinh(c_{\nu}  m_h a)}{c_{\nu}  m_h a}\sum_{PS=\pi,\eta_s} c_{PS}\big(m_{PS}^2 - m_{PS,\mathrm{phys}}^2\big)\,,
\end{align}
and the bands represent the results of \(\nu_0\sinh(c_\nu m_h a) / (c_\nu m_h a)\).

\begin{figure}
    \centering
    \includegraphics[width=0.5\linewidth]{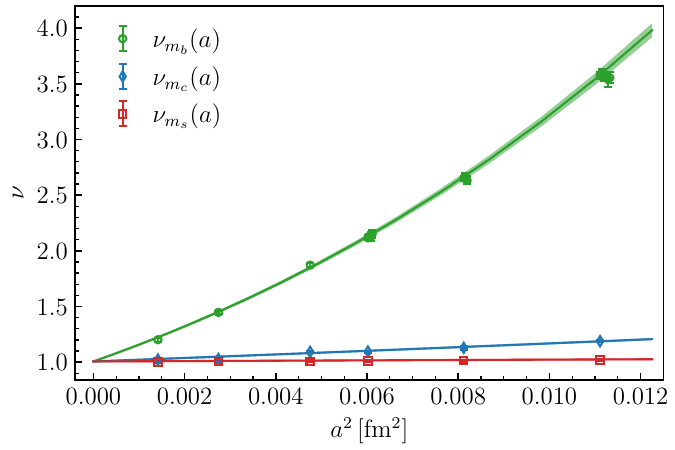}
    \caption{Anisotropy parameter \(\nu\) for the bottom, charm, and strange quarks as a function of \(a^2\). The bands represent the empirical form \(\nu_0\sinh(c_\nu m_h a) / (c_\nu m_h a)\) with \(c_\nu = 0.6119(25)\). The meson mass \(m_h\) is taken as the corresponding quarkonium mass: \(m_\Upsilon\sim m_{\eta_b}\) for bottom, \(m_{J/\psi}\sim m_{\eta_c}\) for charm, and \(\sim1\rm GeV\) for strange.}
    \label{fig:anisotropic_vs_a2}
\end{figure}

The same procedure can be extended to lighter flavors. For example, the anisotropy parameter \(\nu\) for the charm quark can be determined from the \(J/\psi\) mass and its dispersion relation. As shown in Fig.~\ref{fig:anisotropic_vs_a2} and Table~\ref{tab:nu_mc}, the resulting values for charm (blue diamonds) are consistent with the empirical parametrization (blue band) from Eq.~\ref{eq:nu_empirical_sm} and lie significantly closer to unity than those for the bottom quark. Table~\ref{tab:nu_mc} also provides the value of \(\nu\) obtained from the empirical formula in Eq.~(\ref{eq:nu_empirical_sm}) using \(m_{\eta_c}\); the difference between the two determinations is only 1.15\% at the coarsest lattice spacing. Similarly, the difference between using \(m_{\eta_b}\) and \(m_{\Upsilon}\) is 1.34\%, and both approach unity in the continuum limit with \(\mathcal{O}(a^2)\) corrections.

This trend continues for successively lighter quarks. For the strange quark (black triangles), with \(m_{\phi} \sim 1\) GeV, the deviation from \(\nu = 1\) is further reduced to 1.8\% at the coarsest lattice spacing.

\begin{table}[h] 
\centering
\renewcommand\arraystretch{1.3}
\caption{Comparison of the tuned $\nu$ using the $J/\psi$ mass and its dispersion relation, and the value of $\bar{\nu}$ predicted by Eq.~(\ref{eq:nu_empirical_sm}) using $m_{J/\psi}$ and $m_{\eta_c}$ using the fitted parameters obtained from the fit of the $\nu^{\rm tuned}$ of $\Upsilon$.}
\label{tab:nu_mc}
\begin{tabular}{c|cccccc}
\hline\hline
&C24P29 &E32P29 &F32P30  &G36P29 &H48P32 &I64P30 \\
\hline
$\nu^{\rm tuned}$                       &1.2071(49) &1.146(13)  &1.1142(37) &1.1107(66) &1.0495(87) &1.04324(11) \\
$\nu(m_{J/\psi})$ &1.2065(79) &1.155(07) &1.1216(72) &1.1001(70) &1.0705(71) &1.0482(75)\\
$\nu(m_{\eta_c})$ &1.1927(79) &1.145(07)  &1.1145(72) &1.0945(70) &1.0673(71) &1.0466(75) \\
\hline\hline
\end{tabular}
\renewcommand\arraystretch{1.0}
\end{table}

\subsection{Correlation function fits}\label{sec:two-pt_correlator}
The meson decay constants are defined by the matrix elements of local operators between vacuum and meson states,
\begin{align}
    Z_A\langle 0 | A_0(0) |P(p) \rangle = f_{PS} m_{PS}\, ,\\
    Z_V\langle 0 | V_\mu(0) |V(p, \epsilon) \rangle = f_V m_V \epsilon_\mu,
\end{align}
where $A_\mu(x) = \bar{\psi}_{f_1}(x)\gamma_\mu \gamma_5 \psi_{f_2}(x)$ is the axial vector operator, $V_\mu(x) = \bar{\psi}_{f_1}(x)\gamma_\mu \psi_{f_2}(x)$ is the vector operator, $|P(p) \rangle$ is the pseudoscalar meson state with momentum $p_\mu$, and $|V(p, \epsilon) \rangle$ is the vector meson state with polarization vector $\epsilon_\mu$. In addition, the quark mass can be defined through the partially conserved axial current (PCAC) relation~\cite{JLQCD:2007xff, CLQCD:2023sdb},
\begin{align}
     \langle 0 | \partial_\mu A^\mu(x) |P(p) \rangle  = (m^{\mathrm{PC}}_1+m^{\mathrm{PC}}_2)\langle 0 | P(x) |P(p) \rangle\, ,
\end{align}
where $P(x) = \bar{\psi}_{f_1}(x)\gamma_5 \psi_{f_2}(x) $ is the pesudo-scalar operator and $m^{\mathrm{PC}}_{1(2)}$ is the PCAC quark mass for quark field $\psi_{f_1(f_2)}(x)$.
The decay constants and effective masses can be extracted from two-point correlators at large time distances. 

We calculated the Coulomb gauge fixed wall source propagators, and construct the wall-to-point correlation function $C^{\Gamma_1 \Gamma_2}_{2,wp}(t)$ and wall-to-wall correlation function $C^{\Gamma_1 \Gamma_2}_{2,ww}(t)$ as
\begin{subequations}
\label{eq:ww_wp_R}
\begin{align}
    C^{O_1 O_2}_{2,wp}(t) &= \frac{1}{L^3}\sum_{\boldsymbol{x}, \boldsymbol{y}, \boldsymbol{z}}\left\langle \bar{\psi}_{f_1}(\boldsymbol{x}, t_0+t)\Gamma_{O_1}\psi_{f_2}(\boldsymbol{x}, t_0+t) \left[\bar{\psi}_{f_1}(\boldsymbol{y}, t_0)\Gamma_{O_2}\psi_{f_2}(\boldsymbol{z}, t_0)\right]^\dagger \right \rangle~,\label{eq:C2wp}\\
    C^{O_1 O_2}_{2,ww}(t) &= \frac{1}{L^3}\sum_{\boldsymbol{x_1}, \boldsymbol{x_2}, \boldsymbol{y}, \boldsymbol{z}}\left\langle \bar{\psi}_{f_1}(\boldsymbol{x_1}, t_0+t)\Gamma_{O_1}\psi_{f_2}(\boldsymbol{x_2}, t_0+t) \left[\bar{\psi}_{f_1}(\boldsymbol{y}, t_0)\Gamma_{O_2}\psi_{f_2}(\boldsymbol{z}, t_0)\right]^\dagger \right \rangle~,\label{eq:C2ww}
\end{align}
\end{subequations}
where $\Gamma_{O}$ represents the corresponding Dirac gamma matrix for the interpolating field $O$.

Since excitded-state contamination for the two-point functions constructed by meson interpolators with heavy quark fields is non-negligible, the two-state fit is employed. To extract the pseudoscalar meson decay constant $f_{PS}$, mass $m_{PS}$ and PCAC quark mass $m^{\rm PC}$, we parameterize the combinations of the correlation functions into the following ansatz,
\begin{subequations}
\label{eq:C2PS}
\begin{align}
&C_{2, w p}^{\mathrm{P} \mathrm{P}}(t)=\frac{Z_{w p}}{2 m_{PS}}\left(e^{-m_{PS} t} + e^{-m_{PS}(T- t)}\right)\left(1+R^{\mathrm{P} \mathrm{P}}_{wp}(t)\right)\, ,\label{eq:C2PSMPS}\\
&C_{2, wp}^{\mathrm{P} \mathrm{P} } \left(t\,\right)/C_{2, ww}^{\mathrm{P} \mathrm{P} } \left(t\,\right)=\left(\frac{f_{PS} m_{PS}^2}{m_1^{\mathrm{PC}} +m_2^{\mathrm{PC}} }\right)^2 \frac{1}{Z_{w p}}\frac{1+R^{\mathrm{P} \mathrm{P}}_{wp}(t)}{1+R^{\mathrm{P} \mathrm{P}}_{ww}(t)}\, ,\label{eq:C2PSfPS} \\ 
&\left.\left[C_{2, w p}^{\mathrm{A}_4 \mathrm{P}}(t-1)-C_{2, w p}^{\mathrm{A}_4 \mathrm{P}}(t+1)\right]\right/2 C_{2, w p}^{\mathrm{PP}}(t)=\frac{\operatorname{Sinh}\left(m_{PS}\right)}{m_{PS}} (m_1^{\mathrm{PC}} +m_2^{\mathrm{PC}})\frac{1+R^{\mathrm{A_4} \mathrm{P}}_{wp}(t)}{1+R^{\mathrm{P} \mathrm{P}}_{wp}(t)}\, \label{eq:C2PSmq},
\end{align}
\end{subequations}
where $Z_{wp}$, $m_{PS}$, $f_{PS}$, and $m_{1}^{\rm PC}+m_{2}^{\rm PC}$ are fit parameters. $R^{\mathrm{\Gamma_1} \mathrm{\Gamma_2}}_{wp(ww)}$ is parameterized as
\begin{subequations}
\label{eq:R_ex}
\begin{align}
    R^{O_1 O_2}_{wp(ww)}&= r^{O_1 O_2}_{wp(ww)}\left[e^{-(m_{PS}+\delta E)t}+e^{-(m_{PS}+\delta E)(T-t)}\right]/ \left[e^{-m_{PS}t}+e^{-m_{PS}(T-t)}\right]~,\label{eq:Rexcit}
\end{align}
\end{subequations}
where $r^{O_1 O_2}_{wp(ww)}$ and $\delta E$ are additional parameters for the excited state. For the vector meson cases, we use the similar parameterization,
\begin{subequations}
\label{eq:C2VV}
\begin{align}
&C_{2, w p}^{\mathrm{V}_i \mathrm{V}_i}(t)=\frac{Z_{w p}}{2 m_V}\left(e^{-m_{V} t} + e^{-m_{V}(T- t)}\right)\left(1+R^{\mathrm{V}_i \mathrm{V}_i}_{wp}(t)\right)\, , \label{eq:C2VVMV}\\
&C_{2, wp}^{\mathrm{V}_i \mathrm{V}_i } \left(t\,\right)/C_{2, ww}^{\mathrm{V}_i \mathrm{V}_i } \left(t\,\right)=\frac{f^2_{V}m^2_{V}}{Z_{w p}}\frac{1+R^{\mathrm{V}_i \mathrm{V}_i}_{wp}(t)}{1+R^{\mathrm{V}_i \mathrm{V}_i}_{ww}(t)}\, ,\label{eq:C2VVfV}
\end{align}
\end{subequations}
and three spatial directions are averaged to improve signals.

\begin{figure}
    \centering
    \includegraphics[width=1.0\linewidth]{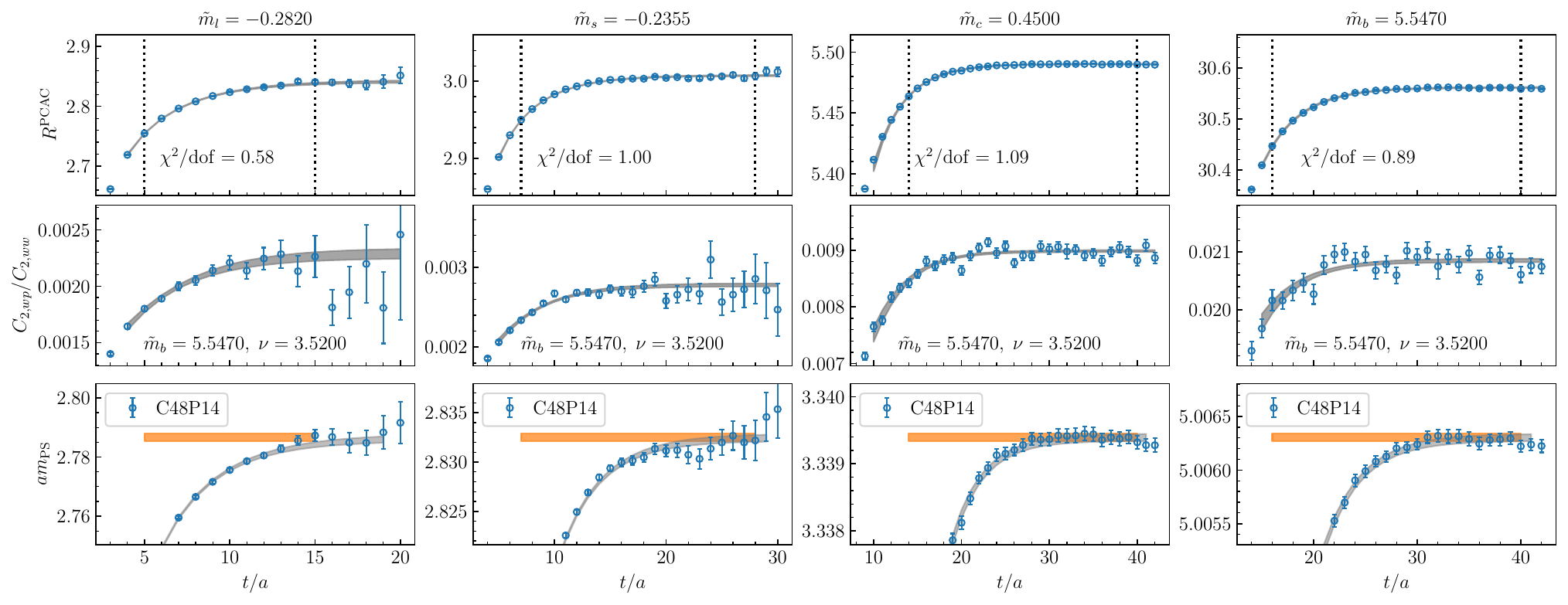}
    \caption{Fit results for Eq.~(\ref{eq:C2PS}) on ensemble C48P14. $R^{\mathrm{PCAC}}$ represents the ratio of Eq.~(\ref{eq:C2PSmq}). The $\chi^2/\mathrm{dof}$ of the fit is displayed in the top-row panels, and the bare quark masses of the lighter quark in the meson (excluding the bottom quark) are listed above the panels in the first three columns. The bare quark mass above the panels in the fourth column is for bottomonium meson. Values of $\tilde{m}_b$ and $\nu$ are shown in the middle-row panels. Data points in the bottom-row panels correspond to the effective masses defined by Eq.~(\ref{eq:eff_mass}), with gray bands representing the corresponding fit results from Eq.~(\ref{eq:C2PSMPS}). Two vertical dashed lines in the top-row panels mark the fit range. Horizontal bands in the bottom-row panels also show the fit range and the bootstrap statistical errors of effective masses.}
    \label{fig:PCAC_twopt_C48P14}
\end{figure}

\begin{figure}
    \centering
    \includegraphics[width=1.0\linewidth]{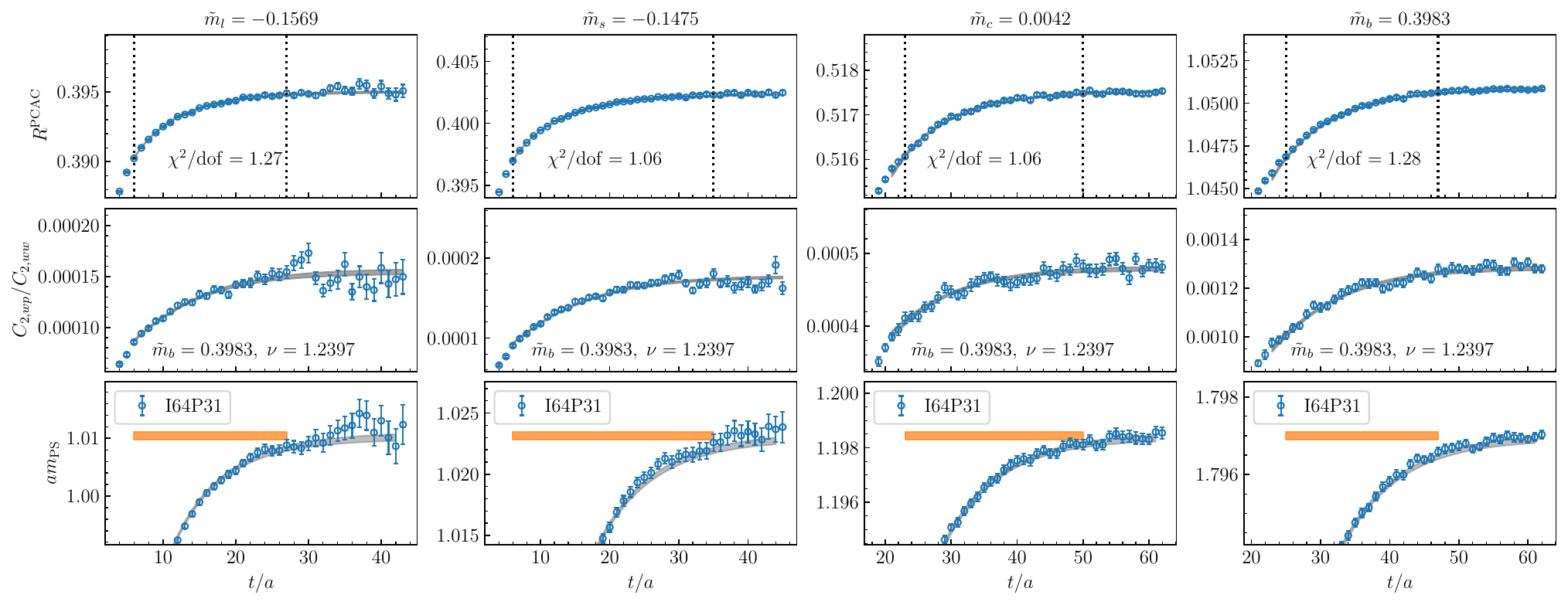}
    \caption{Fit results for Eq.~(\ref{eq:C2PS}) on ensemble I64P31. See the caption of Fig.~\ref{fig:PCAC_twopt_C48P14} for further details.}
    \label{fig:PCAC_twopt_I64P31}
\end{figure}

\begin{figure}
    \centering
    \includegraphics[width=1.0\linewidth]{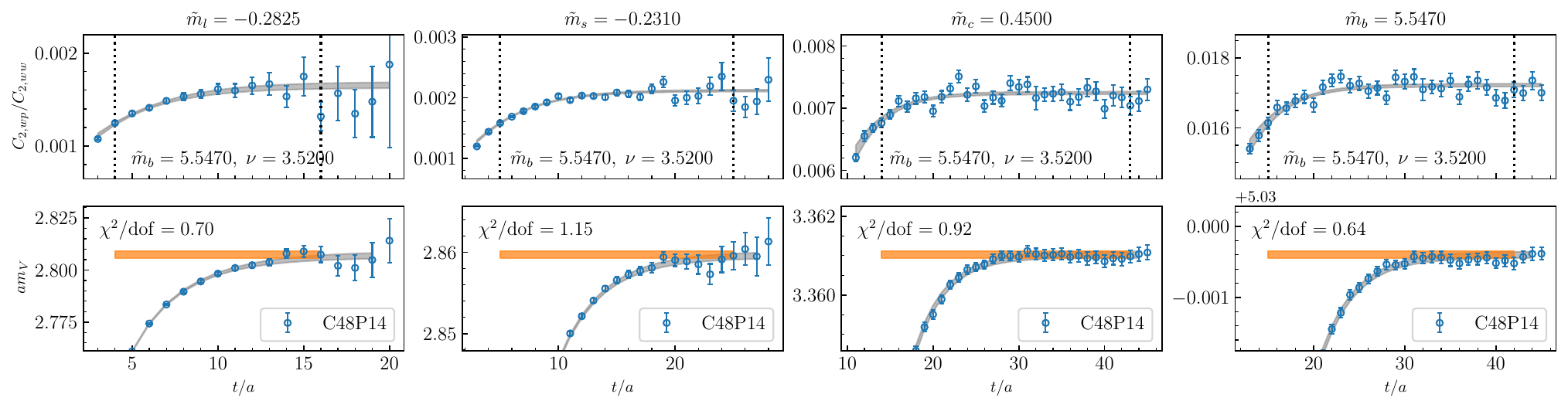}
    \caption{Fit results for Eq.~(\ref{eq:C2VV}) on ensemble C48P14. Values of $\tilde{m}_b$ and $\nu$ are shown in the top-row panels, and the bare quark masses of the lighter quark in the meson (excluding the bottom quark) are listed above the panels in the first three columns. The bare quark mass above the panels in the fourth column is for bottomonium. The $\chi^2/\mathrm{dof}$ of the fit is displayed in the top-row panels. Data points in the bottom-row panels correspond to the effective masses defined by Eq.~(\ref{eq:eff_mass}), with gray bands representing the corresponding fit results from Eq.~(\ref{eq:C2VVMV}). Two vertical dashed lines in the top-row panels mark the fit range. Horizontal bands in the bottom-row panels also show the fit range and the bootstrap statistical errors of effective masses.}
    \label{fig:vector_twopt_C48P14}
\end{figure}

\begin{figure}
    \centering
    \includegraphics[width=1.0\linewidth]{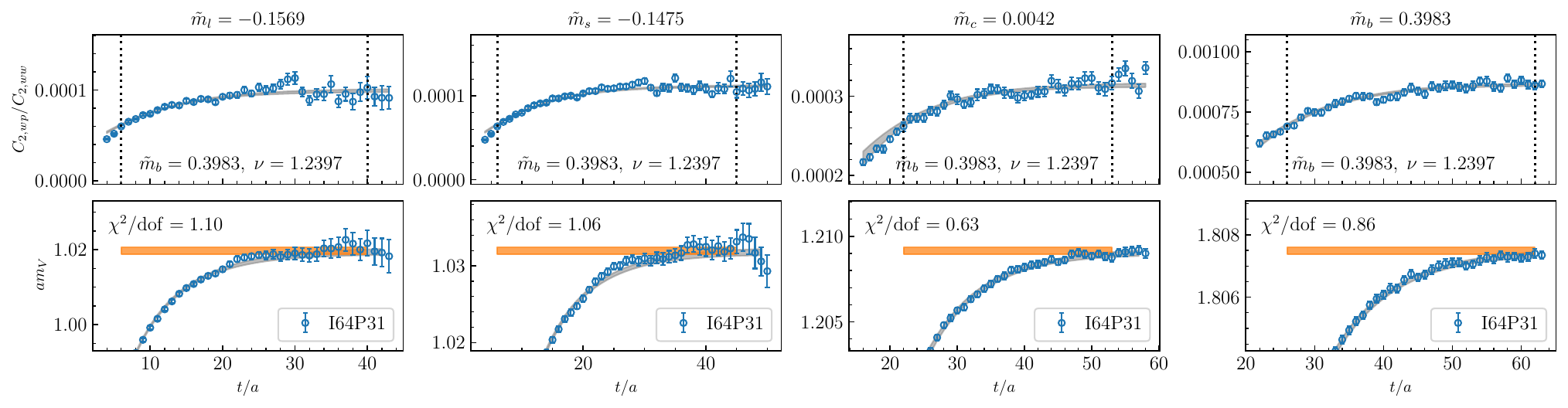}
    \caption{Fit results for Eq.~(\ref{eq:C2VV}) on ensemble I64P31. See the caption of Fig.~\ref{fig:vector_twopt_C48P14} for further details.}
    \label{fig:vector_twopt_I64P31}
\end{figure}

It is observed that for the heavy-light interpolating field, the statistical fluctuation of the two-point function will be severe at relatively large time distances. However, when both quark fields are at the bottom quark mass or when one of the quark field is at the charm quark mass, the signal will still be good even at half the entire temporal length. Figs.~\ref{fig:PCAC_twopt_C48P14} and \ref{fig:PCAC_twopt_I64P31} display the fit results of Eq.~(\ref{eq:C2PS}) for ensembles C48P14 and I64P31, respectively. Figs.~\ref{fig:vector_twopt_C48P14} and \ref{fig:vector_twopt_I64P31} show the fit results of Eq.~(\ref{eq:C2VV}) for ensembles C48P14 and I64P31, respectively. Data points in the plots at the last row of these figures are the effective masses defined by
\begin{align}
    m_{\text{eff}}(t) \equiv \operatorname{arcosh} \left[ \frac{C_{2, wp}(t-1) + C_{2, wp}(t+1)}{2 C_{2, wp}(t)} \right]\,, \label{eq:eff_mass}
\end{align}
and the bands display the fit results for effective masses with bootstrap statistical errors. The analogous plots for other values of $(\tilde{m}_b, \nu)$ and other ensembles look similar.

\subsection{Renormalization}
\subsubsection{Non-singlet operator renormalization}
The decay constants and quark masses extracted from lattice correlation functions are unrenormalized quantities. Appropriate renormalization procedure is therefore required to obtain the physical values. Given that spacetime is discretized on the lattice, the $\overline{{\rm MS}}$ scheme based on the dimensional regularization cannot be directly adopted. Renormalization constants are typically computed in the intermediate regularization independent (RI) SMOM~\cite{Sturm:2009kb} or MOM~\cite{Martinelli:1994ty} scheme, and subsequently converted to the $\overline{{\rm MS}}$ scheme with perturbative matching kernels. The evolution to a specific renormalization scale is then performed in the $\overline{{\rm MS}}$ scheme. 

For a given non-singlet local operator $\mathcal{O}_{f_1 f_2}(x)=\bar{\psi}_{f_1}(x)\Gamma\psi_{f_2}(x)$, $f_1$ and $f_2$ represent different flavors of quark fields, and $\Gamma$ represents a gamma matrix. When the point source propagator is used, the three-point correlation function and propagator in the momentum space are expressed by
\begin{align}
    G_{\mathcal{O}_{f_1f_2}}(p_1, p_2) &= \sum_{x, y} e^{-i(p_1 \cdot x - p_2 \cdot y)} \langle \psi_{f_1}(x) \mathcal{O}_{f_1f_2}(0) \bar{\psi}_{f_2}(y) \rangle\, ,\label{eq:three_pt_G}\\
    S_{f}(p) &= \sum_{x} e^{-ip \cdot x} \langle \psi_f(x) \bar{\psi}_f(0) \rangle\, ,
    \label{eq:prop_in_mom}
\end{align}
where $\langle\dots\rangle$ represents gauge configuration average and repeated indices do not imply summation. The amputated vertex function is defined in the form of 
\begin{align}
    \Lambda_{\mathcal{O}_{f_1f_2}}(p_1, p_2) = S_{f_1}^{-1}(p_1) G_{\mathcal{O}_{f_1f_2}}(p_1, p_2) S_{f_2}^{-1}(p_2)\, .
\end{align}
The kinematical conditions of the SMOM scheme require that $p_1^2=p_2^2=q^2=\mu^2$, where $q=p_1-p_2$ and $\mu$ is the renormalization scale. On a $L^3\times T$ lattice with periodic spatial and antiperiodic temporal boundary conditions, we choose the following momentum:
\begin{align}
    p_1=\left(\frac{2n\pi}{L}, \frac{2n\pi}{L}, \frac{2n\pi}{L}, [T/L]\frac{2n\pi}{T}+\frac{\pi}{T}\right)\,,\quad p_2=\left(-\frac{2n\pi}{L}, \frac{2n\pi}{L}, \frac{2n\pi}{L}, [T/L]\frac{2n\pi}{T}+\frac{\pi}{T}\right)\,,
\end{align}
where $[T/L]$ represents the nearest integer if $T$ is not an integer multiple of $L$. With such momentum form, $q^2$ is not exactly equal to $p_1^2$ and $p_2^2$, however its effects on the renormalization constants are negligible compared with other uncertainties. The renormalization constants are defined by (flavor subscripts are omitted for clarity)
\begin{align}
     \Lambda_{\mathcal{O}, R}(p_1, p_2) = \frac{Z_{\mathcal{O}}}{Z_{q}} \Lambda_{\mathcal{O},B}(p_1, p_2)\,, \quad \mathcal{O}_R=Z_\mathcal{O} \mathcal{O}_B\,, \quad \psi_R(x) = \sqrt{Z_q} \psi_B(x) \,.
\end{align}
In this work, we apply the RI/SMOM scheme and the renormalization conditions for vector(V), scalar(S), axial vector(A), and pseudoscalar(P) local operators are listed in the following:
\begin{align}
\begin{aligned}
\frac{1}{12 q^2} \text{Tr}[q_\mu \Lambda^\mu_{V,B}(p_1, p_2) \slashed{q}] &= \frac{Z_q}{Z_V}\,, \quad
&\frac{1}{12}\, \mathrm{Tr} \left[\Lambda_{S,B}(p_1, p_2) \right] &= \frac{Z_q}{Z_S}\, ,\\
\frac{1}{12 q^2} \text{Tr}[q_\mu \Lambda^\mu_{A,B}(p_1, p_2) \gamma_5 \slashed{q}] &= \frac{Z_q}{Z_A}\,,
\quad
&\frac{1}{12}\, \mathrm{Tr} \left[\Lambda_{P,B}(p_1, p_2)\gamma_5 \right] &= \frac{Z_q}{Z_P}\,,
\end{aligned}
\label{eq:proj_SMOM}
\end{align}
and the quark field renormalization condition is $Z_q=\operatorname{Tr}\left[S^{-1}(p)\slashed{p}\right]/(12p^2)$. We avoid directly calculating $Z_q$ using this definition which suffers from larger discretization errors~\cite{He:2022lse}. Instead, we first calculate the ratios of renormalization constants over $Z_V$, and then determine $Z_V$ from normalization for the matrix element of the vector operator as discussed in Section~\ref{sec:ZV_norm}.

As stated in the main text, two types of propagators are used in the calculation. $S(x;U)$ are calculated using the original gauge field $U$ without stout smearing, while $S(x;V)$ are calculated from the gauge field $V$ with 1-step stout smearing. When $S(p;U)$ and $S(p;V)$ denote the corresponding propagators in the momentum space, there are three types of correlation functions in Eq.~(\ref{eq:three_pt_G}): Type I is constructed for the light-light operator $\bar{\psi}_l(x)\Gamma\psi_l(x)$, where two quark legs are $S(p_1;V)$ and $S(p_2;V)$, and the renormalization constant is labeled by $Z^{s}$; Type II is constructed for the heavy-light operator $\bar{\psi}_h(x)\Gamma\psi_l(x)$, where two quark legs are $S(p_1;V)$ and $S(p_2;U)$, and the renormalization constant is labeled by $Z^{ns,s}$; and Type III is constructed for the heavy-heavy operator $\bar{\psi}_h(x)\Gamma\psi_h(x)$, where two quark legs are $S(p_1;U)$ and $S(p_2;U)$, and the renormalization constant is labeled by $Z^{ns}$. The quark propagators are calculated at the finite quark mass on the lattice. However, for an identical bare quark mass parameter, two-point functions constructed from $S(x;U)$ or $S(x;V)$ yield distinct effective meson masses. To address this, We first calculate propagators $S(x;V)$ at a series of bare quark masses $m_{q_1}$ and extract the corresponding pseudoscalar meson masses $m_{PS}(m_{q_1})$ from two-point functions constructed from $S(x;V)$. Next, we calculate propagators $S(x;U)$ at a set of bare quark masses $m_{q_2}$ and extract the associated pseudoscalar meson masses $m_{PS}(m_{q_2})$. Given that the relation between $m_{q_2}$ and $m_{PS}(m_{q_2})$ can be well described by the expression $m_{PS}^2(m_{q_2})=c_0(m_{q_2}-c_1)$, the bare quark mass parameter $m_{q_2}$ corresponding to $m_{PS}(m_{q_1})$ can be derived via $m_{PS}^2(m_{q_1})/c_0+c_1$. With the bare quark mass parameters properly tuned via the aforementioned procedures, two quark legs for the second type of correlation functions are guaranteed to correspond to the same pseudoscalar meson mass (or equivalently, the same physical quark mass).

Since we are aiming at the mass-independent renormalization scheme, propagators in Eq.~(\ref{eq:prop_in_mom}) are computed at six different valence quark masses corresponding to six different pseudoscalar meson masses, and then the corresponding renormalization constants calculated from Eq.~(\ref{eq:proj_SMOM}) are extrapolated to the chiral limit with 
\begin{align}
    Z(m_{PS}) = Z + c_v m_{PS}^2
\end{align}
In the continuum limit, $Z_V$ and $Z_A$ in the RI/SMOM scheme are equal to $1$, as in the $\overline{\mathrm{MS}}$ shceme~\cite{Sturm:2009kb}, so no conversion factors are required. $Z_S$ and $Z_P$ are equal and the perturbative conversion factor from RI/SMOM to $\overline{\mathrm{MS}}$ scheme is defined as $Z^{\overline{\mathrm{MS}}}_{S(P)}(\mu)=C_{S(P)}^{\overline{\mathrm{MS}}/\mathrm{SMOM}}(\mu)Z_{S(P)}^{\mathrm{SMOM}}(\mu)$, and $C_S^{\overline{\mathrm{MS}}/\mathrm{SMOM}}(\mu)$ up to three loops are available in Refs.~\cite{Bednyakov:2020ugu,Kniehl:2020sgo}. The scale evolution of renormalization constant $Z$ is governed by the equation 
\begin{align}
    \gamma(\alpha_S) = -\left( \mu \frac{dZ}{d\mu} \right) Z^{-1}\,, \quad \beta(\alpha_S) = \mu^2 \frac{d}{d\mu^2} \alpha_S(\mu)~,
    \label{eq:scale_evo}
\end{align}
where $\alpha_S(\mu)$ is the running coupling of QCD (the iteration solution~\cite{Kniehl:2006bg} of $\alpha_S(\mu)$ is used in the calculation), $\beta$ function up to five loops can be found in Ref.~\cite{Baikov:2016tgj}, and the anomalous dimension $\gamma$ for $Z_S$ up to five loops can be found in Ref.~\cite{Baikov:2014qja}. 

\begin{figure}
    \centering
    \includegraphics[width=0.45\linewidth]{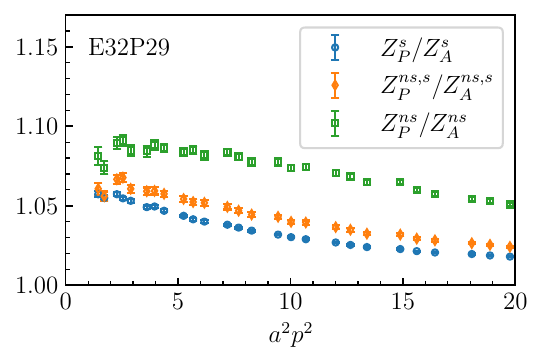}
    \includegraphics[width=0.45\linewidth]{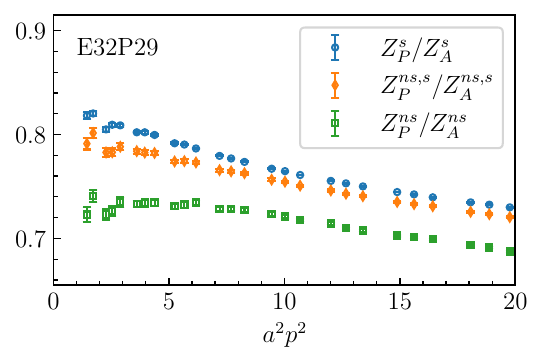}
    \caption{Comparison of three types of ratios for renormalization constants on ensemble E32P29. Left panel shows three types of ratio for $Z_A/Z_V$, and right panel shows three types of ratio for $Z_P/Z_A$. $Z_P$ has been converted to $\overline{\mathrm{MS}}$ scheme and evolved to $\mu=2\,\mathrm{GeV}$.}
    \label{fig:comp_RZ_three_types}
\end{figure}

\begin{figure}
    \centering
    \includegraphics[width=0.45\linewidth]{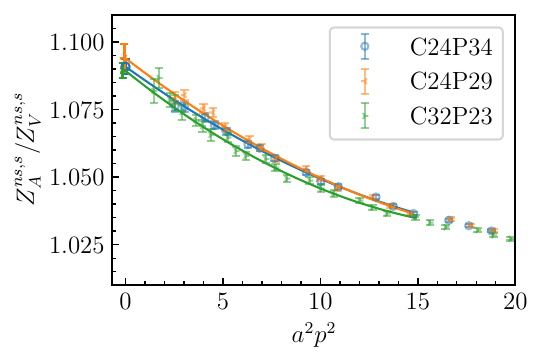}
    \includegraphics[width=0.45\linewidth]{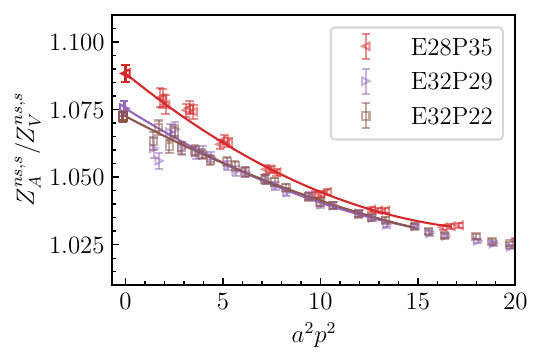}
    \includegraphics[width=0.45\linewidth]{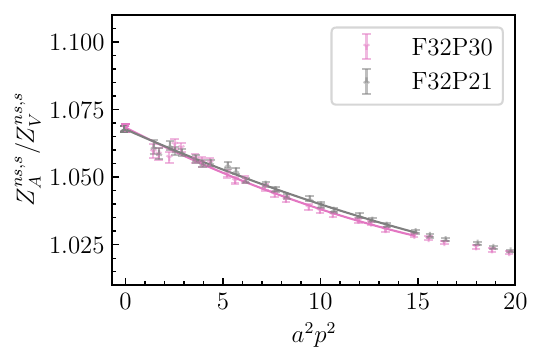}
    \includegraphics[width=0.45\linewidth]{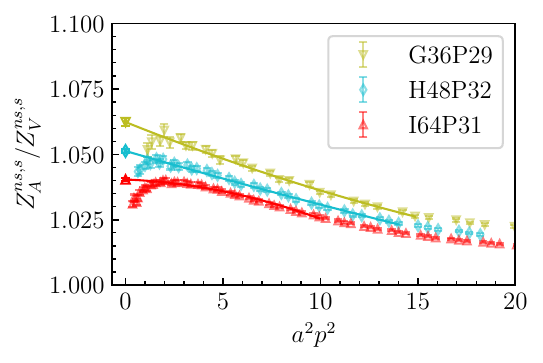}    
    \caption{Ratios of type II renormalization constants $Z^{ns,s}_A/Z^{ns,s}_V$ on different ensembles. Solid curves show the fits extrapolating $a^2p^2$ to zero, with the points at $a^2p^2=0$ corresponding to the extrapolation results (some points are slightly offset for clarity). }
    \label{fig:RZAZV_sns}
\end{figure}

\begin{figure}
    \centering
    \includegraphics[width=0.45\linewidth]{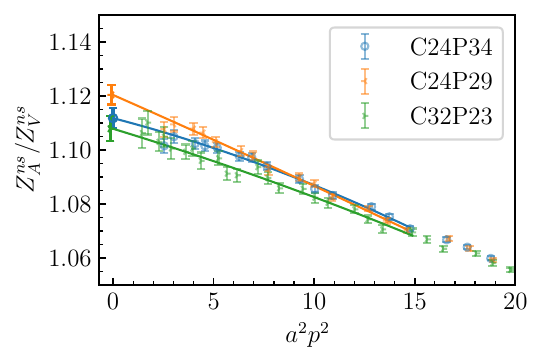}
    \includegraphics[width=0.45\linewidth]{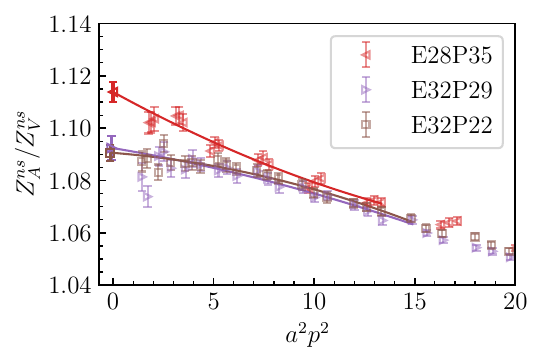}
    \includegraphics[width=0.45\linewidth]{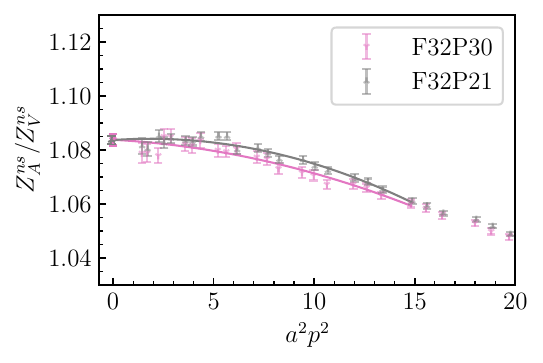}
    \includegraphics[width=0.45\linewidth]{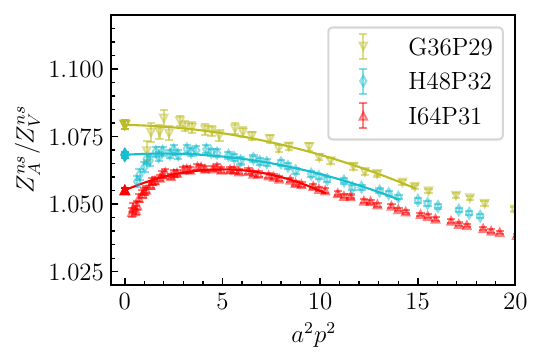}    
    \caption{Ratios of type III renormalization constants $Z^{ns}_A/Z^{ns}_V$ on different ensembles. Solid curves show the fits extrapolating $a^2p^2$ to zero, with the points at $a^2p^2=0$ corresponding to the extrapolation results (some points are slightly offset for clarity).}
    \label{fig:RZAZV_ns}
\end{figure}

\begin{figure}
    \centering
    \includegraphics[width=0.45\linewidth]{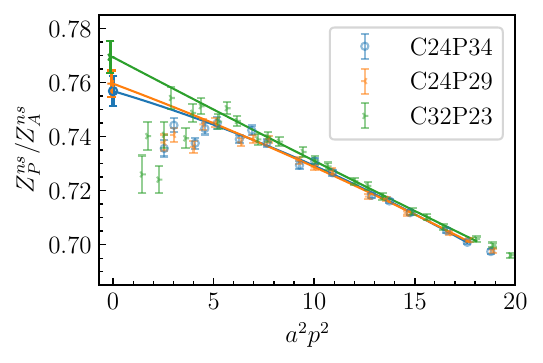}
    \includegraphics[width=0.45\linewidth]{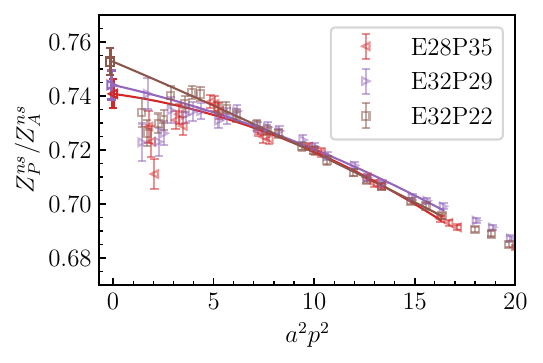}
    \includegraphics[width=0.45\linewidth]{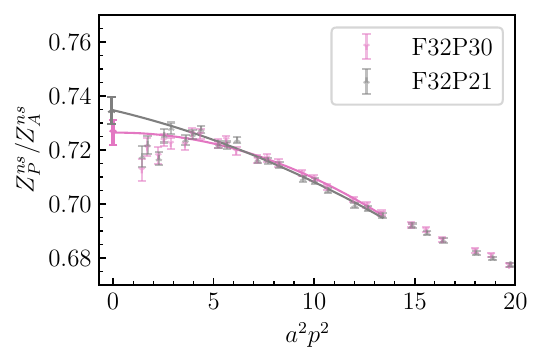}
    \includegraphics[width=0.45\linewidth]{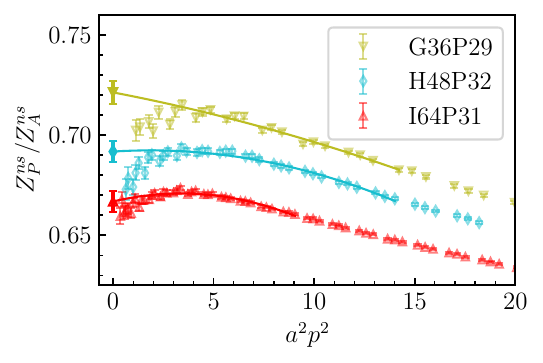}    
    \caption{Ratios of type III renormalization constants $Z^{ns}_P/Z^{ns}_A$ on different ensembles. Solid curves show the fits extrapolating $a^2p^2$ to zero, with the points at $a^2p^2=0$ corresponding to the extrapolation results (some points are slightly offset for clarity). $Z^{ns(ns,s)}_P$ is in $\overline{\mathrm{MS}}$ scheme and evolved to $\mu=2\,\mathrm{GeV}$.}
    \label{fig:RZPZA_ns}
\end{figure}

The left and right panels of Fig.~\ref{fig:comp_RZ_three_types} display the three types of ratios $Z_A/Z_V$ and $Z_P/Z_A$ for ensemble E32P29 at $\overline{\mathrm{MS}}$ scheme, respectively. $Z_P$ has been converted to $\overline{\mathrm{MS}}$ scheme and evolved to the reference scale $\mu=2\,\mathrm{GeV}$. The data points in these figures demonstrate that the ratios corresponding to the three types of renormalization constants are distinct. Fig.~\ref{fig:RZAZV_sns} presents the ratios for $Z^{ns,s}_A/Z^{ns,s}_V$ on different ensembles. Figs.~\ref{fig:RZAZV_ns} and \ref{fig:RZPZA_ns} presents the ratios for $Z^{ns}_A/Z^{ns}_V$ and $Z^{ns}_P/Z^{ns}_A$ on different ensembles, respectively. $Z^{ns(ns,s)}_P$ is also converted to $\overline{\mathrm{MS}}$ scheme and evolved to the reference scale $\mu=2\,\mathrm{GeV}$. Renormalization constants in these plots still have dependence on the scale $\mu$ due to the discretization artifacts. We utilize the following fit formula to extrapolate the renormalization constants to $a^2p^2 = 0$,
\begin{align}
    Z(a^2p^2) = Z(0)+\sum_{n=1}^{2} c_n a^{2n}p^{2n}\,.
\end{align}
The points at $a^2p^2=0$ in the figures show the extrapolation results. The right endpoints of the solid curves indicate the maximum $a^2p^2$ of the fit range, and the initial fit range of $p^2$ is approximately $10\,\mathrm{GeV}^2$. The extrapolated values for ratios relevant to the current work are summarized in Table~\ref{tab:type2_RC} and Table~\ref{tab:type3_RC}. The uncertainty of $Z_A/Z_V$, as well as the first uncertainty of $Z_P/Z_A$, is estimated by $\sqrt{\sigma^2_{\mathrm{stat}}+\sigma^2_{\mathrm{cut}}}$, where $\sigma_{\mathrm{stat}}$ is the uncertainty from the extrapolation for $a^2p^2$ and $\sigma_{\mathrm{cut}}$ is the variance of the central values obtained by excluding the first and the last data points in the extrapolation range. The second uncertainty of $Z_P/Z_A$ originates from the perturbative matching and scale evolution. It is estimated by $\sqrt{\sigma^2_{\Lambda}+\sigma^2_{n_\alpha}+\sigma^2_{n_C}+\sigma^2_{n_\gamma}}$, where $\sigma_{\Lambda}$ is the variance of the central values after rescaling the asymptotic scale parameter $\Lambda_{\mathrm{QCD}}=0.332\,\mathrm{GeV}$ by a factor of $1.05$. In addition, $\sigma_{n_\alpha}$, $\sigma_{n_C}$ and $\sigma_{n_\gamma}$ are the variances of the central values when the number of loops used in the perturbative expressions for the iteration solution of $\alpha_S$, conversion factor $C_{P}^{\overline{\mathrm{MS}}/\mathrm{SMOM}}$ and anomalous dimension $\gamma$, respectively, is reduced by one compared to the maximal available loop order.

With these renormalization constant ratios for different ensembles, a global chiral extrapolation on the sea quark masses is implemented using 
\begin{align}
    R(a,m_\pi)=R(a)(1+c_s m_\pi^2)\,.
    \label{eq:global_chiral}
\end{align}
The final results for the ratios at different lattice spacings are presented in Table~\ref{tab:sea_chiral_RC}. The global extrapolation fit for the ratios $Z_P/Z_A$ yields a reasonable $\chi^2/\mathrm{dof}$. However, since the ratios $Z_A/Z_V$ are determined with high precision and the parameterization in Eq.~(\ref{eq:global_chiral}) is somewhat oversimplified, the fit for $Z_A/Z_V$ results in a $\chi^2/\mathrm{dof}$ greater than unity. To maintain consistency with the fit function used for the ratio $Z_P/Z_A$, we continue to employ Eq.~(\ref{eq:global_chiral}), but we rescale the error of $Z_A/Z_V$ obtained from the extrapolation by a factor of $\sqrt{\chi^2/\mathrm{dof}}$ to provide a more reliable estimate of the uncertainty.

\begin{table}[h] 
\centering
\renewcommand\arraystretch{1.3}
\caption{Type II renormalization constant ratios on each ensemble.}
\label{tab:type2_RC}
\begin{tabular}{c|cccccc}
\hline\hline
   &C24P34 &C24P29 &C32P23 &E28P35 &E32P29 &E32P22 \\
\hline
$Z^{ns,s}_A/Z^{ns,s}_V$ &1.0909(28) &1.0940(52) &1.0893(28) &1.0883(30) &1.0754(26) &1.0725(20)  \\
\hline\hline
&F32P30 &F32P21 &G36P29 &H48P32 &I64P30 \\
\hline
$Z^{ns,s}_A/Z^{ns,s}_V$ &1.0683(15) &1.0677(11) &1.0624(14) &1.05131(76) &1.04040(68)   \\
\hline\hline
\end{tabular}
\end{table}

\begin{table}[h] 
\centering
\renewcommand\arraystretch{1.3}
\caption{Type III renormalization constant ratios on each ensemble.}
\label{tab:type3_RC}
\begin{tabular}{c|cccccc}
\hline\hline
   &C24P34 &C24P29 &C32P23 &E28P35 &E32P29 &E32P22 \\
\hline
$Z^{ns}_A/Z^{ns}_V$ &1.1118(37) &1.1204(35) &1.1079(47) &1.1139(37) &1.0924(46) &1.0906(31)   \\
$Z^{ns}_P/Z^{ns}_A$ &0.7568(40)(40) &0.7596(28)(40) &0.7694(42)(41) &0.7408(35)(41) &0.7441(34)(42) &0.7527(29)(41)  \\
\hline\hline
&F32P30 &F32P21 &G36P29 &H48P32 &I64P30 \\
\hline
$Z^{ns}_A/Z^{ns}_V$ &1.0838(21) &1.0837(15) &1.0793(16) &1.0682(10) &1.05718(69)    \\
$Z^{ns}_P/Z^{ns}_A$ &0.7265(23)(41) &0.7348(26)(43) &0.7213(34)(47) &0.6918(13)(50) &0.6668(10)(50)   \\
\hline\hline
\end{tabular}
\end{table}

\begin{table}[h] 
\centering
\renewcommand\arraystretch{1.3}
\caption{Renormalization constants at each lattice spacing.}
\label{tab:sea_chiral_RC}
\begin{tabular}{c|ccccccc}
\hline \hline
& $\hat{\beta}=6.200$  & $\hat{\beta}=6.308$ & $\hat{\beta}=6.410$ & $\hat{\beta}=6.498$ & $\hat{\beta}=6.720$ & $\hat{\beta}=7.020$ &$c_s\,(\mathrm{GeV}^{-2}$) \\
\hline
$Z^{ns,s}_A/Z^{ns,s}_V$ &1.0839(52) &1.0711(43) &1.0629(33) &1.0556(48) &1.0437(48) &1.0331(46) &0.074(45)  \\
\hline\hline
$Z^{ns}_A/Z^{ns}_V$ &1.1039(94) &1.0896(85) &1.0766(60) &1.0693(85) &1.0570(88) &1.0464(84) &0.107(82)    \\
$Z^{ns}_P/Z^{ns}_A$ &0.7766(41)(38) &0.7606(39)(39) &0.7423(33)(40) &0.7365(51)(43) &0.7082(41)(48) &0.6823(38)(47) &-0.233(54)  \\
\hline \hline
\end{tabular}
\end{table}

\subsubsection{Normalization constant $Z_V$} \label{sec:ZV_norm}
The renormalization constant $Z_V$ is determined from the normalization condition for the matrix element of the vector current operator between pseudoscalar meson states $|H(p)\rangle$ 
\begin{align}
    Z_{V_\mu}\langle H(p)|V_\mu|H(p)\rangle=2p_\mu\,,
    \label{eq:ZV_norm}
\end{align}
where $V^\mu=\bar{\psi}_f\gamma^\mu\psi_f$ is the vector current operator. Since we are using the anisotropic fermion action, $Z_{V_t}$ is not equal to $Z_{V_i}$ on the finite lattice spacing. We calculate $Z_{V_t}$ and $Z_{V_i}$ with the temporal index operator $V_t$ and spatial index operator $V_i$, respectively. The matrix element $\langle H(p)|V_\mu|H(p)\rangle$ can be extracted from the ratio of three-point and two-point functions. As stated in the main text, we target for a heavy quark mass dependent $Z_V$ to improve the discretization error of renormalized quantities, so the propagators in the three-point and two-point functions are computed at the corresponding bottom quark masses on each ensemble. Compared with two-point functions in Section~\ref{sec:two-pt_correlator}, the excited state contamination for the three-point functions is more complicated. We utilize the current sequential technique to construct a summed ratio, which is expressed in the form of
\begin{align}
    R^{\mathrm{sum}}(t_f)=\frac{\sum_t C_3(t_f,t;\boldsymbol{p})}{C_2(t_f;\boldsymbol{p})}=\frac{c_0t_f+(c_1+c_2t_f)e^{-\Delta Et_f}+c_3}{1+d_1e^{-\Delta Et_f}}\,,
    \label{eq:zv_sum_ratio}
\end{align}
if the lowest excited state is considered, and $\Delta E$ is viewed as the energy gap between the ground and lowest excited states. The two-point and three point functions in the expression are
\begin{equation}
    \begin{gathered}
    C_2(t_f;\boldsymbol{p}) = \sum_{\boldsymbol{y}}e^{-i\boldsymbol{p}\cdot\boldsymbol{y}}\left\langle \bar{\psi}(\boldsymbol{y}, t_f)\gamma_5\psi(\boldsymbol{y}, t_f) \left[\bar{\psi}(\boldsymbol{0}, 0)\gamma_5\psi(\boldsymbol{0}, 0)\right]^\dagger  \right\rangle\,,\\
        C_3(t_f,t;\boldsymbol{p}) = \sum_{\boldsymbol{y}}e^{-i\boldsymbol{p}\cdot\boldsymbol{y}}\sum_{\boldsymbol{z}}\left\langle \bar{\psi}(\boldsymbol{y}, t_f)\gamma_5\psi(\boldsymbol{y}, t_f) \bar{\psi}(\boldsymbol{z}, t)\gamma^\mu\psi(\boldsymbol{z}, t)  \left[\bar{\psi}(\boldsymbol{x}, 0)\gamma_5\psi(\boldsymbol{x}, 0)\right]^\dagger  \right\rangle\,.
    \end{gathered}
    \label{eq:3pt_func}
\end{equation}
To suppress the excited-state effects, we fit the difference between $R^{\mathrm{sum}}(t_f+1)$ and $R^{\mathrm{sum}}(t_f)$ with
\begin{align}
    \delta R^{\mathrm{sum}}(t_f)=R^{\mathrm{sum}}(t_f+1)-R^{\mathrm{sum}}(t_f)=c_0+(c'_1+c'_2t_f)e^{-\Delta Et_f}+O(e^{-2\Delta Et_f})
    \label{eq:delta_sum_ratio}
\end{align}
to extract the value of $c_0$. $Z_{V_\mu}$ is then obtained by $Z_{V_\mu}=p_\mu/(E_{\boldsymbol{p}} \,c_0)$. Eq.~(\ref{eq:ZV_norm}) illustrates that the extraction of $\langle H(p)|V_i|H(p)\rangle$ require a meson state with nonzero spatial momentum $\boldsymbol{p}$, so we use $\boldsymbol{p}=2\pi/L(0,0,1)$ in Eq.~(\ref{eq:3pt_func}) to project a state with $\boldsymbol{p}$, and the point source is used. For the extraction of $\langle H(p)|V_t|H(p)\rangle$, we utilize the rest meson state with $\boldsymbol{p}=\boldsymbol{0}$ at the sink and the coulomb wall source. Figs.~\ref{fig:zvi_delta_R} and \ref{fig:zvt_delta_R} display the fits to Eq.~(\ref{eq:delta_sum_ratio}) on ensembles C32P29, F32P21, and G36P29. The three point function constructed from the spatial index vector current operator is used in the $R^{\mathrm{sum}}(t_f)$ for Fig.~\ref{fig:zvi_delta_R}, and the three point function constructed from the temporal index vector current operator is used in the $R^{\mathrm{sum}}(t_f)$ for Fig.~\ref{fig:zvt_delta_R}. A clear quark mass dependence of $Z_V$ is observed. The analogous plots for other ensembles look similar.

\begin{figure}[pt] 
   \centering
    \includegraphics[width=0.325\textwidth]{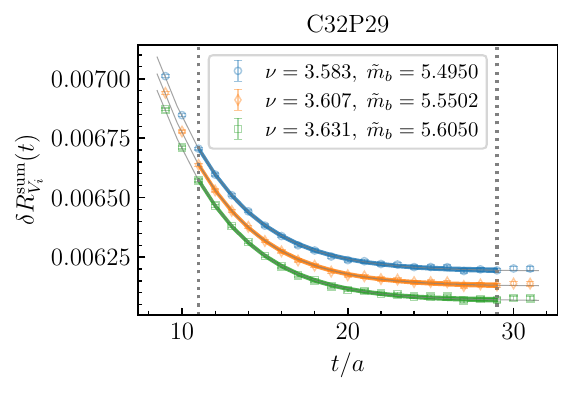} 
    \includegraphics[width=0.325\textwidth]{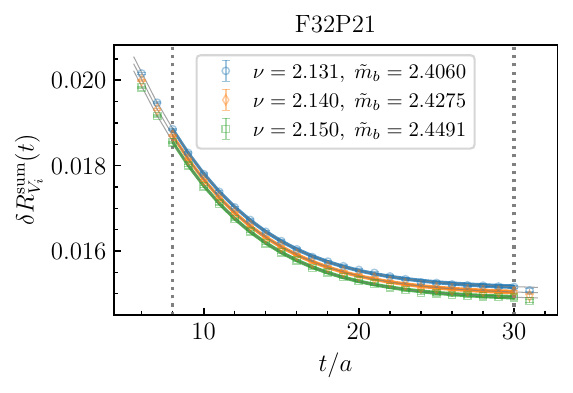} 
    \includegraphics[width=0.325\textwidth]{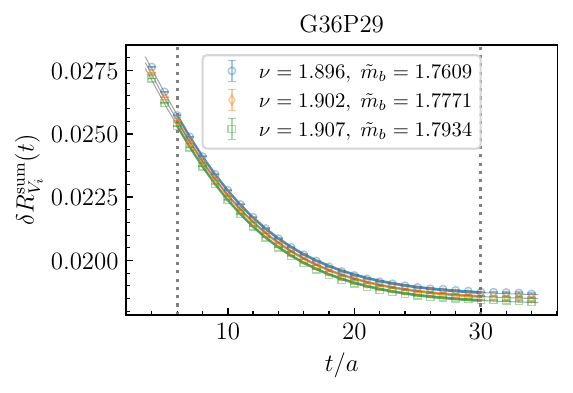} 
   \caption{Fits to Eq.~(\ref{eq:delta_sum_ratio}) on ensembles C32P29, F32P21, and G36P29. The three point function constructed from the spatial index vector current operator is used in the $R^{\mathrm{sum}}(t_f)$. Two vertical dashed lines show the fit range.}
   \label{fig:zvi_delta_R}
\end{figure}

\begin{figure}[pt] 
   \centering
    \includegraphics[width=0.325\textwidth]{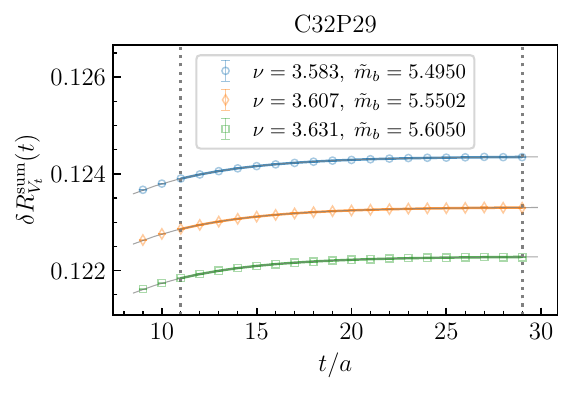} 
    \includegraphics[width=0.325\textwidth]{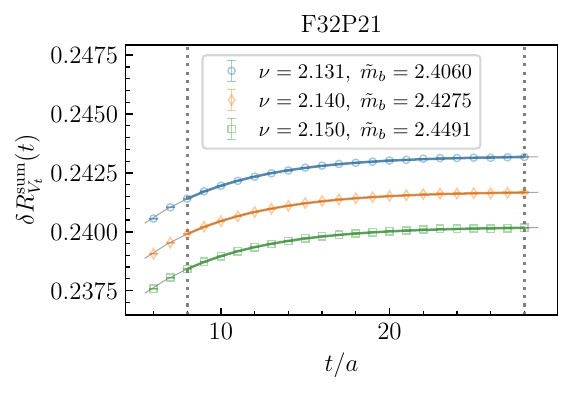} 
    \includegraphics[width=0.325\textwidth]{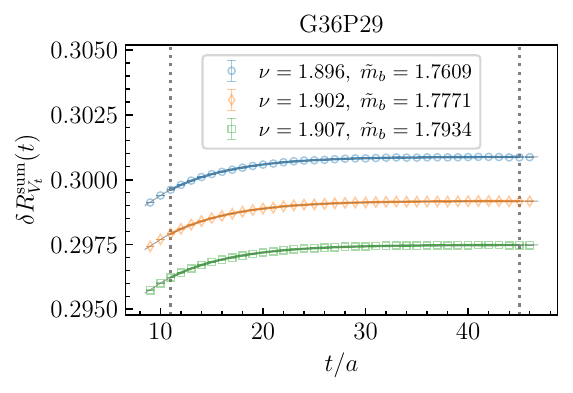} 
   \caption{Fits to Eq.~(\ref{eq:delta_sum_ratio}) on ensembles C32P29, F32P21, and G36P29. The three point function constructed from the temporal index vector current operator is used in the $R^{\mathrm{sum}}(t_f)$. Two vertical dashed lines show the fit range.}
   \label{fig:zvt_delta_R}
\end{figure}

 The difference between $Z_{V_t}$ and $Z_{V_i}$ is a discretization artifact and would vanish in the continuum. We extrapolated ratio $R(Z_V)\equiv Z_{V_t}/Z_{V_i}$ to the continuum by
\begin{align}
    R(m_{\pi}, m_{\eta_s}, a) &= R(m_{\pi}^{\text{phys}}, m_{\eta_s}^{\text{phys}}, 0) + c_{m_\pi} \left( m_{\pi}^2 - m_{\pi,\text{phys}}^2 \right)
    + c_{m_{\eta_s}}\left( m_{\eta_s}^2 - m_{\eta_s,\text{phys}}^2 \right) + c_{a^2} a^2 + c_{a^4} a^4
\end{align}
where $R(m_{\pi}^{\text{phys}}, m_{\eta_s}^{\text{phys}}, 0)=0.996(15)$, $c_{m_\pi}=0.208(65)\,\mathrm{GeV}^{-2}$, $c_{m_{\eta_s}}=0.015(59)\,\mathrm{GeV}^{-2}$, $c_{a^2}=1.34(16)\,\mathrm{GeV}^{2}$, and $c_{a^4}=-1.50(40)\,\mathrm{GeV}^{4}$. We have rescaled the uncertainty of $Z_{V_t}/Z_{V_i}$ by a factor of $2.7$ to bring the $\chi^2/\mathrm{dof}$ of the fit to approximately unity. Fig.~\ref{fig:zv_ratio} shows the variance of the ratio $Z_{V_t}/Z_{V_i}$ with the lattice spacings. Points labeled by $Z^b_{V_t}/Z^b_{V_i}(a,m^{\mathrm{sea}}_q)$ are the ratios corresponding to the physical bottom quark mass on various ensembles. Points labeled by $Z^b_{V_t}/Z^b_{V_i}(a)$ are the values after the extrapolation to physical $\pi$ and $\eta_s$ meson masses, which are explicitly expressed as
\begin{align}
    Z^b_{V_t}/Z^b_{V_i}(a) &= R(m_{\pi}, m_{\eta_s}, a) - c_{m_\pi} \left( m_{\pi}^2 - m_{\pi,\text{phys}}^2 \right)
    - c_{m_{\eta_s}}\left( m_{\eta_s}^2 - m_{\eta_s,\text{phys}}^2 \right)~,
\end{align}
and the gray band shows the results of $R(m_{\pi}^{\text{phys}}, m_{\eta_s}^{\text{phys}}, 0)+ c_{a^2} a^2 + c_{a^4} a^4$.
\begin{figure}
    \centering
    \includegraphics[width=0.5\linewidth]{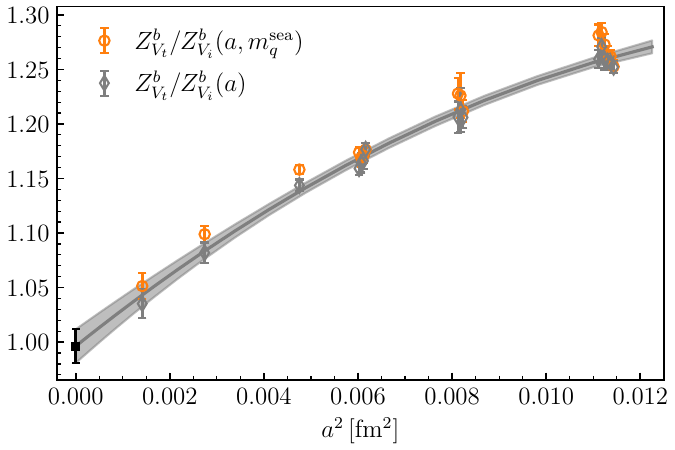}
    \caption{Ratios $Z_{V_t}/Z_{V_i}$ on different ensembles. $Z_{V_t}$ and $Z_{V_i}$ are different for the anisotropic fermion action at finite lattice spacing, but the extrapolated ratio to the continuum limit is consistent with unity .}
    \label{fig:zv_ratio}
\end{figure}

We use the isotropic clover action for the light, strange, and charm quarks, so $Z^q_{V}$ in Eq.~(\ref{eq:ZO_ZV}) calculated from the temporal or spatial index vector operator is the same. We use the quark mass dependent $Z_V(m_c)$ for the vector current operators with charm quark field and the chiral limit $Z_V$ for the vector current operators with light or strange quark field from the appendix of Ref.~\cite{CLQCD:2024yyn}. For the new ensembles used in the present work, we follow the procedure in Ref.~\cite{CLQCD:2023sdb,CLQCD:2024yyn} to calculate the corresponding $Z_V(m_c)$ and chiral limit $Z_V$. $Z_{V_t}(\tilde{m}_b)$ ($Z_{V_i}(\tilde{m}_b)$) interpolated to the physical bottom quark mass, $Z_{V}(\tilde{m}_c)$ interpolated to the physical charm quark mass, and $Z_V$ at the chiral limit for each ensemble are summarized in Table~\ref{tab:ZVt_ZVi_ZV}.

\begin{table}[h] 
\centering
\renewcommand\arraystretch{1.3}
\caption{Renormalization constants for vector current on each ensemble. $Z_{V_t}(\tilde{m}^{\mathrm{phys}}_b)$ and $Z_{V_i}(\tilde{m}^{\mathrm{phys}}_b)$ are values corresponding to the physical bottom quark mass, $Z_{V}(\tilde{m}^{\mathrm{phys}}_c)$ is the value corresponding to the physical charm quark mass, and $Z_V$ is the value at the chiral limit.}
\label{tab:ZVt_ZVi_ZV}
\begin{tabular}{c|cccccccc}
\hline\hline
         &C24P34 &C24P29 &C32P29 &C32P23 &C48P23 &C48P14 &E28P35 &E32P29 \\
         \hline
$Z_{V_t}(\tilde{m}^{\mathrm{phys}}_b)$  &8.10418(41) &8.11443(19) &8.14500(18) &8.16620(22) &8.19018(25) &8.20058(12) &5.58701(58) &5.59514(41) \\
$Z_{V_i}(\tilde{m}^{\mathrm{phys}}_b)$  &6.323(10) &6.3196(98) &6.3991(63) &6.4611(78) &6.489(11) &6.5463(32) &4.549(16) &4.563(26)\\
$Z_V(\tilde{m}^{\mathrm{phys}}_c)$ &1.5593(23) &1.5685(20) &1.5675(20) &1.5735(20) &1.5750(19) &1.5752(19) &1.4102(11) &1.4157(11) \\
$Z_V$ &0.79676(32) &0.79814(23) &0.79810(13) &0.79957(13) &0.799540(50) &0.799570(60) &0.817680(40) &0.818794(83) \\
\hline\hline
         &E32P22 &F32P30 &F32P21 &F48P21 &F64P14 &G36P29 &H48P32 &I64P30 \\ 
         \hline
$Z_{V_t}(\tilde{m}^{\mathrm{phys}}_b)$  &5.61617(31) &4.135822(74) &4.14402(14) &4.143561(80) &4.136908(34) &3.348299(46) &2.283337(58) &1.650967(30) \\
$Z_{V_i}(\tilde{m}^{\mathrm{phys}}_b)$  &4.636(11) &3.5233(33) &3.5441(56) &3.5385(60) &3.5155(23) &2.8914(30) &2.0772(50) &1.5697(64) \\
$Z_V(\tilde{m}^{\mathrm{phys}}_c)$ &1.4188(11) &1.30518(83) &1.30720(81) &1.30695(82) &1.30681(82) &1.23831(65) &1.13246(34) &1.0499(17) \\
$Z_V$ &0.81958(13) &0.83548(12) &0.835790(90) &0.835670(50) &0.835942(31) &0.846360(90) &0.868550(40) &0.889241(34) \\
\hline\hline
\end{tabular}
\end{table}

\subsection{Continuum and chiral extrapolation}
Section~\ref{sec:aniso_action} introduces that the quantities extracted from two-point functions are calculated at three sets of parameters $(\nu, \tilde{m}_b)$ and then interpolated to the value corresponding to the physical bottom quark masses $\tilde{m}_b(m_\Upsilon)$ on each ensemble. Since the decay constants extracted from the two-point correlators are bare quantities and we employ quark mass dependent $Z_{V_{t(i)}}(\tilde{m}_q)$ to improve discretization errors, before the linear interpolation to the values corresponding the physical quark masses, we first multiply the bare decay constants by the quark mass dependent renormalization constants. $\sqrt{Z_{V_{t(i)}}(\tilde{m}_b)}$ is multiplied for $f^{\mathrm{bare}}_{B^{(*)}}$ and $f^{\mathrm{bare}}_{B^{(*)}_s}$. $\sqrt{Z_{V_{t(i)}}(\tilde{m}_b)Z_V(\tilde{m}_c)}$, $Z_{V_{t}}(\tilde{m}_b)$ and $Z_{V_{i}}(\tilde{m}_b)$ are multiplied for $f^{\mathrm{bare}}_{B^{(*)}_c}$, $f^{\mathrm{bare}}_{\eta_b}$ and $f^{\mathrm{bare}}_\Upsilon$, respectively. In our renormalization strategy, renormalization constant ratio $Z_A/Z_V$ is extrapolated to the chiral limit, so it is multiplied just before the continuum extrapolation procedure for convenience. Fig.~\ref{fig:interp_mb_f_F64P14} shows the interpolation of meson masses and decay constants to the values corresponding to the physical bottom quark mass on ensemble F64P14. Panel titles indicate the respective quantities, and the bare mass of the lighter valence quark in the meson (excluding the bottom quark) is labeled in the first three columns of panels. The fourth column of panels presents bottomonium results. We find that upon multiplication by the appropriate bottom quark mass dependent renormalization constants, the decay constants vary only slightly with the bare bottom quark mass.

\begin{figure}
    \centering
    \includegraphics[width=1.0\linewidth]{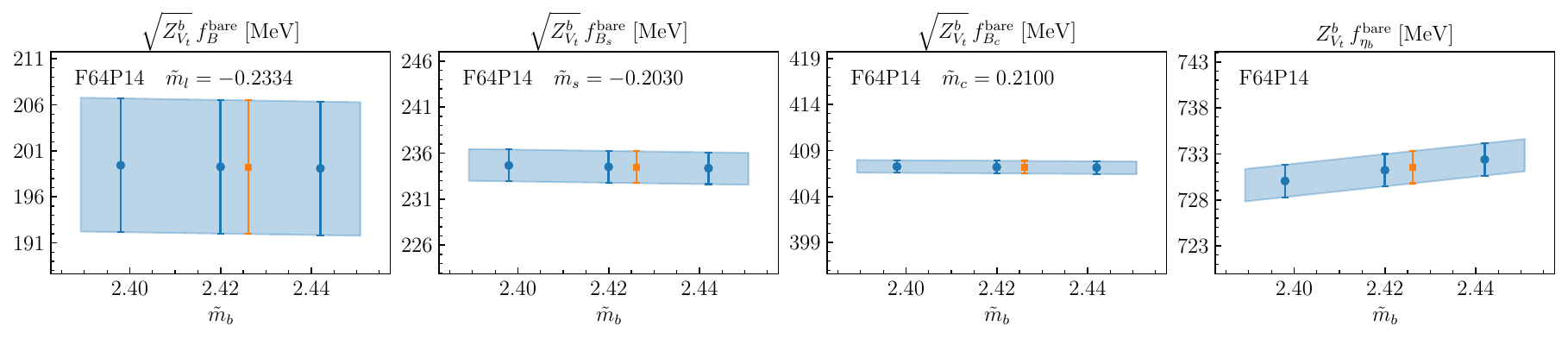}
    \includegraphics[width=1.0\linewidth]{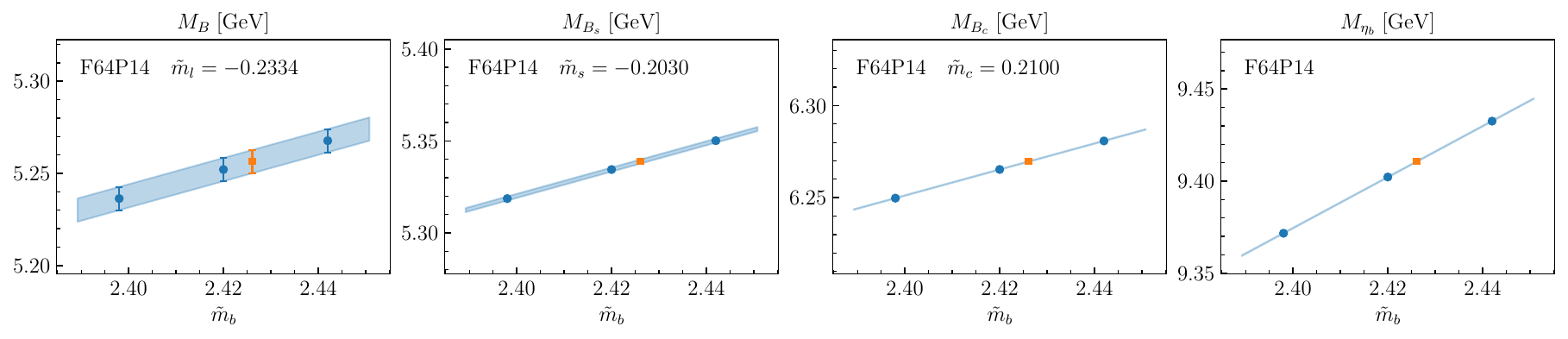}
    \includegraphics[width=1.0\linewidth]{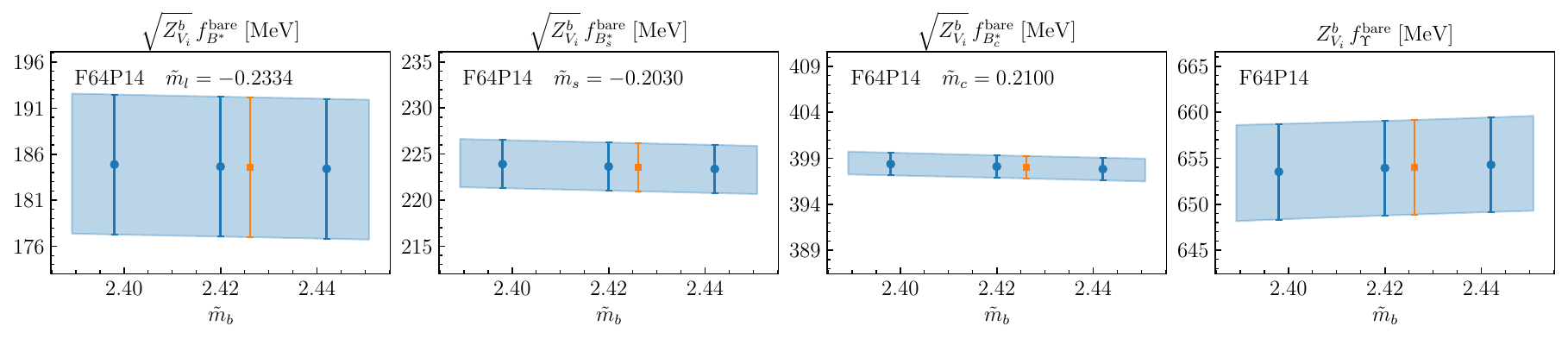}
    \includegraphics[width=1.0\linewidth]{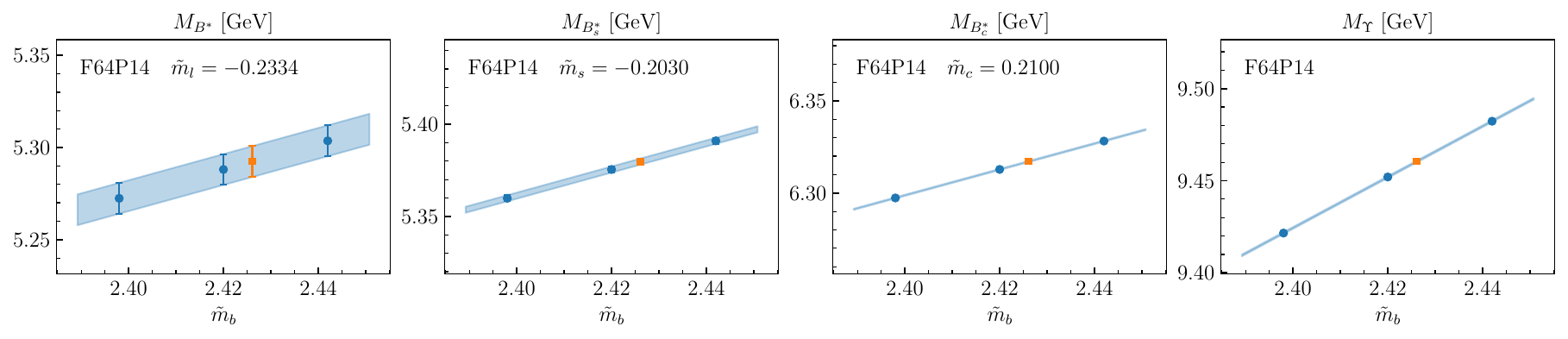}
    \caption{Interpolation for meson masses and decay constants to the values corresponding to the physical bottom quark mass on ensemble F64P14.}
    \label{fig:interp_mb_f_F64P14}
\end{figure}

In addition, the quantities related to $B^{(*)}$, $B^{(*)}_s$ and $B^{(*)}_c$ are calculated at three light quark masses around the sea quark mass $\tilde{m}^b_l$, three quark masses around the strange quark mass $\tilde{m}^b_s$ and three quark masses around the charm quark mass, respectively. In the following, $X_A(\tilde{m}_1,\tilde{m}_2)$ is used to represent a physical quantity, such as mass or lepton decay constant, for meson $A$, where $\tilde{m}_1$ and $\tilde{m}_2$ are the bare masses for the two valence quarks of meson $A$. Thus, the quantities interpolated to the value corresponding to the physical bottom quark masses is represented by $X_A(\tilde{m}_b(m_\Upsilon), \tilde{m})$. 
For $X_{B_s^{(*)}}(\tilde{m}_b(m_\Upsilon), \tilde{m}_s)$ and $X_{B_c^{(*)}}(\tilde{m}_b(m_\Upsilon), \tilde{m}_c)$, we first linearly interpolate them to the values corresponding to the physical strange quark mass $\tilde{m}_s(m_{\eta_s})$ and charm quark mass $\tilde{m}_c(m_{D_s})$, respectively, a single physical $X$ value for $B^{(*)}$ or $B_c^{(*)}$ on each ensemble is then used in the combined continuum and chiral fit. 


A similar procedure used to determine $\tilde{m}_b(m_\Upsilon)$ is applied to obtain $\tilde{m}_s(m_{\eta_s})$. On each ensemble, pseudoscalar meson masses $m_{\mathrm{PS}}(\tilde{m}_s)$ are extracted from two-point correlation functions constructed from propagators at three values of $\tilde{m}_s$ around the sea strange quark mass $\tilde{m}^b_s$. It is assumed that $m^2_{\rm PS}$ varies linearly with $\tilde{m}_s$. The physical strange quark mass, $\tilde{m}_s(m_{\eta_s})$, is then defined as the value of $\tilde{m}_s$ for which $m_{\mathrm{PS}}$ matches the target value $m_{\eta_s}=689.89(49)\,\mathrm{MeV}$~\cite{Borsanyi:2020mff}. For the determination of charm quark mass $\tilde{m}_c(m_{D_s})$, $m_{D_s}(\tilde{m}_s,\tilde{m}_c)$ are evaluated at three values of $\tilde{m}_s$ and three values of $\tilde{m}_c$, then a linear interpolation over $\tilde{m}_s$ is first performed to obtain the mass of $D_s$ corresponding to the physical strange quark mass $\tilde{m}_s(m_{\eta_s})$, denoted as $m_{D_s}(\tilde{m}_s(m_{\eta_s}),\tilde{m}_c)$. It is also assumed that $m_{D_s}(\tilde{m}_s(m_{\eta_s}),\tilde{m}_c)$ varies linearly with $\tilde{m}_c$. Then, the physical charm quark mass $\tilde{m}_c(m_{D_s})$ is defined as the value of $\tilde{m}_c$ for which the meson mass matches the target value $m_{D_s}=1968.35(7)\,\mathrm{MeV}$~\cite{ParticleDataGroup:2024cfk}. In Table~\ref{tab:msmcphys}, we give the physical values of the valence strange quark mass $\tilde{m}_s(m_{\eta_s})$ and charm quark mass $\tilde{m}_c(m_{D_s})$ on each ensemble.

\begin{table}[h] 
\centering
\renewcommand\arraystretch{1.3}
\caption{$\tilde{m}^{\mathrm{phys}}_s$ and $\tilde{m}^{\mathrm{phys}}_c$ (uncertainties are from pseudoscalar meson effective masses).  }
\begin{tabular}{c|cccccccc}
\hline\hline
                               &C24P34 &C24P29 &C32P29 &C32P23 &C48P23 &C48P14 &E28P35 &E32P29 \\
                               \hline
$\tilde{m}^{\mathrm{phys}}_s$  &-0.23959(13) &-0.23567(11) &-0.23582(9) &-0.23379(9) &-0.23384(11) &-0.23351(9) &-0.21936(12) &-0.21794(12) \\
$\tilde{m}^{\mathrm{phys}}_c$  &0.4111(4) &0.4202(2) &0.4192(2) &0.4233(2) &0.4250(3) &0.4248(2) &0.2879(7) &0.2921(3)  \\
\hline\hline
&E32P22 &F32P30 &F32P21 &F48P21 &F64P14 &G36P29 &H48P32 &I64P30 \\
\hline
$\tilde{m}^{\mathrm{phys}}_s$  &-0.21654(10) &-0.20389(10) &-0.20234(9) &-0.20266(7) &-0.20226(6) &-0.19281(5) &-0.17004(4) &-0.14687(5) \\
$\tilde{m}^{\mathrm{phys}}_c$  &0.2945(4) &0.1994(3) &0.2015(2) &0.2014(2) &0.2011(2) &0.1449(2) &0.0589(1) &0.0023(1) \\
\hline\hline
\end{tabular}
\label{tab:msmcphys}
\end{table}

We apply bootstrap method in the analysis of the correlation functions, so the correlation among quantities are kept within the same ensemble. When generating the bootstrap samples for lattice spacings, it is assumed that the uncertainty $\sigma_{w_0}$ is correlated, and the uncertainty $\sigma_{\mathrm{stat}+\mathrm{ansatz}}$ is independent. When generating the bootstrap samples for renormalization constant ratios, the statistical errors are assumed to be independent, but the systematical errors arising from the perturbative matching and scale evolution are regarded as correlated across the ensembles at different lattice spacings. For different ensembles, the bootstrap number is matched, and 1000 times of fitting are performed in the continuum and chiral extrapolation to estimate the uncertainties. The following fit ansatz is utilized in the continuum and chiral extrapolation:
\begin{align}
    X(m_{\pi},m_{\eta_s},a) = X(m_{\pi,\,\mathrm{phys}},m_{\eta_s,\,\mathrm{phys}},0)\left[1 + c^X_{\mathrm{val}}\delta m^2_{\mathrm{PS},\,\mathrm{val}}+c^X_{\mathrm{sea}}\delta m^2_{\mathrm{PS},\,\mathrm{sea}}+d_a(a^2,a^2\delta m^2_{\mathrm{PS},\,\mathrm{sea}})\right]\, ,
    \label{eq:extrapolation_continuum}
\end{align}
where $\delta m^2_{\mathrm{PS},\,\mathrm{val}}$ accounts for the deviation of the valence quark masses from the physical values, $\delta m^2_{\mathrm{PS},\,\mathrm{sea}}=2\left( m_{\pi,\,\mathrm{sea}}^2 - m_{\pi,\,\mathrm{phys}}^2 \right)+\left( m_{\eta_s,\,\mathrm{sea}}^2 - m_{\eta_s,\,\mathrm{phys}}^2 \right)$ quantifies the deviation of the sea quark masses from their physical values, and $d_a(a^2,a^2\delta m^2_{\mathrm{PS},\,\mathrm{sea}})$ parametrizes the discretization effects. As discussed above, we interpolated quantities associated with $B^{(*)}_s$, $B^{(*)}_c$, $\eta_b$, and $\Upsilon$ to the values corresponding to the physical valence quark masses, such that $c^X_{\mathrm{val}}\delta m^2_{\mathrm{PS},\,\mathrm{val}}$ vanishes for these states. For the quantities related to $B^{(*)}$, since two additional partial quenched light valence quark masses are used and the unitary light quark masses are not exactly at the physical point, $c^X_{\mathrm{val}}\left( m^2_{\pi,\,\mathrm{val}} - m^2_{\pi,\,\mathrm{phys}}\right)$ contributes to the extrapolation. The discretization error term $d_a(a^2, a^2\delta m^2_{\mathrm{PS},\,\mathrm{sea}})$ is taken as $c_{a^2}a^2 + c_{a^4}a^4$ in most cases. For the extrapolations of observables associated with bottomonium ($\eta_b$ and $\Upsilon$), however, the discretization effects are more complex, and we therefore include an additional term $c^X_{a^2,\,\mathrm{sea}}\,a^2\delta m^2_{\mathrm{PS},\,\mathrm{sea}}$ to yield a better $\chi^2/\mathrm{dof}$. We also consider the potential $O(a\alpha_S)$ discretization effects in the extrapolation of the renormalized decay constants, decay constant ratios and bottom quark mass by including an additional term $c_a\Lambda_{\chi} a \operatorname{log}(u_0)$, where $c_a$ is assigned with a prior of $[-1,1]$ with $\Lambda_{\chi}=1\,\mathrm{GeV}$. We treat the variance of the central value as a systematic error. 

Table.~\ref{tab:ori_meson_mass} presents the values of $M_{B^{(*)}}$, $M_{B^{(*)}_s}$, $M_{B^{(*)}_c}$ and $m_{\eta_b}$ at the unitary light, physical valence strange, charm and bottom quark masses on each ensemble, respectively. The uncertainties for masses include contributions from two-point function fits, lattice spacings and the interpolation to the physical quark masses. Table.~\ref{tab:ori_meson_mass} also lists the bottom quark mass $m_b$ at the scale $\mu=2\,\mathrm{GeV}$ in $\overline{\mathrm{MS}}$ scheme, whose uncertainty additionally incorporates that from renormalization constants. Table.~\ref{tab:ori_meson_decay_constant} displays the values of $f_{B^{(*)}}$, $f_{B^{(*)}_s}$, $f_{B^{(*)}_c}$, $f_{\eta_b(\Upsilon)}$ at the unitary light, physical valence strange, charm, and bottom quark masses on each ensemble, respectively. The uncertainties for the decay constants similarly include those from renormalization constants. Tables.~\ref{tab:ori_meson_mass} and \ref{tab:ori_meson_decay_constant} also show the extrapolation parameters and $\chi^2/\mathrm{dof}$ for the corresponding quantities. The uncertainties for these parameters also include those from the $\pi$ and $\eta_s$ masses in the extrapolation procedure. Since the extrapolation of $B^{(*)}_{c}$ mass using original lattice data gives $\chi^2/\mathrm{dof}$ significantly larger than unity, we rescale the error of the data by a factor of $4$ to obtain a reasonable uncertainty estimate. To obtain the bottom quark mass at the scale $\mu=m_b$, we solve the equation $m_b(m_b)=Z_m(\mu=m_b)/Z_m(\mu=2\,\mathrm{GeV})m_b(2\,\mathrm{GeV})$ for each bootstrap fit result. The scale evolution of $Z_m$ is governed by Eq.~\ref{eq:scale_evo}, with its anomalous dimension being the negative of that for $Z_S$.

In Figs.~\ref{fig:masses}, \ref{fig:decay_constants}, \ref{fig:RfVfPS_mhs} and \ref{fig:Rf_deltaM}, the data points show the values obtained by the extrapolation to the physical point, which are expressed explicitly as 
\begin{align}
    X(a) = X^{\mathrm{latt}}(m_{\pi},m_{\eta_s},a) -   X(m_{\pi,\,\mathrm{phys}},m_{\eta_s,\,\mathrm{phys}},0)\left[c^X_{\mathrm{val}}\delta m^2_{\mathrm{PS},\,\mathrm{val}}+c^X_{\mathrm{sea}}\delta m^2_{\mathrm{PS},\,\mathrm{sea}}+d_a(0,a^2\delta m^2_{\mathrm{PS},\,\mathrm{sea}})\right]\, ,
\end{align}
where $X^{\mathrm{latt}}(m_{\pi},m_{\eta_s},a)$ denotes the original lattice data, while $X(m_{\pi,\,\mathrm{phys}},m_{\eta_s,\,\mathrm{phys}},0)$, $c^X_{\mathrm{val}}$, $c^X_{\mathrm{sea}}$, and $d_a(0,a^2\delta m^2_{\mathrm{PS},\,\mathrm{sea}})$ are the parameters from the fits of Eq.~(\ref{eq:extrapolation_continuum}). The optical bands in these figures display the results acquired from $X(m_{\pi,\,\mathrm{phys}},m_{\eta_s,\,\mathrm{phys}},0)\left[1 +d_a(a^2,0)\right]$, which show the lattice spacing dependence of the respective quantities. Figs.~\ref{fig:masses} and \ref{fig:decay_constants} depict the fit results for meson masses, decay constants and the bottom quark mass. Since we use $\Upsilon$ meson mass to tune the bare bottom quark mass parameter, $m_\Upsilon$ is just the experiment value on each ensemble and the mass splitting curve in Fig.~\ref{fig:comp_mhfs} can indicate the lattice spacing dependence of $m_{\eta_b}$. The curves for the masses of $\eta_b$ and $\Upsilon$ are not shown in Fig.~\ref{fig:comp_mhfs}. We also calculate the mass splitting for other three pairs of bottom mesons, and these results are displayed in the left panels of Fig.~\ref{fig:RfVfPS_mhs}. Our results are consistent with the experiment values from PDG within $1\,\sigma$. The fits for the ratios of the decay constants of vector and pseudoscalar meson are displayed in the right panels of Fig.~\ref{fig:RfVfPS_mhs}, and our results are consistent with the previous lattice calculations. We gives the most precise prediction for $f_{B^*_c}/f_{B_c}$ from lattice calculation at present. Ratios $f_{B_s}/f_B$ and $f_{B^*_s}/f_{B^*}$ are shown in the left panels of Fig.~\ref{fig:Rf_deltaM}. We also calculate the ratios of $(m_{B^*_s}-m_{B^*})/(m_{B_s}-m_{B})$ and $(m_{B^*_s}-m_{B_s})/(m_{B^*}-m_{B})$, and their fit results are displayed in the right panels of Fig.~\ref{fig:Rf_deltaM}. The ratios of the mass difference are consistent with the results from PDG.

\begin{figure}
    \centering
    \includegraphics[width=0.465\linewidth]{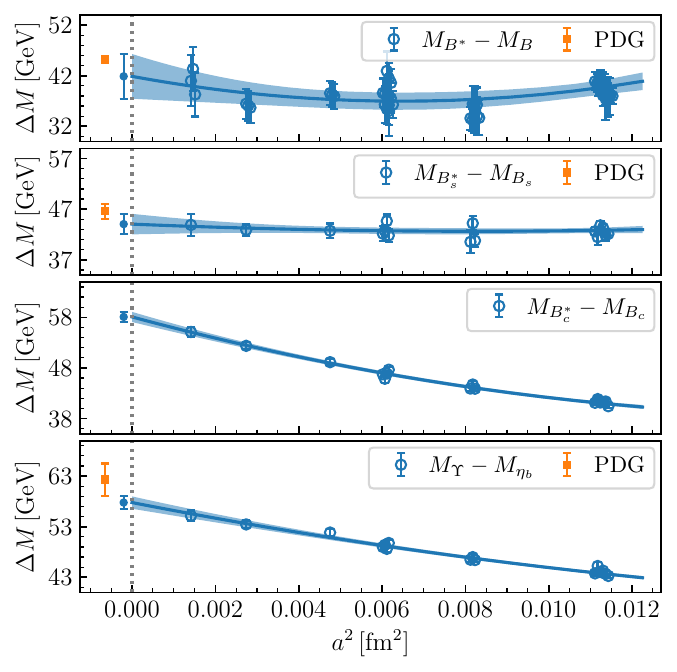}
    \includegraphics[width=0.48\linewidth]{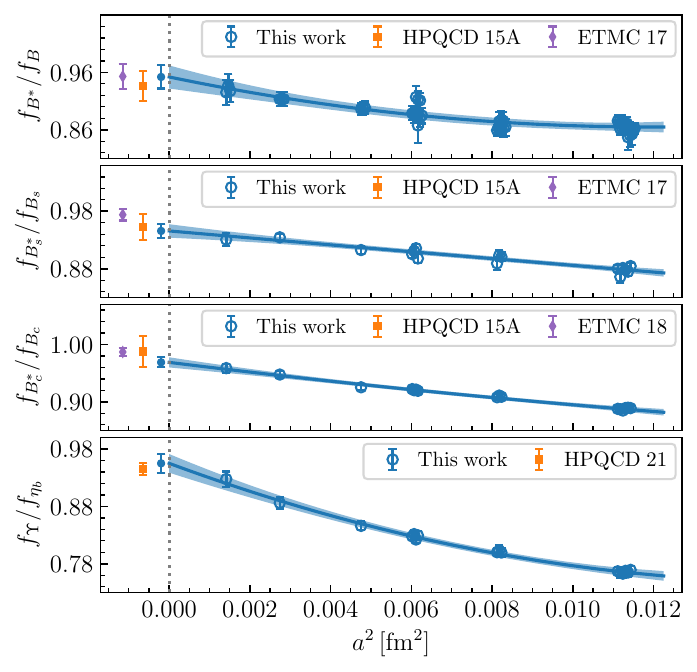}
    \caption{Left panels display the mass splitting between vector and pseudoscalar mesons. Results from PDG~\cite{ParticleDataGroup:2024cfk} are plotted at $a<0\,\mathrm{fm}$ for comparison. Right panels show the ratios of decay constants. Results from HPQCD~\cite{Colquhoun:2015oha} and ETM collaborations~\cite{Lubicz:2017asp,Becirevic:2018qlo} are plotted at $a<0\,\mathrm{fm}$ for comparison. }
    \label{fig:RfVfPS_mhs}
\end{figure}

\begin{figure}
    \centering
    \includegraphics[width=0.47\linewidth]{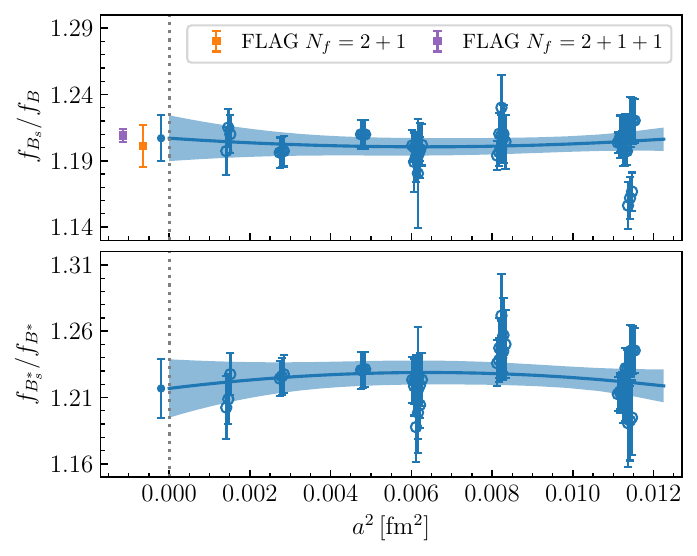}
    \includegraphics[width=0.48\linewidth]{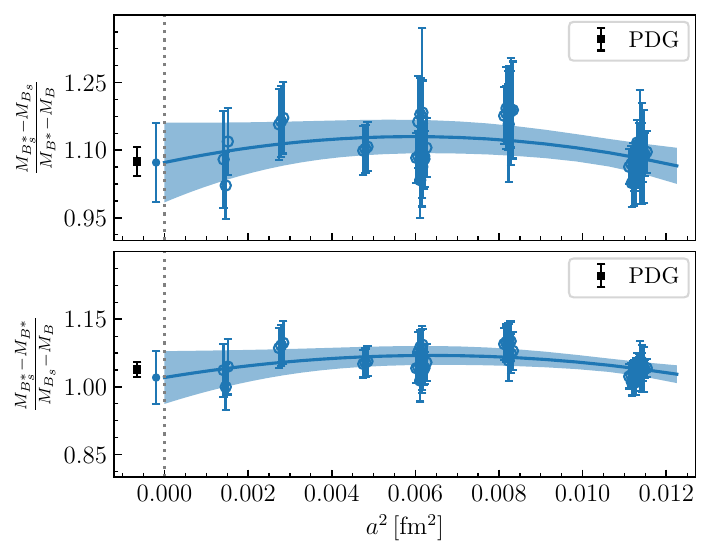}
    \caption{Left panels show the ratios of $f_{B_s}/f_B$ and $f_{B^*_s}/f_{B^*}$, and FLAG~\cite{FlavourLatticeAveragingGroupFLAG:2024oxs} average results are plotted at $a<0\,\mathrm{fm}$ for comparison. Right panels show the ratios of mass differences, and PDG~\cite{ParticleDataGroup:2024cfk} results are plotted at $a<0\,\mathrm{fm}$ for comparison.}
    \label{fig:Rf_deltaM}
\end{figure}

To decompose the total uncertainty of a given quantity into its constituent components, we assume that the square of the total uncertainty $\sigma_{\mathrm{tot}}$ equals the sum of the squares of the component uncertainties, i.e., $\sigma^2_{\mathrm{tot}} = \sum_i \sigma_i^2$. when generating the bootstrap samples, if we rescale one of the uncertainties by $\sqrt{\lambda_k}$, it gives $\sigma^2_{\lambda_k,\mathrm{tot}} = \sum_{i\ne k} \sigma_i^2 +\lambda_k \sigma^2_k$, then $\sigma_k$ can be separated via $\sigma_k=\sqrt{(\sigma^2_{\lambda_k,\mathrm{tot}}-\sigma^2_{\mathrm{tot}})/(\lambda_k-1)}$. Following the above procedure, we decompose the total uncertainty into statistical and systematic uncertainties. The statistical uncertainties for the quantities related to meson masses contain the uncertainty from two-point function fits, the combined statistical and fit ansatz uncertainty of lattice spacing ($\sigma_{\mathrm{stat}+\mathrm{ansatz}}$) and the uncertainty arising from the $\pi$ and $\eta_s$ masses used in the extrapolation procedure. The statistical uncertainties for the quantities associated with meson decay constants additionally incorporate contributions from renormalization constants. Systematic uncertainties for the meson masses related quantities includes $\sigma_{w_0}$ of lattice spacing as well as the uncertainty from the interpolation to the physical quark masses. For the quantities associated with meson decay constants, systematic uncertainties also includes the effect of $a\alpha_S$ term in the extrapolation. The detailed decomposition of the statistical and systematical uncertainties for the quantities computed in this work are summarized in Tables~\ref{tab:meson_mass}-\ref{tab:mhsR_dmR_error_budget}.

Eventually, the comparison of our results with those in the literatures are shown in Figs.~\ref{fig:comp_fPS_B_Bs}-\ref{fig:comp_RfVPS}.

\begin{table}[h] 
\renewcommand\arraystretch{1.3}
\caption{Meson masses on different ensembles (in $\mathrm{GeV}$), together with the fit parameters and $\chi^2/\mathrm{dof}$ for the chiral and continuum extrapolation.}
\begin{tabular}{c|cccccccc}
\hline\hline
       &$m_B$      &$m_{B^*}$  &$m_{B_s}$  &$m_{B^*_s}$&$m_{B_c}$   &$m_{B^*_c}$ &$m_{\eta_b}$ &$m_b$\\
\hline
C24P34 &5.2635(29) &5.3135(37) &5.3307(19) &5.3795(26) &6.23650(57) &6.28007(57) &9.41430(45) &4.917(36) \\
C24P29 &5.2488(29) &5.2932(40) &5.3253(22) &5.3685(24) &6.24044(57) &6.28309(56) &9.41439(77) &4.924(36) \\
C32P29 &5.2523(32) &5.2972(42) &5.3281(20) &5.3736(25) &6.24163(29) &6.28362(42) &9.41555(43) &4.938(36) \\
C32P23 &5.2408(28) &5.2802(39) &5.3262(17) &5.3692(22) &6.24366(45) &6.28501(37) &9.41602(48) &4.948(36) \\
C48P23 &5.2470(34) &5.2856(55) &5.3282(13) &5.3696(22) &6.24669(69) &6.28810(44) &9.41670(56) &4.960(36) \\
C48P14 &5.2393(32) &5.2786(38) &5.3295(16) &5.3717(21) &6.24728(40) &6.28794(36) &9.41691(45) &4.965(36) \\
E28P35 &5.2684(39) &5.3102(51) &5.3377(24) &5.3833(38) &6.2459(10) &6.2919(12) &9.41262(64) &4.949(36) \\
E32P29 &5.2564(39) &5.2979(51) &5.3344(22) &5.3814(30) &6.24957(64) &6.29549(75) &9.41264(60) &4.956(36) \\
E32P22 &5.2495(43) &5.2856(57) &5.3365(24) &5.3780(30) &6.25333(63) &6.29767(66) &9.41370(51) &4.974(36) \\
F32P30 &5.2689(35) &5.3126(47) &5.3445(21) &5.3892(28) &6.25367(74) &6.30153(63) &9.41094(54) &4.996(36) \\
F32P21 &5.2677(55) &5.3051(59) &5.3437(24) &5.3862(30) &6.25574(91) &6.30172(95) &9.41094(62) &5.005(36) \\
F48P21 &5.2627(30) &5.3068(49) &5.3435(16) &5.3880(24) &6.25575(48) &6.30277(54) &9.41171(62) &5.006(36) \\
F64P14 &5.2565(66) &5.2924(87) &5.3409(19) &5.3820(26) &6.25321(51) &6.30076(47) &9.41065(61) &4.997(36) \\
G36P29 &5.2673(31) &5.3112(41) &5.3442(21) &5.3899(28) &6.25303(77) &6.30342(53) &9.40819(59) &4.934(45) \\
H48P32 &5.2880(32) &5.3306(37) &5.3578(19) &5.4041(26) &6.26000(65) &6.31380(63) &9.40683(66) &4.961(45) \\
I64P30 &5.2895(39) &5.3382(55) &5.3608(23) &5.4073(32) &6.2659(10) &6.3222(11) &9.40531(85) &4.934(44) \\
\hline\hline
$m\,[\mathrm{GeV}]$ &5.2835(46) &5.3252(58) &5.3673(23) &5.4112(33) &6.2715(45) &6.3297(45) &9.4026(12) &4.936(52) \\
$c_{a^2}\,[\mathrm{GeV}^2]$ &-0.0421(87) &-0.052(11) &-0.0355(44) &-0.0378(59) &-0.0172(68) &-0.0318(77) &0.0067(10) &0.162(61) \\
$c_{\mathrm{sea}}\,[\mathrm{GeV}^{-2}]$ &0.0040(13) &0.0081(19) &0.00132(62) &0.00546(93) &-0.00452(97) &-0.0032(11) &0.00022(52) &-0.0362(69) \\
$c_{a^4}\,[\mathrm{GeV}^4]$&0.042(23) &0.070(30) &0.034(11) &0.038(15) &0.009(17) &0.028(19) &-0.0053(23) &-0.52(18) \\
$c_{\mathrm{val}}\,[\mathrm{GeV}^{-2}]$ &0.0319(29) &0.0369(37) &- &- &- &- &- &- \\
$c_{a^2,\mathrm{sea}}$ &- &- &- &- &- &- &-0.0038(19) &- \\
$\chi^2/\mathrm{dof}$ &0.93 &0.82 &0.67 &0.53 &0.72 &0.60 &0.78 &0.15 \\
\hline\hline
\end{tabular}
\label{tab:ori_meson_mass}
\end{table}

\begin{table}[h] 
\renewcommand\arraystretch{1.3}
\caption{Decay constants on different ensembles (in $\mathrm{GeV}$), together with the fit parameters and $\chi^2/\mathrm{dof}$ for the chiral and continuum extrapolation.}
\begin{tabular}{c|cccccccc}
\hline\hline
&$f_B$ &$f_{B^*}$ &$f_{B_s}$ &$f_{B^*_s}$ &$f_{B_c}$ &$f_{B^*_c}$ &$f_{\eta_b}$ &$f_{\Upsilon}$\\
\hline
C24P34 &0.2040(24) &0.1823(27) &0.2309(19) &0.2033(22) &0.5209(29) &0.4597(21) &0.7604(73) &0.5723(43) \\
C24P29 &0.1948(26) &0.1697(30) &0.2273(28) &0.1968(19) &0.5189(29) &0.4595(28) &0.7637(73) &0.5802(62) \\
C32P29 &0.1961(28) &0.1728(32) &0.2267(21) &0.1998(21) &0.5208(29) &0.4599(22) &0.7716(76) &0.5848(46) \\
C32P23 &0.1872(24) &0.1628(27) &0.2243(19) &0.1969(18) &0.5186(28) &0.4611(20) &0.7733(72) &0.5933(37) \\
C48P23 &0.1978(37) &0.1687(49) &0.2254(17) &0.1972(21) &0.5221(30) &0.4645(19) &0.7857(68) &0.6007(33) \\
C48P14 &0.1857(33) &0.1610(29) &0.2259(18) &0.1998(16) &0.5234(28) &0.4651(16) &0.7893(73) &0.6062(32) \\
E28P35 &0.2053(33) &0.1797(35) &0.2309(24) &0.2054(34) &0.5061(28) &0.4573(32) &0.7804(70) &0.6131(61) \\
E32P29 &0.1953(32) &0.1731(38) &0.2280(24) &0.2059(28) &0.5052(27) &0.4590(38) &0.7796(70) &0.6192(69) \\
E32P22 &0.1888(39) &0.1644(42) &0.2250(27) &0.2026(29) &0.5058(25) &0.4592(24) &0.7930(68) &0.6315(45) \\
F32P30 &0.1981(28) &0.1784(34) &0.2288(21) &0.2072(24) &0.4941(23) &0.4545(19) &0.7918(52) &0.6483(32) \\
F32P21 &0.1934(46) &0.1724(41) &0.2271(22) &0.2069(24) &0.4962(43) &0.4565(42) &0.7893(81) &0.6558(76) \\
F48P21 &0.1926(25) &0.1778(44) &0.2271(16) &0.2080(22) &0.4958(20) &0.4567(22) &0.7996(49) &0.6566(43) \\
F64P14 &0.1936(73) &0.1688(71) &0.2284(22) &0.2051(27) &0.4928(20) &0.4532(20) &0.7876(52) &0.6540(61) \\
G36P29 &0.1970(25) &0.1790(27) &0.2292(22) &0.2092(24) &0.4866(26) &0.4490(19) &0.7791(64) &0.6508(31) \\
H48P32 &0.2031(25) &0.1882(25) &0.2319(19) &0.2168(19) &0.4733(25) &0.4470(24) &0.7687(67) &0.6705(40) \\
I64P30 &0.1979(31) &0.1878(40) &0.2304(22) &0.2145(26) &0.4594(35) &0.4394(39) &0.7455(73) &0.6828(59) \\
\hline\hline
$f_H\,[\mathrm{GeV}]$ &0.1898(34) &0.1812(41) &0.2291(24) &0.2168(27) &0.4510(40) &0.4379(41) &0.726(10) &0.701(11) \\
$c_{a^2}\,[\mathrm{GeV}^2]$ &-0.05(17) &-0.61(24) &-0.07(10) &-0.32(13) &0.713(91) &0.32(10) &0.99(14) &-0.30(12) \\
$c_{\mathrm{sea}}\,[\mathrm{GeV}^{-2}]$ &0.146(37) &0.195(50) &0.083(17) &0.080(26) &-0.002(10) &-0.028(13) &0.008(52) &-0.116(67) \\
$c_{a^4}\,[\mathrm{GeV}^4]$&-0.04(47) &0.78(66) &0.02(28) &0.02(36) &-0.60(25) &-0.42(25) &-2.55(38) &-0.71(28) \\
$c_{\mathrm{val}}\,[\mathrm{GeV}^{-2}]$ &0.496(79) &0.493(95) &- &- &- &- &- &- \\
$c_{a^2,\mathrm{sea}}$ &- &- &- &- &- &- &-0.44(19) &-0.13(27) \\
$\chi^2/\mathrm{dof}$ &0.64 &0.54 &0.14 &0.43 &0.33 &0.58 &0.75 &1.14 \\
\hline\hline
\end{tabular}
\label{tab:ori_meson_decay_constant}
\end{table}

\begin{table}[h] 
\renewcommand\arraystretch{1.3}
\caption{Error budget for meson masses, in units of $\mathrm{GeV}$.}
\begin{tabular}{c|ccccccc}
\hline\hline
&$m_B$ &$m_{B^*}$ &$m_{B_s}$ &$m_{B^*_s}$ &$m_{B_c}$ &$m_{B^*_c}$ &$m_{\eta_b}$ \\
\hline
central value &5.2835(46) &5.3252(58) &5.3673(23) &5.4112(33) &6.2715(45) &6.3297(45) &9.4026(12) \\
\textbf{statistical} (total) &0.0042 &0.0053 &0.0019 &0.0027 &0.0022 &0.0036 &0.0006 \\
lattice spacing &0.0009 &0.0011 &0.0007 &0.0009 &0.0008 &0.0007 &0.0004 \\
two-point function &0.0041 &0.0052 &0.0017 &0.0025 &0.0021 &0.0036 &0.0005 \\
$\pi$ and $\eta_s$ mass uncertainty &0.0003 &0.0004 &0.0001 &0.0002 &0.0004 &0.0003 &$<0.0001$ \\
\textbf{systematic} (total) &0.0018 &0.0023 &0.0014 &0.0019 &0.0039 &0.0027 &0.0011 \\
$w_0$ scale &0.0017 &0.0023 &0.0012 &0.0018 &0.0024 &0.0007 &0.0005 \\
interpolation &0.0004 &0.0004 &0.0006 &0.0007 &0.0031 &0.0026 &0.0009 \\
\hline\hline
\end{tabular}
\label{tab:meson_mass}
\end{table}

\begin{table}[h] 
\renewcommand\arraystretch{1.3}
\caption{Error budget for decay constants and bottom quark mass, in units of $\mathrm{MeV}$.}
\begin{tabular}{c|ccccccccc}
\hline\hline
&$f_B$ &$f_{B^*}$ &$f_{B_s}$ &$f_{B^*_s}$ &$f_{B_c}$ &$f_{B^*_c}$ &$f_{\eta_b}$ &$f_{\Upsilon}$ &$m_b(2\,\mathrm{GeV})$\\
\hline
central value &189.8(3.5) &181.2(4.2) &229.1(2.4) &216.8(2.7) &451.0(5.2) &437.9(4.3) &726.4(10.5) &701.4(11.3) &4936.1(51.8) \\
\textbf{statistical} (total) &2.9 &3.9 &2.0 &2.4 &3.4 &3.9 &9.5 &8.8 &51.1 \\
lattice spacing &0.5 &0.6 &0.5 &0.5 &0.8 &1.0 &1.9 &2.2 &2.2 \\
two-point function &2.6 &3.8 &1.5 &2.3 &1.9 &3.5 &5.3 &6.3 &0.4 \\
$\pi$ and $\eta_s$ mass uncertainty &0.2 &0.3 &0.1 &0.1 &$<0.1$ &0.1 &0.5 &0.7 &$<0.1$ \\
renormalization constants &1.1 &0.5 &1.2 &0.7 &2.7 &1.5 &7.6 &5.8 &51.1 \\
\textbf{systematic} (total) &1.9 &1.4 &1.4 &1.3 &3.9 &1.7 &4.4 &7.0 &8.2 \\
$w_0$ scale &0.9 &1.1 &0.7 &1.2 &0.7 &1.4 &2.2 &6.4 &7.3 \\
interpolation &1.7 &0.8 &1.2 &0.4 &1.9 &0.9 &3.5 &2.4 &1.9 \\
$O(a\alpha_S)$ &0.4 &0.4 &0.2 &0.3 &3.3 &0.1 &1.3 &1.2 &3.1 \\
\hline\hline
\end{tabular}
\label{tab:meson_decay_constant}
\end{table}

\begin{table}[h] 
\renewcommand\arraystretch{1.3}
\caption{Error budget for the hyperfine splittings, in units of $\mathrm{MeV}$.}
\begin{tabular}{c|cccc}
\hline\hline
&$\Delta m_{\bar{b}u}$ &$\Delta m_{\bar{b}s}$ &$\Delta m_{\bar{b}c}$ &$\Delta m_{\bar{b}b}$ \\
\hline
central value &41.9(4.4) &43.6(2.0) &58.1(1.0) &57.8(1.2) \\
\textbf{statistical} (total) &4.2 &1.7 &0.8 &1.1 \\
lattice spacing &0.1 &0.1 &0.2 &0.4 \\
two-point function &4.2 &1.7 &0.7 &1.0 \\
$\pi$ and $\eta_s$ mass uncertainty &0.1 &0.1 &0.0 &0.0 \\
\textbf{systematic} (total) &1.5 &1.1 &0.6 &0.6 \\
$w_0$ scale &0.4 &0.4 &0.6 &0.5 \\
interpolation &1.4 &1.0 &0.2 &0.3 \\
\hline\hline
\end{tabular}
\label{tab:mhs_error_budget}
\end{table}

\begin{table}[h] 
\renewcommand\arraystretch{1.3}
\caption{Error budget for the ratios of vector and pseudoscalar meson decay constants.}
\begin{tabular}{c|cccc}
\hline\hline
&$f_{B^*}/f_B$ &$f_{B^*_s}/f_{B_s}$ &$f_{B^*_c}/f_{B_c}$ &$f_{\Upsilon}/f_{\eta_b}$\\
\hline
central value &0.9570(204) &0.9458(117) &0.9689(94) &0.9551(165)  \\
\textbf{statistical} (total) &0.0200 &0.0114 &0.0087 &0.0148 \\
two-point function &0.0188 &0.0091 &0.0057 &0.0086 \\
$\pi$ and $\eta_s$ mass uncertainty &0.0005 &$<0.0001$ &$<0.0001$ &0.0005 \\
renormalization constants &0.0068 &0.0069 &0.0066 &0.0120 \\
\textbf{systematic} (total) &0.0037 &0.0026 &0.0035 &0.0073 \\
interpolation &0.0020 &0.0026 &0.0017 &0.0066 \\
$O(a\alpha_S)$ &0.0031 &0.0002 &0.0031 &0.0029 \\
\hline\hline
\end{tabular}
\label{tab:RfVPS_error_budget}
\end{table}

\begin{table}[h] 
\renewcommand\arraystretch{1.3}
\caption{Error budget for pseudoscalar meson decay constant ratios.}
\begin{tabular}{c|cccccc}
\hline\hline
&$f_{B_s}/f_{B}$ &$f_{B_c}/f_{B}$ &$f_{\Upsilon}/f_{B}$ &$f_{B_c}/f_{B_s}$ &$f_{\eta_b}/f_{B_s}$ &$f_{\eta_b}/f_{B_c}$\\
\hline
central value &1.207(17) &2.390(43) &3.902(115) &1.967(20) &3.178(52) &1.625(15)  \\
\textbf{statistical} (total) &0.017 &0.041 &0.099 &0.019 &0.044 &0.014 \\
two-point function &0.017 &0.041 &0.098 &0.019 &0.042 &0.011 \\
$\pi$ and $\eta_s$ mass uncertainty &0.001 &0.003 &0.007 &0.001 &0.003 &$<0.001$ \\
renormalization constants &0.000 &0.003 &0.012 &0.002 &0.013 &0.009 \\
\textbf{systematic} (total) &0.004 &0.011 &0.060 &0.006 &0.027 &0.006 \\
interpolation &0.002 &0.011 &0.059 &0.006 &0.027 &0.005 \\
$O(a\alpha_S)$ &0.003 &0.003 &0.003 &0.001 &0.005 &0.003 \\
\hline\hline
\end{tabular}
\label{tab:RfPS_error_budget}
\end{table}

\begin{table}[h] 
\renewcommand\arraystretch{1.3}
\caption{Error budget for vector meson decay constant ratios.}
\begin{tabular}{c|cccccc}
\hline\hline
&$f_{B_s^*}/f_{B^*}$ &$f_{B_c^*}/f_{B^*}$ &$f_{\Upsilon}/f_{B^*}$ &$f_{B^*_c}/f_{B^*_s}$ &$f_{\Upsilon}/f_{B^*_s}$ &$f_{\Upsilon}/f_{B^*_c}$\\
\hline
central value &1.217(22) &2.459(59) &3.767(139) &2.019(30) &3.226(78) &1.604(21)  \\
\textbf{statistical} (total) &0.022 &0.057 &0.104 &0.029 &0.057 &0.019 \\
two-point function &0.022 &0.057 &0.102 &0.029 &0.055 &0.017 \\
$\pi$ and $\eta_s$ mass uncertainty &0.001 &0.004 &0.010 &0.001 &0.006 &0.001 \\
renormalization constants &0.000 &0.003 &0.018 &0.002 &0.014 &0.008 \\
\textbf{systematic} (total) &0.005 &0.015 &0.092 &0.008 &0.053 &0.009 \\
interpolation &0.003 &0.013 &0.090 &0.008 &0.052 &0.008 \\
$O(a\alpha_S)$ &0.003 &0.008 &0.021 &0.002 &0.006 &0.004 \\
\hline\hline
\end{tabular}
\label{tab:RfV_error_budget}
\end{table}

\begin{table}[h] 
\renewcommand\arraystretch{1.3}
\caption{Error budget for the ratios $\frac{m_{B^*_s}-m_{B_s}}{m_{B^*}-m_B}$ and $\frac{m_{B^*_s}-m_{B^*}}{m_{B_s}-m_{B}}$,  as well as the mass splittings $m_{B_s}-m_B$ and $m_{B^*_s}-m_{B^*}$, in units of $\mathrm{MeV}$.}
\begin{tabular}{c|cccc}
\hline\hline
&$\frac{m_{B^*_s}-m_{B_s}}{m_{B^*}-m_B}$ &$\frac{m_{B^*_s}-m_{B^*}}{m_{B_s}-m_{B}}$ &$m_{B_s}-m_B$ &$m_{B^*_s}-m_{B^*}$ \\
\hline
central value &1.073(88) &1.021(59) &0.0856(37) &0.0896(43) \\
\textbf{statistical} (total) &0.0789 &0.0544 &0.0035 &0.0040 \\
lattice spacing &0.0000 &0.0000 &0.0001 &0.0001 \\
two-point function &0.0789 &0.0544 &0.0035 &0.0040 \\
$\pi$ and $\eta_s$ mass uncertainty &$<0.0001$ &$<0.0001$ &0.0001 &0.0002 \\
\textbf{systematic} (total) &0.0401 &0.0216 &0.0011 &0.0014 \\
$w_0$ scale &0.0000 &0.0000 &0.0004 &0.0002 \\
interpolation &0.0401 &0.0216 &0.0011 &0.0014 \\
\hline\hline
\end{tabular}
\label{tab:mhsR_dmR_error_budget}
\end{table}

\begin{figure}
    \centering
    \includegraphics[width=0.48\linewidth]{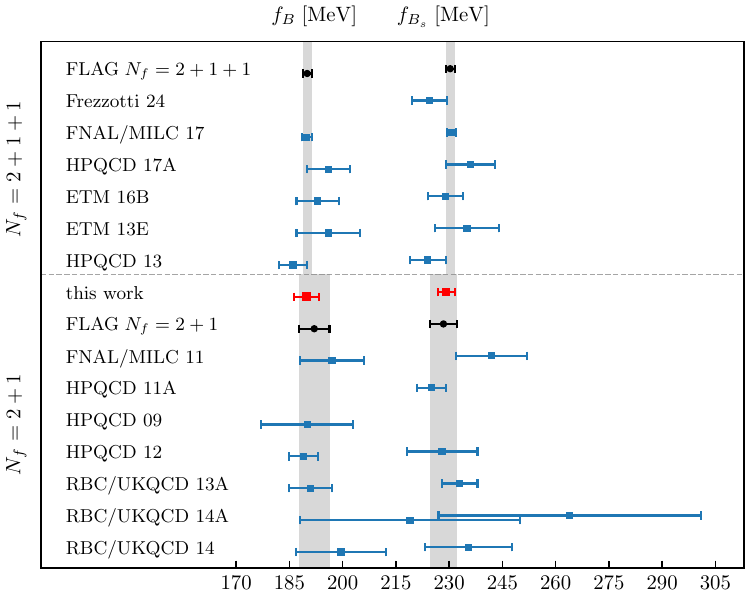}
    \caption{Comparison of our results for $f_B$ and $f_{B_s}$ with those from FLAG average~\cite{FlavourLatticeAveragingGroupFLAG:2024oxs}, HPQCD 09~\cite{Gamiz:2009ku}, FNAL/MILC 11~\cite{FermilabLattice:2011njy}, HPQCD 11A~\cite{McNeile:2011ng}, HPQCD 12~\cite{Na:2012kp}, RBC/UKQCD 13A~\cite{Witzel:2013sla}, RBC/UKQCD 14A~\cite{Aoki:2014nga}, RBC/UKQCD 14~\cite{Christ:2014uea}, HPQCD 13~\cite{Dowdall:2013tga}, ETM 13E~\cite{Carrasco:2013naa}, ETM 16B~\cite{ETM:2016nbo}, HPQCD 17A~\cite{Hughes:2017spc}, FNAL/MILC 17~\cite{Bazavov:2017lyh}, and Frezzotti 24~\cite{Frezzotti:2024kqk}.}
    \label{fig:comp_fPS_B_Bs}
\end{figure}

\begin{figure}
    \centering
    \includegraphics[width=0.5\linewidth]{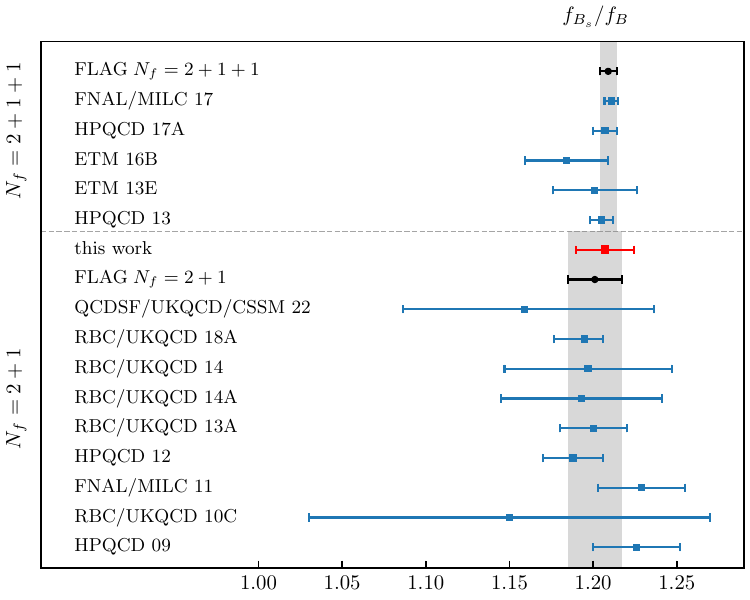}
    \caption{Comparison of our results for $f_{B_s}/f_B$ with those from FLAG average~\cite{FlavourLatticeAveragingGroupFLAG:2024oxs}, HPQCD 09~\cite{Gamiz:2009ku}, RBC/UKQCD 10C~\cite{Albertus:2010nm}, FNAL/MILC 11~\cite{FermilabLattice:2011njy}, HPQCD 12~\cite{Na:2012kp}, RBC/UKQCD 13A~\cite{Witzel:2013sla}, RBC/UKQCD 14A~\cite{Aoki:2014nga}, RBC/UKQCD 14~\cite{Christ:2014uea}, RBC/UKQCD 18A~\cite{Boyle:2018knm}, QCDSF/UKQCD/CSSM 22~\cite{Hollitt:2022exk}, HPQCD 13~\cite{Dowdall:2013tga}, ETM 13E~\cite{Carrasco:2013naa}, ETM 16B~\cite{ETM:2016nbo}, HPQCD 17A~\cite{Hughes:2017spc}, and FNAL/MILC 17~\cite{Bazavov:2017lyh}.}
    \label{fig:comp_RfBsB}
\end{figure}

\begin{figure}
    \centering
    \includegraphics[width=0.5\linewidth]{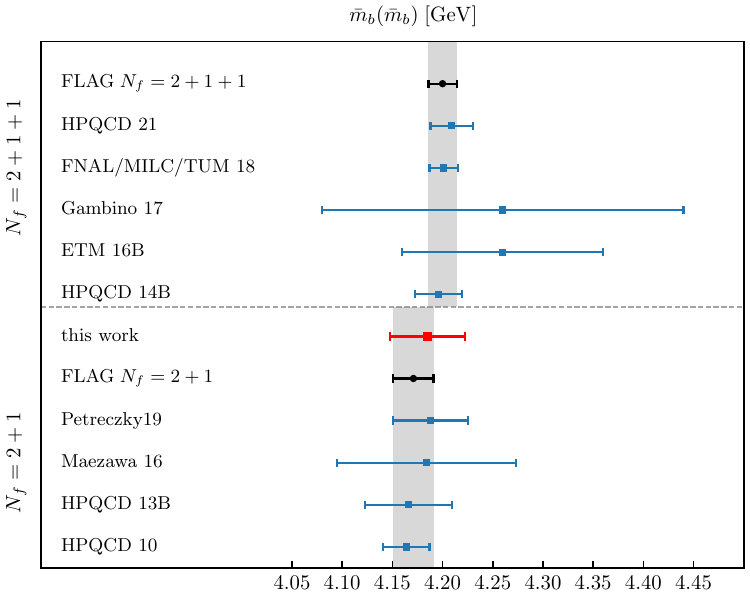}
    \caption{Comparison of our results for $\bar{m}_b(\bar{m}_b)$ with those from FLAG average~\cite{FlavourLatticeAveragingGroupFLAG:2024oxs}, HPQCD 10~\cite{McNeile:2010ji}, HPQCD 13B~\cite{Lee:2013mla}, Maezawa 16~\cite{Maezawa:2016vgv}, Petreczky19~\cite{Petreczky:2019ozv}, HPQCD 14B~\cite{Colquhoun:2014ica}, ETM 16B~\cite{ETM:2016nbo}, Gambino 17~\cite{Gambino:2017vkx}, FNAL/MILC/TUM 18~\cite{FermilabLattice:2018est}, and HPQCD 21~\cite{Hatton:2021syc}.}
    \label{fig:comp_mb}
\end{figure}

\begin{figure}
    \centering
    \includegraphics[width=0.5\linewidth]{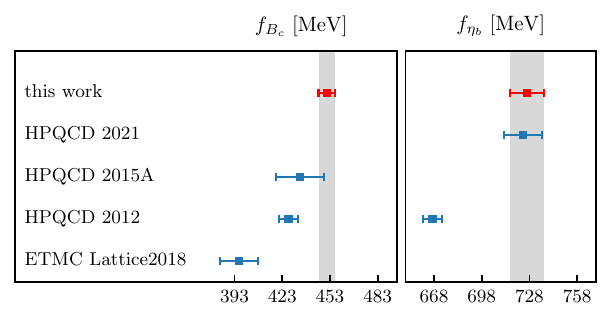}
    \caption{Comparison of our results for $f_{B_c}$ and $f_{\eta_b}$ with those from  HPQCD2021~\cite{Hatton:2021dvg}, HPQCD 2015A~\cite{Colquhoun:2015oha}, HPQCD 2012~\cite{McNeile:2012qf}, and ETMC Lattice2018~\cite{Becirevic:2018qlo}.}
    \label{fig:comp_fPS_Bc_etab}
\end{figure}

\begin{figure}
    \centering
    \includegraphics[width=1.0\linewidth]{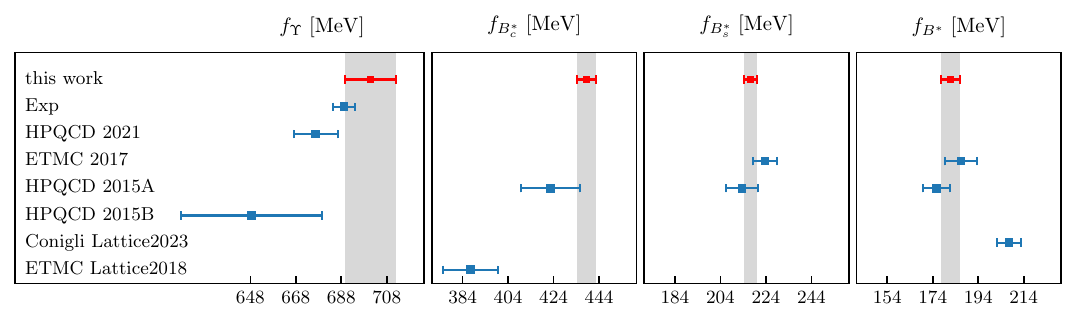}
    \caption{Comparison of our results for $f_{B^*}$, $f_{B_s^*}$, $f_{B_c^*}$, and $f_{\Upsilon}$ with those from HPQCD 2021~\cite{Hatton:2021dvg}, ETMC 2017~\cite{Lubicz:2017asp}, HPQCD 2015A~\cite{Colquhoun:2015oha}, HPQCD 2015B~\cite{Colquhoun:2014ica}, ETMC Lattice2018~\cite{Becirevic:2018qlo}, and Conigli Lattice2023~\cite{Conigli:2023rod}. The data labeled by "Exp" is estimated from the experimental value of $\Upsilon$ leptonic decay width~\cite{ParticleDataGroup:2024cfk} in Ref.~\cite{Hatton:2021dvg}. }
    \label{fig:comp_fV_B_Bs}
\end{figure}

\begin{figure}
    \centering
    \includegraphics[width=1.0\linewidth]{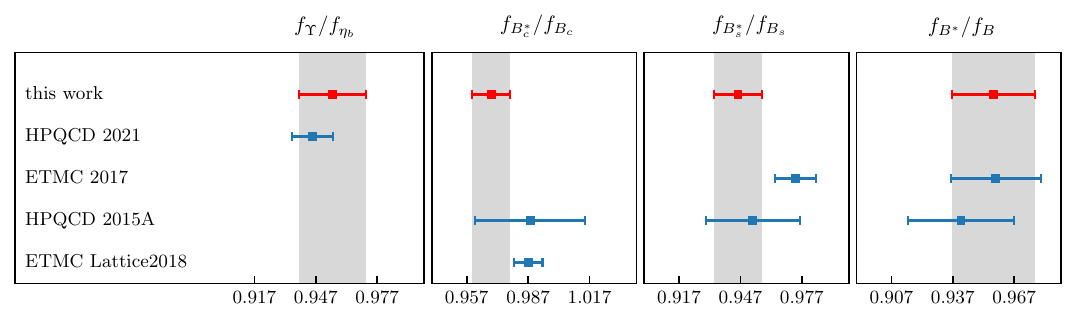}
    \caption{Comparison of our results for $f_{B^*}/f_{B}$, $f_{B_s^*}/f_{B_s}$, $f_{B_c^*}/f_{B_c}$, and $f_{\Upsilon}/f_{\eta_b}$ with those from HPQCD 2021~\cite{Hatton:2021dvg}, ETMC 2017~\cite{Lubicz:2017asp}, HPQCD 2015A~\cite{Colquhoun:2015oha}, and ETMC Lattice2018~\cite{Becirevic:2018qlo}.}
    \label{fig:comp_RfVPS}
\end{figure}

\end{widetext}



\end{document}